\documentclass[manuscript,nonacm]{acmart}

\AtBeginDocument{%
  }

\setcopyright{none}
\acmDOI{}

\usepackage{adjustbox}
\usepackage{pdfpages}

\begin{document}

\title[Theory of Developer Momentum: Progress and the Craft of Maintaining Motion]{Theory of Developer Momentum: \\The Developer's Experience of Progress and the Craft of Maintaining Motion}

\author{Arty Starr}
\email{artystarr@uvic.ca}
\orcid{0009-0007-4025-1516}
\affiliation{%
  \institution{University of Victoria}
  \city{Victoria}
  \state{BC}
  \country{Canada}
}

\author{Keith Mann}
\email{keith.mann@harness.io}
\orcid{0009-0009-4697-1471}
\affiliation{%
\institution{Harness Inc.}
  \state{Ontario}
  \country{Canada}
}

\author{Margaret-Anne Storey}
\email{mstorey@uvic.ca}
\orcid{0000-0003-2278-2536}
\affiliation{%
  \institution{University of Victoria}
  \city{Victoria}
  \state{BC}
  \country{Canada}
}

\begin{abstract}
This paper introduces a Theory of Developer Momentum that helps us see the activity of developing software in motion and the skilled craft through which progress unfolds.  We define \textit{Developer Momentum} as the developer's experience of progress and forward motion toward the goal of a task.  Our theory disentangles the related constructs of flow state, momentum, and motivation, so we can more clearly see the relationships between them and how they interweave to support progress.  In order to make progress, developers build up working context in memory, become immersed in a cognitive puzzle, and gain clarity through motion in uncertain terrain.  The theory provides researchers with a rich experiential lens to investigate how developers' cognitive experiences are shifting in emerging AI workflows.  Building on prior grounded theory work, we conducted in-depth interviews with a theoretically sampled subset of 12 professional developers, and eight feedback interviews to evaluate resonance and investigate how the theory translates to agentic AI workflows.  We followed a Constructivist Grounded Theory approach to construct a theory grounded in empirical data.  Our theory contributes a novel paradigm for seeing the motion of developing software at multiple scales and resolutions, seeing the living capacities that support this motion, and expanding what becomes available for improvement, skilled judgment, and craft.
\end{abstract}



\maketitle

\section{Introduction}

Prior research across Developer Experience, productivity, and satisfaction all highlight \textit{flow} as critical to effective software work \cite{noda2023devex, forsgren2021space, storey2019towards}. The DevEx framework highlights flow as a central underpinning that supports productivity \cite{noda2023devex}. The SPACE framework proposes a more holistic set of factors contributing to developer productivity, and similarly identifies flow as an important component  \cite{forsgren2021space}.  From Csikszentmihalyi’s research on the \textit{flow state}, which originated the concept, we know that flow also has the potential to create enjoyment and positive feelings \cite{csikszentmihalyi1990flow}. The importance of flow is widely recognized---yet, beyond giving developers large blocks of uninterrupted time to focus, there are very few ways to operationalize this insight in practice.

Flow explains immersion, but not \textit{forward progress}. It does not account for how developers get started, sustain momentum across days, or recover momentum while working on complex software tasks. Developers in our study often described their progress through a task using motion metaphors: going uphill, hitting a wall, feeling sticky and slow, and getting over the hump. The developers crafted strategies to maintain forward motion in their work, yet these practices remained largely implicit.  Our theory contributes conceptual foundations that explain the developer's experience of progress and help us see the activity of developing software in motion---surfacing practical strategies that help us recognize, reason about, and deliberately support developers in the craft through which progress unfolds.

The software industry is also undergoing a period of profound change with the widespread adoption of AI. Foundational assumptions that have held for decades---how software is constructed by writing code, and how developers come to understand that software by exploring code in IDEs---are being uprooted. While research shows that developers are perceiving productivity benefits from AI, the developer's experience may be getting worse \cite{vella2026impact}. In particular, deteriorating flow states appear to be a significant challenge in emerging AI workflows \cite{vella2026impact, beck_air_2026}. Without theoretical models that help us understand how the developer's experience is shifting, the ways we can respond are limited.

In the study presented in this paper, we construct a Theory of Developer Momentum grounded in empirical data using a Constructivist Grounded Theory (CGT) approach \cite{charmaz_constructing_2014}. In a prior study with 27 professional developers, we constructed a Theory of Troubleshooting that explains the developer's cognitive experience of overcoming confusion \cite{starr2026theory}. Building on this work, we conducted follow-up interviews with a theoretically sampled subset of 12 participants whose earlier accounts discussed experiences of flow and sustaining development work. As a central research question, we asked: "How do developers experience, build, and sustain momentum while working on complex software tasks?"  

Our findings reveal that maintaining motion in complex software work is a \textit{craft} that developers learn across years of experience.  It reflects an experiential intuition for how to navigate the work and make everyday decisions.  Developers have to clarify the goal, form a plan, identify steps, and establish feedback loops to iteratively refine a solution. \textit{\textbf{Developer Momentum} is the developer's experience of progress and forward motion toward the goal of a task.}  Our theory disentangles the related constructs—flow state, momentum, and motivation—and shows how each shapes the experience of progress.  Within the overarching theory, we construct interrelated theoretical accounts of the temporal dynamics of flow, the craft of maintaining motion, caring as the source of motivating energy, and the living capacities of the place (the habitat) needed to sustain motion.  Because the theory centers the motion of problem-solving toward a goal, the foundational constructs can translate across changing workflows.  Whether a developer is prompting and reviewing AI-generated code, orchestrating agents, or writing code by hand, there is a human trying to make progress toward a goal. Our theory provides a rich experiential lens for investigating how this experience is being reshaped.

The paper begins with related work on Developer Experience, the flow state, productivity, and forward motion. Next, we describe our methodology and the CGT approach and data analysis. We then present our Theory of Developer Momentum. The discussion explores implications for research and industry in the AI era, offers a novel paradigm that helps us see the activity of developing software at multiple scales and resolutions, and illuminates new opportunities for craft. We conclude by evaluating the theoretical contributions and outlining directions for future work.

\section{Background}

The developer's experience of progress and momentum intersects with a variety of different lenses and overlapping constructs: Developer Experience, flow, perceived productivity, motivation, and progress toward goals.  As background context, we examine the evolution and conceptualization of Developer Experience, frameworks and ways of seeing developer experience, and how the activity is shifting in the AI era. Next, we review Csikszentmihalyi's concept of flow state and flow research in software engineering. Finally, we turn to perceived productivity, developer motivation, and goal progress to establish the conceptual foundations for understanding forward motion.

\subsection{Developer Experience}

Software development is fundamentally a human activity. While software engineering has traditionally focused on languages, tools, architectures, and processes, a growing body of research recognizes that the lived experiences of developers play a central role in shaping the outcomes and long-term success of software projects \cite{hoda2021socio, noda2023devex}.

\subsubsection{Conceptualizing Developer Experience}

The strategy of understanding and improving developer experience emerged within industry practice. There were many communities arriving at a similar insight: improving the experiences of developers would lead to better outcomes on software projects. In the late 90s, human-centered programming language design was already considering the developer's experience with languages like Ruby, created in 1995 by Yukihiro "Matz" Matsumoto, where happiness while programming was an explicit design goal \cite{matsumoto2001ruby}.  In 2008, Hans Dockter, founder of Gradle, pioneered efforts to reimagine build infrastructure with a vision of reducing developer pain and creating an experience that developers would love \cite{rimple2010gradle}—a philosophy he later described as "Build Happiness" \cite{gradle_build_happiness}.  At the time, many developers were engineering complex build infrastructures using Ant that often became frustrating and too hard to maintain---motivating new innovations.  Efforts like Continuous Delivery, pioneered by Jez Humble and David Farley in 2010, inspired a vision for what reliable build and deployment infrastructure could look like \cite{humble2010continuous}. Questions about developer experience naturally arose as practitioners inquired about the experiences of other developers and designed improvements.  By 2011, "Developer Experience" had emerged as a recognizable term within industry discourse \cite{cohick_effective_2011}.  

\begin{figure}[h]
\centering
\includegraphics[scale=0.33]{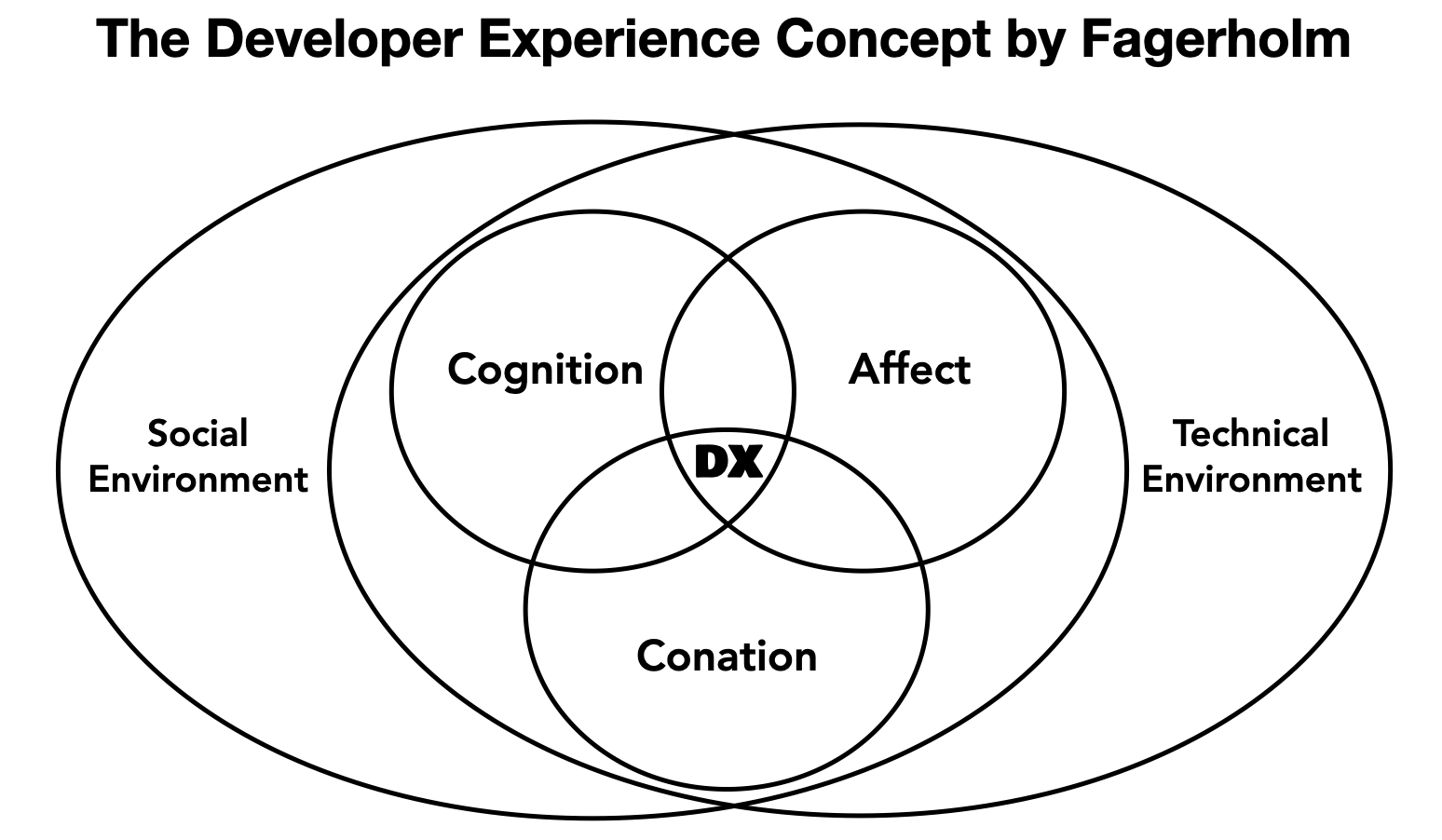}
\caption{The Developer Experience concept redrawn from Fagerholm's theory of software developer experience \cite{fagerholm2015software}}
\label{fig:fagerholm_dx}
\end{figure}

Simultaneously, researchers sought conceptual frameworks and empirical evidence for understanding developers' experiences \cite{fagerholm2015software, razzaq2024systematic} and began collaborating with industry to develop actionable frameworks that could help \cite{forsgren2021space, greiler2022actionable, noda2023devex}. The Developer Experience conversation gained significant momentum with the popular DevEx Triangle by Noda et al. which builds on the DX Framework by Greiler et al. \cite{noda2023devex, greiler2022actionable}. DevEx, as conceptualized by Noda et al., focuses on understanding the lived experience of developers and the points of friction they encounter in everyday work. The DX Framework defines Developer Experience as "how developers think about, feel about, and value their work" \cite{greiler2022actionable}.  This definition is inspired by the earlier work of Fagerholm and Münch, which conceptualizes Developer Experience as an interplay of cognitive (thinking, decision-making), affective (feeling, emotion), and conative (striving, volition) dimensions of human experience within a social and technical environment \cite{fagerholm2012developer}. This perspective aligns with a broader body of research that frames software development as an inherently sociotechnical activity \cite{hoda2021socio, baxter2011socio, storey2020software}. This conceptualization is presented more fully in Fagerholm's doctoral thesis \cite{fagerholm2015software} and is redrawn in Figure~\ref{fig:fagerholm_dx}.

Fagerholm developed a theory of software developer experience that centers the \textit{activity} of developing software \cite{fagerholm2015software}.
The theory highlights that the activity should be considered a design process rather than a manufacturing process. Unlike manufacturing, the decision-making in a software design process requires collaboration across many stakeholders and brings the full complexity of human social interaction into the work \cite{fagerholm2015software}.  Fagerholm and Münch also distinguish between the verb \textit{experiencing}—a person’s stream of perceptions, interpretations, and resulting emotions that occur during interaction—and the noun \textit{experience}, a person’s encounter with the system that has a beginning and an end \cite{fagerholm2012developer}.  Fagerholm's conceptualization of the activity---and emphasis on how developers are experiencing that activity throughout his thesis---establishes a verb-oriented lens for understanding Developer Experience. Building on this conceptualization, our theory centers the activity of developing software.  We use the term \textit{cognitive experience} to refer to the developer’s internal experience of thinking, feeling, and striving over time as an integrated whole.  

\subsubsection{Frameworks for Understanding What Matters}

As Developer Experience research matured, research and industry collaborations led to new actionable frameworks to help organizations make improvements.  The SPACE framework highlights what matters most to productivity across five complementary dimensions: \textbf{S}atisfaction and well-being, \textbf{P}erformance, \textbf{A}ctivity, \textbf{C}ommunication and collaboration, and \textbf{E}fficiency and flow \cite{forsgren2021space}.  The DevEx Triangle sees Developer Experience through three core dimensions: feedback loops, cognitive load, and flow state \cite{noda2023devex}. Developer Productivity Engineering (DPE), pioneered by Gradle, includes three pillars for improving Developer Experience: Faster Feedback Cycles, Efficient Failure Troubleshooting, and Reliable Builds \& Tests \cite{peerless_making_2023}.

\subsubsection{Ways of Seeing Developer Experience}

In his keynote at the first DPE Summit in 2022, Dockter highlighted a core belief of “joy == productivity," expressing how joy and productivity are in alignment with one another: with smart investments in engineering capabilities, we can unlock innovation, productivity, and joyful developer experiences at the same time \cite{gradle_dpe_2022}. He also highlighted the challenges that make it difficult to justify improvements in Developer Experience \cite{gradle_dpe_2022}.  He described a persistent communication breakdown: developers often have a strong intuitive sense of when their tools, systems, or processes are creating drag, but struggle to explain this “gut feel" to managers who do not share the lived experience of writing code. He argued that the software industry urgently needs better conceptual foundations that could make the developers' experiences more legible to others.  This gap motivated our decision to construct a Theory of Troubleshooting that could explain the developer's cognitive experience of overcoming confusion and how a loss of understanding led to a loss of control \cite{starr2026theory}.  Our Theory of Developer Momentum contributes another way of seeing: the developer's experience of progress and the craft of maintaining motion.

Developer Experience has increasingly been understood through multiple conceptual lenses, each illuminating different aspects of the activity of developing software.  Some of these lenses are implicit—for example, the longstanding influence of manufacturing metaphors—while others have been proposed explicitly through theory and conceptual work. Table~\ref{tab:lenses} summarizes conceptual lenses that shape our understanding of Developer Experience.

\begin{table}[t]
\centering
\caption{Conceptual lenses that shape our understanding of Developer Experience.}
\label{tab:lenses}
\begin{tabular}{|p{0.28\linewidth}|p{0.62\linewidth}|}
\hline
\textbf{Conceptual Lens} & \textbf{Helps us see...} \\
\hline

Manufacturing (implicit) &
The outputs of a production and delivery pipeline, emphasizing building things, speed, efficiency, reliability, utilization, and output.
\\
\hline

Activity (Fagerholm) &
The way developers are experiencing the activity of developing software as a design process involving the collaboration of many stakeholders rather than a manufacturing process. The experience includes cognitive, affective, and conative dimensions, social and technical environments \cite{fagerholm2015software}.
\\
\hline

Developer Happiness Principle (joy==productivity, DPE; Build Happiness, Gradle; Programmer Happiness, Ruby )&
Developer happiness as a first principle for human-centered design, suggesting that designing engineering tools to make the developer's experience a happy one will naturally support productivity.
\\
\hline

Flow Theory (Csikszentmihalyi) &
The experience of immersion, focused attention, uninterrupted time to "get into the zone", challenge-skill balance, immediate feedback, intrinsic motivation, and optimal conditions for joyful, rewarding experiences \cite{csikszentmihalyi1990flow}.
\\
\hline

Cognitive Load Theory (Sweller) &
The experience of increasing mental effort, working memory load, selective attention and processing capacity constraints during problem-solving, the role of schema construction in acquiring problem-solving skills, and the three types of cognitive load (intrinsic, extraneous, and germane) \cite{sweller1988cognitive}.
\\
\hline

Habitat Thinking (Miles) &
The whole place---the people, the agents, the relationships between them, code, conventions, and feedback loops they share---as a habitat, and asks the oldest question habitability ever posed: is this a good place to live \cite{miles_habitat_manifesto}?

\\
\hline

Triple Debt Model (Storey)     

extended by Starr 
&
Three team-level properties of software health: cognitive debt, intent debt, and technical debt. In the humans, in the artifacts, in the code. With relationships and interactions that reinforce one another \cite{storey_technical_2026}.  

"Software as an Activity" (Starr) extends this model with a paradigm shift that asks, how are the three types of knowledge (cognitive, intent, and technical) supporting the activity of developing software   \cite{starr_activity_2026}?
\\
\hline

Theory of Troubleshooting

(Starr and Storey)
&
The experience of puzzling events, overcoming confusion, hunches and experiential intuition, poking and seeing, figuring it out, cognitive fatigue, and the loss of understanding that leads to a loss of control \cite{starr2026theory}.
\\
\hline

Theory of Developer Momentum 

(this work)
&
The experience of progress toward the fuzzy goal of a task, gaining clarity, getting momentum going, and the craft of maintaining motion across task sessions or days—the activity of developing software in motion.
\\
\hline

\end{tabular}
\end{table}

\subsubsection{The Landscape of Developer Experience in the AI Era}

The major shifts in the AI era can be seen in the language researchers use to describe their impact on software engineering: reshaping developer experience \cite{houck_space_of_ai}, fundamentally altering interactions with IDEs \cite{sergeyuk_humanai_2026}, fundamentally changing how work gets done \cite{houck2026engthrive}, reshaping the sequence, depth, and regulation of problem-solving activities \cite{anderson_chatgpt_2026}, redistributing the challenges of software engineering \cite{storey_technical_2026}. These are fundamental shifts in the activity of developing software.

We bring another concept into this discussion: \textit{craft}. When the fundamental shape of an activity is changing, people need new ways of orienting themselves to the craft in that activity.  In his book, \textit{The Sovereign Engineer: AI Literacy for Software Professionals}, Miles described the transition to AI-assisted software development as a period of profound disorientation for many senior engineers that froze at the sight of a chatbot writing code \cite{miles_sovereign_2026}. Some dismissed AI as "just autocomplete," while others questioned the future of their careers.  Miles identified a primary reason for the panic—not having a \textit{framework for thinking} about what was happening.  He also described a craft and orientation of teams that have been successful with AI. These teams understand that artificial intelligence (AI) and human intelligence (HI) are two different types of cognition, not the same thing at different speeds. They deliberately design the working environment in which they collaborate with AI: the context they provide, the architectural boundaries they enforce, the feedback loops that catch drift before it compounds. The collaboration space itself becomes an engineering artifact worthy of the same care as the code. \textit{The craft of developing software has not gone away---it has just changed shape.}

One of the most significant shifts is that developers can now generate large amounts of code very quickly with AI, without the corresponding process of a human gaining an understanding of it. The activity of \textbf{creating code and understanding the code are no longer naturally coupled}. This missing understanding can immediately become consequential when the developer needs to troubleshoot the generated code and is missing an intuitive hunch about what is wrong and where to look \cite{starr2026theory}.  In a lab study by Vaithilingam et al. (n=24) comparing completion of a task with and without GitHub Copilot, participants faced difficulties in understanding, editing, and debugging code generated by Copilot, which significantly hindered their effectiveness \cite{vaithilingam_expectation_2022}.  All 12 of the developers who attempted to repair the generated code reported finding it more difficult than code they wrote themselves. Three of the developers were unable to complete their task due to a time-consuming debugging effort \cite{vaithilingam_expectation_2022}.  While developers can read the generated code as a means to understand it, these difficulties suggest this missing understanding is not that easy to overcome.  This phenomenon of missing understanding has been identified as \textit{cognitive debt} \cite{storey_technical_2026}.  As the cognitive debt accumulates over time and across the team, the mental models that developers rely on become increasingly inadequate, making it difficult for the developers to reason about the software and change it safely \cite{storey_technical_2026}.

In an in-depth qualitative lab study, Anderson et al. investigated differences in problem-solving approaches between coding with AI assistance and coding by hand \cite{anderson_chatgpt_2026}. They used Polya's problem-solving phases as a lens to better see the temporal unfolding of the developer's process: (1) understanding the problem, (2) devising a plan, (3) carrying out the plan, and (4) looking back to evaluate the results \cite{polya1945solve}. They identified several new patterns in the ways developers were getting stuck and captured the stuck moments as stories \cite{anderson_chatgpt_2026}.  For example, when the AI has a flawed understanding and is confidently wrong about something, it can be difficult for the developer to challenge the AI's direction, and then they get stuck. These simple stories are a powerful tool that helps illuminate what this new developer experience is like.

As these examples illustrate, the outcomes depend on more than the capabilities of the human or AI; they are also shaped by the relationships and environment through which they work together.  Miles describes this broader orientation toward designing the whole place where collaboration happens as \textit{Habitat Thinking} \cite{miles_sovereign_2026}.  The habitat includes the people, the agents, the relationships between them, the code, conventions, and feedback loops they share, inviting us to design a place where all the different types of cognition in the room can think well and thrive together.  Rather than seeing the sociotechnical system from the outside, Habitat Thinking invites us to see the place we are inhabiting from the inside and asks: \textit{"is this a good place to live?"}  \cite{miles_habitat_manifesto}  Because we live there as inhabitants, we are designing the place in which we exercise our own cognition and craft---supporting and empowering ourselves to do our best work.

\subsection{The Flow State}

What is happening in those unusually good moments of human experience when we become deeply engaged in something for its own sake?  This question was at the center of Mihaly Csikszentmihalyi's research program on optimal experience, through which he developed the concept of the \textit{flow state}---where a person experiences enjoyment and joy \cite{csikszentmihalyi1990flow}.  In \textit{Flow: The Psychology of Optimal Experience}, he provides a qualitative description of the flow state and the conditions that enable it.  Later work has operationalized the flow state in different ways, distinguishing various sets of dimensions, characteristics, and enabling conditions \cite{rissler2018got}. Rather than center these abstractions, we characterize the richer qualitative account of the flow state experience in his book.

\subsubsection{Characteristics of the Flow State}

One of the most salient qualities of the flow state is the experience of being fully absorbed in an activity. Csikszentmihalyi characterizes this as a \textit{merging of action and awareness}, when a person becomes so involved in what they are doing that they stop being aware of themselves, the person's attention becomes completely absorbed, and their actions spontaneous and almost automatic \cite{csikszentmihalyi1990flow}.  The flow state concept centers an "optimal peak state" where several experiential qualities and enabling conditions come together.  At the center of this is an alignment of the goal of the activity with intrinsic motivation. Csikszentmihalyi also characterizes the flow state as an \textit{autotelic experience}---where a person is doing the activity for its own sake (enjoying), rather than for an outcome of it. The flow state occurs within a goal-directed context with immediate feedback that is clear and unambiguous.

The activity needs to also be challenging. When all the person's relevant skills are needed to overcome a challenging situation, the person's attention becomes fully absorbed and engaged.  In the optimal peak flow state, the body or mind is stretched to its limit as the person tries to accomplish something both difficult and worthwhile. The person must also perceive that they have a chance of completing the task, and be free to concentrate on what they are doing in order to get into a flow state \cite{csikszentmihalyi1990flow}.  Once they enter this state, they act with a deep but effortless involvement, feel in control and fully engaged by the activity, and worries and frustrations fade from awareness.  The person's sense of time also changes---where hours can pass by without notice, or a moment can expand to feel like a long time.

\subsubsection{Deep Engagement with a Hard Challenge}

The flow state describes a deeply engaged mode of experience in which a person voluntarily directs their capacities toward a meaningful, goal-directed and hard challenge \cite{csikszentmihalyi1990flow}.  The challenge calls upon their relevant skills, their attention becomes fully invested, and the activity feels meaningful and satisfying to accomplish. Importantly, Csikszentmihalyi's notion of enjoyment does not imply that every moment of the activity is pleasurable or easy. He describes how even moments of intense exertion and difficulty can be remembered among the most meaningful and enjoyable experiences of a person's life:

\begin{quote}
\textit{“The swimmer’s muscles might have ached during his most memorable race, his lungs might have felt like exploding, and he might have been dizzy with fatigue—yet these could have been the best moments of his life.”}
\end{quote}

The challenge must meet the person at the edge of their capacities.  If the challenge is perceived as too difficult relative to the person's skills, they can feel anxious; if the challenge is too easy, they can feel bored \cite{csikszentmihalyi1990flow}.  As the person gains skill, what was once difficult becomes easier, requiring a more difficult challenge to return to a deeply engaged state of flow. Flow-producing activity repeatedly stretches the person's skills through increasingly difficult challenges.  Csikszentmihalyi connects the experience of enjoyment with forward movement, novelty, and accomplishment, and over the long run, a sense of mastery---describing the person as becoming more complex through these experiences.

\subsubsection{Flow in Software Engineering}

Brown et al. investigated the flow experiences of developers with activity log data combined with a diary study and found that flow did not only happen while writing code, but developers experienced flow during a variety of activities, such as responding to emails, drafting design docs, or reading documentation \cite{brown_using_2023}.  Once flow is established, they found flow to be more resilient to interruptions.

Müller and Fritz explored developers' experiences of flow in a lab study where 17 developers wore biometric sensors while coding and periodically assessed their emotions and perceived progress \cite{muller2015stuck}.  Developers frequently commented on emotions and progress together, suggesting a strong connection between affective state and perceived progress. The developers also distinguished between feeling “completely stuck” and feeling “less stuck,” and most reported less than maximum flow, indicating that stuckness and flow are not binary but exist on a continuum. We extend Csikszentmihalyi's Flow Theory by providing greater resolution into how developers transition into flow, experience variations in the flow state, and how these dynamics relate to forward progress. We return to more recent work on the flow state in an AI-assisted development context in Discussion in light of our findings.

\subsection{Forward Motion: Productivity, Motivation, and Progress Toward Goals}

Productivity has long been the central lens for “getting stuff done” in software development, providing a way of conceptualizing progress toward desired outcomes. Traditional measures of productivity have often focused on observable outputs of developer activity, such as lines of code, commits, pull requests, or other artifacts of production. Yet developing software involves many contributions that may not “look like” productivity. In this section, we focus on existing conceptualizations closely related to our concept of progress.

\subsubsection{Perceived Productivity}

Storey et al. highlighted two examples of valuable contributions that the productivity lens makes it hard to see: Mentoring a newcomer may reduce visible productivity for an individual while increasing the long-term productivity of the team.  Learning a new skill may not produce visible output, but contributes to software quality over a longer period of time \cite{storey2019towards}.  What constitutes a meaningful contribution is broader than what can readily be observed through artifacts of production.  These challenges motivated researchers to consider \textit{perceived productivity} as another strategy for understanding productivity beyond its visible artifacts \cite{storey2019towards}.

Meyer et al. investigated when developers perceive themselves as productive versus unproductive \cite{meyer2014software}. The most common reasons developers gave for a "productive workday" included completing tasks or goals, few interruptions or distractions, no meetings, clear goals and requirements, and planning their workday. Many activities, such as meetings, had mixed responses as to whether they were considered productive or unproductive.  Meetings were productive when they included decision making, had clear focus and goals, improved relationships between developers, or helped others.  These examples highlight the range of activities that developers perceive as contributing productively to their work.

\subsubsection{Motivation Toward a Goal}

Motivation has also been identified as an important influence on developer productivity. Beecham et al.'s literature review identified motivation as having the single largest impact on practitioner productivity while also being difficult to manage \cite{beecham2008motivation}. They identified aspects of the work that motivated developers: problem solving, challenge, experimentation, teamwork, and creating something that benefits others \cite{beecham2008motivation}.  Franca et al. highlight that most results from surveys on developer motivation point to the software engineering job itself as the main motivator \cite{francca2012towards}.  

Sharp et al. describe motivation as the \textit{initiation, direction, intensity and persistence of behavior} \cite{sharp2009models}.  Franca et al. use a definition inspired by Steers et al., characterizing motivation as the \textit{set of factors or events that energize, channel, and sustain human behavior over time} \cite{francca2012towards}.  Franca et al. also conducted a grounded theory study of motivation asking developers, managers, and directors about their perceptions and meaning of the concept \cite{francca2012towards}.  Three participants agreed that motivation is \textit{a strong internal feeling of wellbeing or contentment for doing good work}.  Eight participants defined motivation as \textit{an individual's willingness to apply effort or keep high interest for the work}.  The varied definitions help us conceptualize motivation as both a felt experience and an energetic willingness to act and apply effort toward a goal.

\subsubsection{Conceptualizing Goal Progress}

The goal-directed nature of motivation connects with Carver and Scheier's Control Theory, which conceptualizes the human notion of progress toward a goal within the context of a self-regulating feedback system \cite{carver1982control}. Intentional behavior is seen as a feedback process in which a perceived present condition is compared against a desired reference, and the behavior acts to reduce the discrepancy between them.  As the present condition changes, it is perceived and compared again. In later work, Carver and Scheier describe the \textit{rate of progress} toward a goal as the rate at which discrepancy is reduced over time \cite{carver1990origins}. When discrepancy is reduced rapidly, progress toward the goal is perceived as high; when it is reduced slowly, progress is lower; and when discrepancy is not reduced, progress is zero. They draw an analogy between this representation of progress and motion: if the action-guiding system represents distance and direction toward a goal, its rate of progress represents the psychological equivalent of velocity \cite{carver1990origins}.  In other words, the human experiences themselves as moving—or not moving—toward the goal of the task.  This fits our conceptualization of \textit{momentum}: the developer's experience of progress toward the goal of a task.

\section{Methodology} \label{methodology}

To design the study, we followed Charmaz's philosophy and approach to CGT \cite{charmaz_constructing_2014} to develop an empirically-grounded theory. CGT provides an iterative method of data collection and theoretical sampling, with initial grounded coding, constant comparison, and memo writing used to achieve theoretical saturation and construct a grounded theory. The study was carried out with the approval of the Ethics Committee at University of Victoria and with the informed consent of participants (Ethics Certificate \#25-0141). The primary question in a grounded theory study is, ``what is happening in the scene?'' The researcher interprets what is happening, and through coding makes the implicit processes and structures visible \cite{charmaz_constructing_2014}. Rather than treating the researcher as an objective, neutral observer, the constructivist philosophy recognizes the researcher's perspective, privileges, and interactions as part of the research context \cite{charmaz_constructing_2014}.

\subsection{Perspective of the Researchers} \label{perspective}

Charmaz discusses the influence of the researcher when conducting interviews, and highlights that the way participants identify the researcher influences what they will share \cite{charmaz_constructing_2014}.  The interviews and primary data analysis were all done by the first author as part of a PhD dissertation. The second and third authors met with the first author regularly to review and discuss the emerging theoretical categories and data analysis.

The first author comes with over 20 years of experience as a software engineer, is an industry conference and keynote speaker \cite{vmware_tanzu_arty_2021}, and author of the book, \textit{Idea Flow} \cite{starr_idea_2012}.  This influential position allowed us to recruit many senior software engineers with substantial experience, build a peer-level rapport as a fellow software engineer, and construct models that reflect an ``insider's view'' of the world of developers.  Our momentum study leverages the long-term relationships and deep engagement with participants through our studies from 2023 to 2026. The author's book, \textit{Idea Flow}, discusses flow, and her developer tool, FlowInsight, includes metrics for momentum \cite{flowinsight}, introducing potential biases that informed careful design of the study. The author's perspective has also been shaped as a research assistant and PhD student at CHISEL (Computer Human Interaction and Software Engineering Lab) at University of Victoria.

The second author is Head of Research at Harness, with more than twenty-five years as a software architect and engineer followed by a decade as an industry analyst covering software delivery and DevSecOps across many engineering organizations. His research centers the transmission of programming theory and cognitive costs of AI-assisted development. His current role at a vendor of AI-assisted software delivery tools provides close contact with industry practice and represents a commercial interest and potential source of bias.

The third author is a research professor at the University of Victoria studying human-centered software engineering, with two decades of empirical and collaborative research with software professionals across academia and industry.  She brings a lens informed by developer experience, socio-technical systems, and a commitment to understanding the cognitive and collaborative challenges developers face.  Her ongoing collaborations with industry partners have deepened her insights into the practical realities of developer work, especially in the context of AI-assisted workflows.

Most data analysis decisions were made by the first author with input from the second and third authors. As CGT is inherently interpretive, different researchers reviewing the same interviews in a different sequence and noticing different patterns would likely have constructed a different theory.  We adopted a posture of ``strong beliefs, loosely held,'' and used prepared semi-structured interview questions to reduce researcher influence on participant responses.  Feedback interviews provided an additional layer of validation and iterative refinement of our theory.  While these protocols mitigate bias, they may not fully eliminate it. CGT does not call for agreement sessions commonly used in some qualitative traditions; instead, deep solo engagement is typical in CGT, where the researcher's interpretive role, sustained immersion, and reflexive memoing are central to theory development.

\subsection{Recruiting Participants} \label{recruiting}

The initial pool of 27 participants from our troubleshooting study was recruited using convenience sampling. Eight participants were recruited through an invitation on Mastodon, with 19 recruited through targeted LinkedIn searches to increase diversity. We searched for developers with five or more years of experience, primarily in the United States and Canada with some in Europe, striving for balance across front-end, backend, and tools developers, and ensuring representation of women and non-binary developers. The 27 participants represented 570 total years of experience, ranging from five to 42 years, with most having over a decade of experience. From this pool, we theoretically sampled 12 participants who had previously discussed flow and sustaining development work. We searched the troubleshooting transcripts for "flow", "interruptions", and "context", 
then manually reviewed the interviews for relevance, prioritized developers who explicitly discussed flow, and ensured our sample included representation across gender and roles.

Interviews and analysis proceeded iteratively across four rounds of 5, 2, 3, and 2 participants, alternating interviews with initial grounded coding and memo writing. Of 15 participants invited, 12 completed an interview. To evaluate our emerging theory, we invited 9 participants to feedback interviews, with 8 completing an interview.  Table~\ref{tab:participant-demographics} summarizes their demographics, with additional detail in the Appendix. Seven participants were from the US, four from Canada, and one from Europe. No compensation was offered to participants.

\begin{table}[h]
\centering
\renewcommand{\arraystretch}{1.3} 
\setlength{\tabcolsep}{4.5pt}     
\small
\begin{tabular}{|l|c|l|c|l|c|l|c|l|c|l|c|}
\hline
\textbf{Experience} & \textbf{N} &
\textbf{Education} & \textbf{N} &
\textbf{Company Size} & \textbf{N} &
\textbf{Team Size} & \textbf{N} &
\textbf{Gender*} & \textbf{N} &
\textbf{Work Env.} & \textbf{N} \\
\hline
$\geq$ 20 yrs & 7 & Post-graduate & 2 & > 1000 people & 6 & > 10 people & 1 & Woman & 4 & Remote & 10 \\
13–19 yrs   & 3  & Bachelor’s & 9 & 51–1000 people & 3 & 6–10 people & 6 & Man   & 8 & Co-located & 1 \\
10-12 yrs    & 2  & Some college & 1 & $\leq$ 50 people & 3 & $\leq$  5 people & 5 &   &   & Hybrid & 1 \\
\hline
\end{tabular}
\caption{Summary of participant demographics (N = 12). *Gender based on LinkedIn profiles; not self-reported.}
\label{tab:participant-demographics}
\end{table}
 \vspace{-3em} 
 
\subsection{Developer Interview Process} \label{interviews}

We designed a semi-structured, 60-minute interview protocol with open-ended questions. The questions followed a four-part structure, where we asked the developers about 1) getting started first thing in the morning, 2) getting momentum going with coding, 3) the impact of distractions, interruptions, or meetings, and 4) sustaining momentum across multiple days.  All interviews were conducted by the first author over Zoom, with participants encouraged to speak in ``familiar developer terms'' given the interviewer's software engineering background.

We followed prepared questions to reduce the likelihood of leading participants, and we occasionally asked clarifying follow-ups, taking care to avoid suggestive phrasing. For example, although we had our own understanding of momentum, we did not give a definition to participants, but instead asked developers about their experience of getting momentum going, and let them define their own meanings.  Similarly, we did not assume that developers knew Csikszentmihalyi's definition of flow, but rather asked developers how absorbed they were or how much they were enjoying a task, and let developers describe their experiences in their own words. We structured our interviews around a consistent prompt, asking participants to walk through a recent experience of getting momentum going on a task, and how they sustained momentum across days. This approach enabled us to capture diverse instantiations of the phenomena of interest across a wide range of developer contexts. 

After analyzing the first seven interviews, a theory of developer momentum was beginning to emerge. We revised our interview protocol to add questions that helped fill in missing details, and removed questions that were no longer yielding new information. We used the revised protocol for the remaining five interviews, at which point we had sufficient data to support theoretical saturation.  We discuss this process of adjusting interview questions  in Section \ref{category_sort}.  Both versions of the interview protocol are provided in the Appendix. To evaluate resonance and mitigate the risk of incomplete theoretical development, we conducted an additional eight feedback interviews within the same set of 12 developers.  Because the initial interviews occurred in 2025 and feedback interviews in 2026, when many participants were working with agentic AI, these interviews also helped us investigate how the theory translated across changing workflows. We discuss the feedback interviews in Section~\ref{followups}.

\subsection{Data Analysis} \label{data-analysis}

\begin{figure*}[h]
\centering
\includegraphics[scale=0.40]{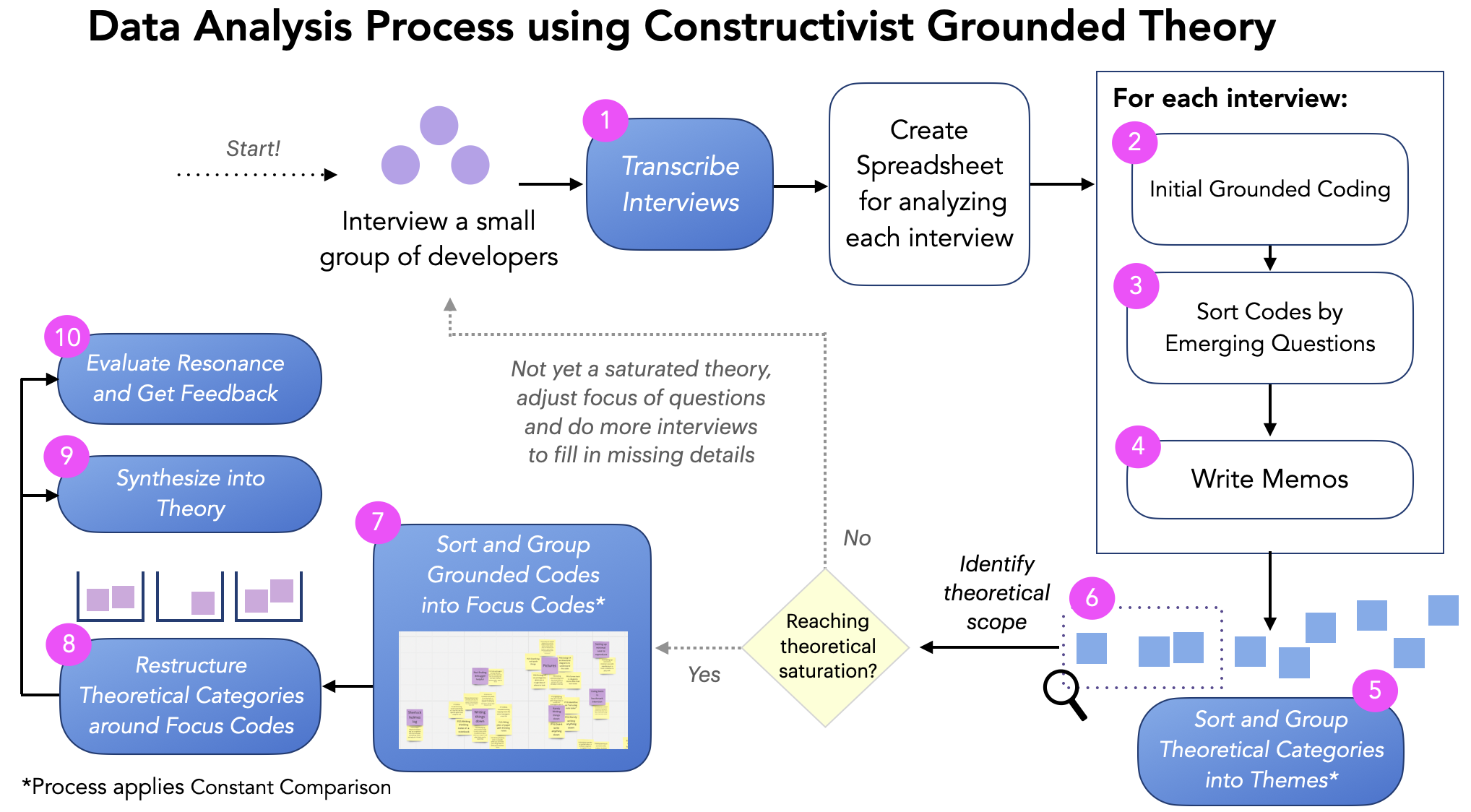}
\caption{Overview of the data analysis process used in applying Constructivist Grounded Theory}
\label{fig:data_analysis}
\end{figure*}

We followed CGT methods from Charmaz \cite{charmaz_constructing_2014}, iteratively conducting rounds of interviews interleaved with data analysis. 
In this section, we walk through the data analysis process as illustrated in Figure~\ref{fig:data_analysis}, and demonstrate how we applied the CGT research method to analyze the data.

\subsubsection{Transcribing Interviews}

We recorded the Zoom interviews on video and transcribed the recordings locally using MacWhisper 4 with OpenAI's Whisper Medium (English) model.  As we watched the videos, we fixed any errors we found in the transcripts.  The resulting line-by-line transcripts (Step 1 in Figure~\ref{fig:data_analysis}) were loaded into a spreadsheet with additional columns to track the initial grounded codes.

\subsubsection{Initial Grounded Coding}

To create the initial grounded codes (Step 2 in Figure~\ref{fig:data_analysis}), we watched the video while reading the transcript, bolding key phrases to aid comprehension and make the transcript easier to scan during repeated analysis. We then coded line-by-line with high-resolution descriptive codes of what was happening. Following Charmaz, we began codes with gerunds (-ing words) to focus attention on actions and processes \cite{charmaz_constructing_2014}. Where gerunds felt awkward or we needed to describe something other than an action or process, we pragmatically designed codes that made sense for the context. We also allowed multiple codes per line to capture the complexity of what was happening. An example of the initial grounded coding is provided in Figure~\ref{fig:grounded_coding} in the Appendix.

Because our aim was to understand micro-level shifts in the developer's cognitive experience, we intentionally used high-resolution descriptive codes to preserve fidelity to the lived experience. Unlike shorter codes more typical in grounded theory, our initial grounded codes were often longer and more detailed, functioning as atomic units of meaning that could be understood independently of the original transcript context.  This yielded 1469 initial grounded codes across the 12 interviews, most of which were unique. This high-resolution approach enabled fine-grained analysis during constant comparison to support theory development. Later in the process, we sorted and grouped similar initial grounded codes into focus codes to raise the level of abstraction.

\subsubsection{Sorting Codes by Emerging Questions}

As we reviewed and compared the initial grounded codes, we identified 12 emerging questions arising from the data. For each interview, we sorted the codes by the emerging questions they helped answer (Step 3 in Figure~\ref{fig:data_analysis}), filtering codes that did not contribute to any question, and duplicating codes that contributed to multiple questions. As new questions arose, we added them and re-analyzed prior interviews as needed for consistency. Sorting by emerging questions enabled us to compare related sets of codes on Miro boards (Section~\ref{sorting}) and narrow our theoretical scope for later analysis (Section~\ref{narrowing}). The emerging questions are available in the Appendix.

\subsubsection{Writing Memos for Emerging Theoretical Categories}

As theoretical categories emerged throughout the analysis process, we created memos for each theoretical category noting relevant insights from the developers along with our thoughts (Step 4 in Figure \ref{fig:data_analysis}). We created memos while listening to interviews, during initial grounded coding, and during constant comparison—the moment a key concept emerged.  We continued adding to these memos whenever the same concept emerged in subsequent interviews, helping us connect different perspectives across developers.

\subsubsection{Sorting and Grouping Theoretical Categories into Themes} \label{category_sort}

After the first seven interviews, we had 98 theoretical categories spanning a wide range of topics and needed to narrow our theoretical focus.  We created a Miro board with each theoretical category mapped to a sticky note, then sorted and grouped them into 13 themes (Step 5 in Figure \ref{fig:data_analysis}). Sorting the theoretical categories illuminated a temporal model of the developer's experience of a task from start to finish.  This temporal perspective became a central pillar of the emerging theory and helped reveal gaps where evidence was still insufficient.  In light of these gaps, we revised the interview protocol to explore how advancing through different stages of a task affected the developer's experience and strategies they used, and how they conceptualized momentum.  The adjustments to our interview protocol helped us fill in missing details and more rapidly reach a saturated theory.  The revised interview protocol, and a link to the Miro board can be found in the Appendix.

\subsubsection{Identifying and Narrowing Theoretical Scope} \label{narrowing}

As we conducted additional interviews with increasingly focused questions, the structure and scope of the theory became clearer. After 12 interviews, we narrowed our theoretical scope to three interrelated concepts: momentum, flow, and motivation, along with practical strategies developers used for maintaining motion. While AI use was emerging among five developers in our study, we additionally considered how developers were using AI to support momentum. The narrowed scope allowed for a more detailed analysis of five of the 12 emerging questions. For these five areas, we had 123 initial grounded codes related to the developer's notion of momentum, 207 related to flow, 156 related to motivation, 341 related to developers' strategies for maintaining motion, and 37 related to how developers were using AI to support momentum. While the AI-related data was more limited, it provided early insights into how these practices were evolving. When no new theoretical categories were emerging within the core scope of momentum, flow, motivation, and strategies for maintaining motion, and the theory had stabilized with substantial supporting data, we determined that we had reached theoretical saturation. While further interviews might reveal variations of existing themes or new theoretical categories outside this scope, the emerging theory was well-grounded in the data.

\subsubsection{Sorting and Grouping Grounded Codes into Focus Codes} \label{sorting}

To support further analysis, we loaded the coded interview transcripts into a Postgres database and generated 12 reports of initial grounded codes organized by emerging question with corresponding participant numbers.  For the five emerging questions selected for detailed analysis, we imported the report rows into Miro as sticky notes (Step 7 in Figure~\ref{fig:data_analysis}).   We then compared the sticky notes and sorted and grouped them around focus codes that represented the common idea or concept.  We refined the groups until the focus codes were clearly named and the grounded codes within each group had good cohesion. This led to 58 focus codes related to the notion of momentum, 76 related to flow, 86 related to motivation, 167 related to developers' strategies for maintaining motion, and 21 related to AI strategies.  Links to the Miro boards and a data archive that includes the 12 emerging question reports (2059 codes) can be found in the Appendix.

\subsubsection{Restructuring Theoretical Categories around Focus Codes}

At this point, we had both a bottom-up process of sorting initial grounded codes into focus codes and a top-down process of theoretical categories emerging through memo writing. While each illuminated the emerging theory, the two structures did not yet connect. Further, the emerging theory of developer momentum was at a higher level of abstraction than our existing theoretical categories, leading us to restructure new theoretical categories around our focus codes (Step 8 in Figure~\ref{fig:data_analysis}).  The emerging theory increasingly reflected the structure of our core emerging questions.  For each corresponding Miro board, the research team discussed the stickies, and highlighted focus codes and initial grounded codes that illuminated key theoretical insights.  We then created a new set of theoretical category memos corresponding to the emerging structure, and populated each memo with relevant focus codes, initial grounded codes, interview quotes, and our own reflections on patterns in the data.  These data-rich theoretical category memos became the basis for grounding our theory.

\subsubsection{Synthesizing into a Grounded Theory}

To synthesize the restructured theoretical categories into a grounded theory (Step 9 in Figure~\ref{fig:data_analysis}), we leveraged the writing process as suggested by Charmaz \cite{charmaz_constructing_2014}. We explained the findings within each theoretical category, created models and diagrams to describe the phenomena, then integrated them into a cohesive theory. Throughout the writing process, we reviewed the data in the restructured theoretical category memos, revisited the transcripts and Miro boards, and considered our research question as we synthesized the ideas.  While all the initial grounded coding and constant comparison were done by hand, we had many conversations with ChatGPT during this stage to explore connections and collaborate in the synthesis and theorizing.

\subsubsection{Evaluating Resonance and Getting Feedback} \label{followups}

To evaluate how our emerging theory resonated with developers and get feedback (Step 10 in Figure~\ref{fig:data_analysis}), we conducted eight feedback interviews over Zoom. Following Charmaz's guidance on re-interviewing with a focus on theoretical categories \cite{charmaz_constructing_2014}, we prioritized participants who had named or richly described the tacit phenomena underlying our theoretical categories, while seeking representation across roles and genders to ensure a broad range of perspectives. 

During the feedback interviews, we first asked developers to describe what they saw in each diagram before offering any explanation. We then asked whether the model resonated with their experience, what they would change, and to rate its resonance from one to five, providing both qualitative and quantitative feedback. We also asked how the models fit their current workflows, which were often using agentic AI. We explicitly told participants that disagreement was welcomed and that our goal was to refine the models, helping reduce response bias.  This led to a series of diagram iterations and rich discussions about how their cognitive experiences were changing and how some aspects remained the same.  Between interviews, we emailed with participants about diagram iterations. We continued refining the models until feedback stabilized with high resonance, with subsequent refinements validated by at least two participants. The feedback interview questions with our initial diagrams can be found in the Appendix, demonstrating our process and how the models evolved. We also emailed participants a draft of the paper for feedback. The feedback interviews and written reflections provided further evidence of the theory's resonance and supported our assessment that we had reached theoretical saturation.

\subsection{Limitations} \label{limitations}

This study built upon our prior troubleshooting study and inherits limitations from its convenience-sampled participant pool. Mastodon recruitment drew from developers connected to the first author's network, while LinkedIn recruitment was geographically limited to the United States, Canada, and Europe.  For the momentum study, we theoretically sampled 12 participants from this pool based on prior discussions of flow and sustaining development work. Although we sought representation across genders and roles, the original convenience-sampled pool and this theoretically sampled subset may have missed important perspectives. While our interview-based approach was appropriate for eliciting developers’ internal experiences, it did not incorporate direct observation of developer behavior during real software work. Future work that complements these accounts with observational or screen-recorded studies could corroborate and deepen understanding of the developers' experiences related to our theory.

\section{Theory of Developer Momentum} \label{findings}

\begin{figure*}[h]
\centerline{\includegraphics[scale=0.37]{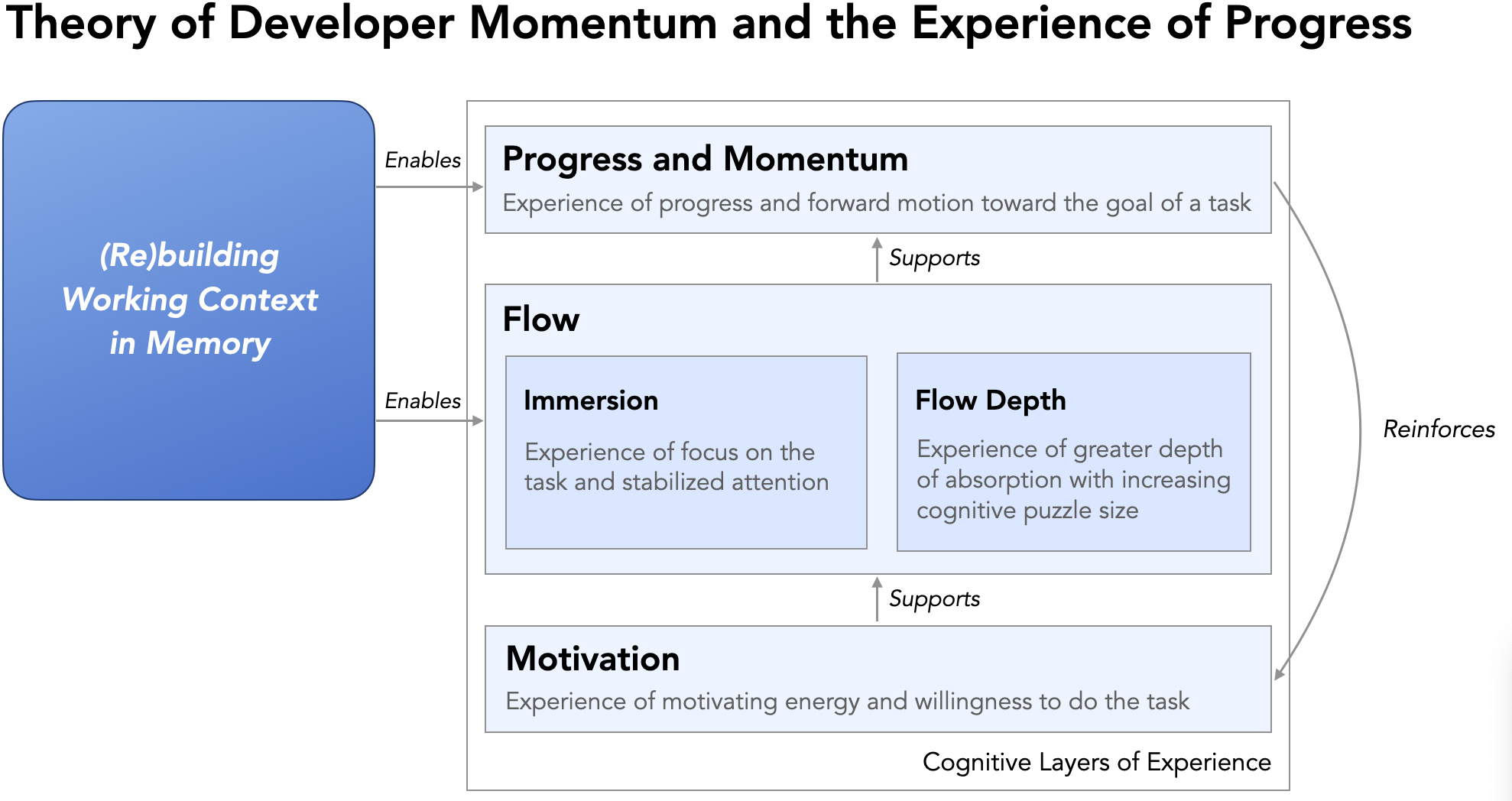}}
   \caption{Developer Momentum is the developer’s experience of progress and forward motion toward the goal of a task. Our theory describes three interrelated cognitive layers of experience (progress and momentum, flow, and motivation), together with a process of (re)building working context in memory. Our theory also highlights how motivation is reinforced as the developer gains momentum.}
  \label{fig:theory_of_momentum}
\end{figure*}

In this section, we introduce the Theory of Developer Momentum that followed from our analysis: \textit{the developer’s experience of progress and forward motion toward the goal of a task}.  Our theory deconstructs the cognitive experience into three cognitive layers (progress and momentum, flow, and motivation), together with a process of (re)building working context in memory that enables developers to sustain motion. Figure~\ref{fig:theory_of_momentum} provides an overview of the theory.

We begin with \textit{the experience of progress}, examining how developers experience forward motion toward the goal of a task. We then explore \textit{strategies to get momentum going}, describing how developers actively construct a viable path forward or create opportunities for nonlinear breakthroughs of insight.  Next, we examine the process of \textit{building up working context} to enable momentum and flow, followed by \textit{the craft of maintaining motion}, which describes the practices developers use to sustain momentum across task sessions, including work that spans multiple days.  We then turn to \textit{what developers mean by "flow"}, examining how developers experience immersion in their work and distinguishing these qualities from the notion of momentum.  Finally, we examine the \textit{motivating energy that supports forward motion} with caring as its source, and then turn outward to consider \textit{how the habitat affects internal motivation} through heartening and disheartening forces that can strengthen or injure caring.

\subsection{The Experience of Progress} \label{experience_progress}

Although "momentum'' is not a commonly used term in software development, when we asked developers about their experiences of getting momentum going on a task, all participants related the concept to making progress.  P01, P07, and P11 described starting at zero or very slow momentum, and characterized the early part of the task as data gathering, exploring, and figuring out a plan. P02 further clarified that momentum does not require having clarity of the entire plan; rather, developers may only have clarity of the immediate next steps, and as they progress, iteratively evaluate whether they are getting closer to the goal. These accounts suggest that the experience of progress is not simply a matter of coding, but one of resolving uncertainty and making decisions within an unclear and complex terrain.

\begin{figure*}[h]
\centerline{\includegraphics[scale=0.39]{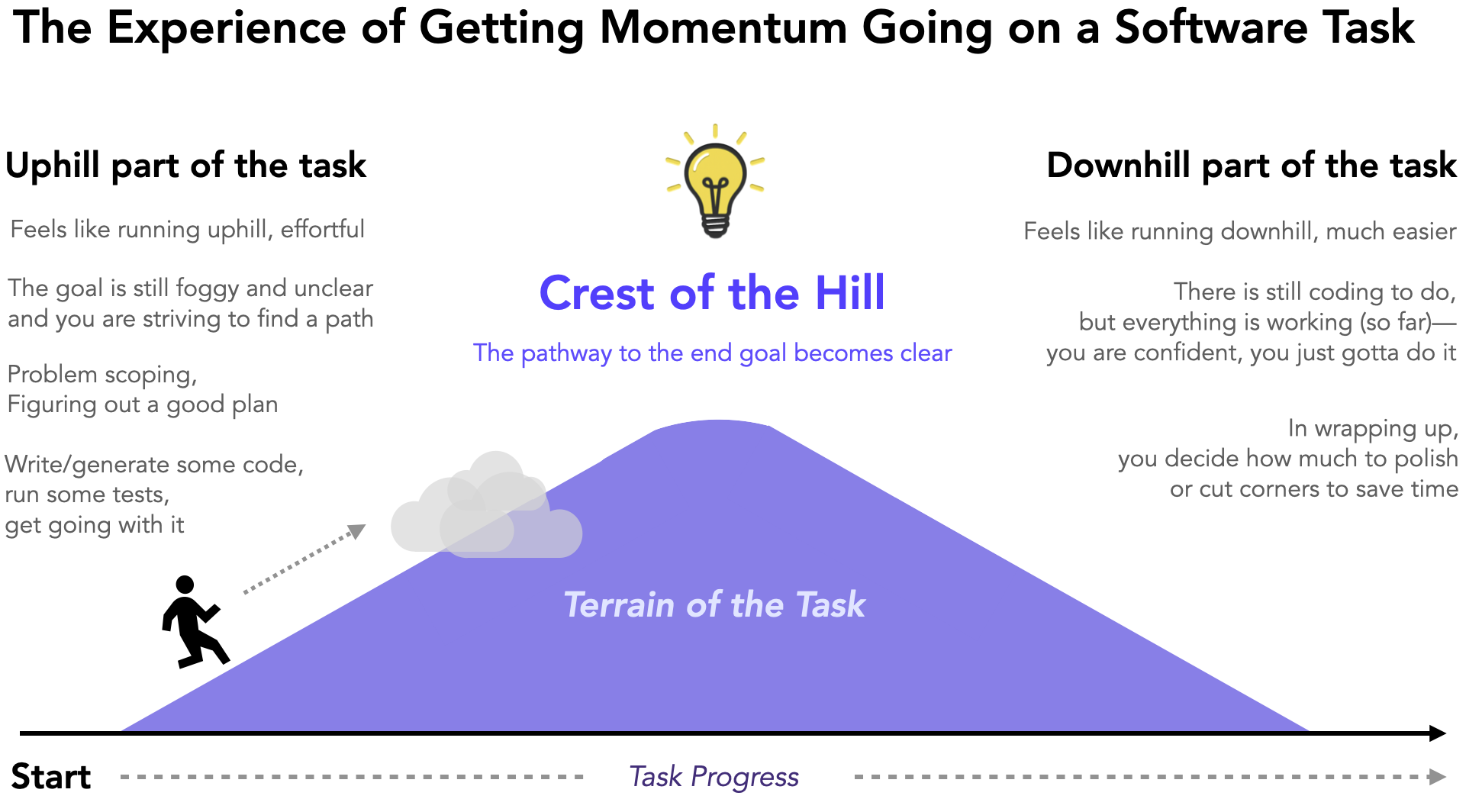}}
   \caption{Developers described the experience of getting momentum going on a software task as analogous to climbing a hill.}
  \label{fig:momentum_hill}
\end{figure*}

Both P05 and P06 independently used the same metaphor of climbing a hill to describe the experience of getting momentum going on a software task, with similar correspondences in meaning. In this metaphor, the uphill portion corresponds to progress feeling more effortful, while the downhill portion corresponds to progress becoming much easier. "Getting momentum going'' refers to the process of overcoming initial inertia and investing the effort required to get steady progress going, both at the start of a task and when re-engaging with it after interruptions or breaks. Figure~\ref{fig:momentum_hill} provides an overview of the metaphor.

During the "uphill'' portion of the task, the goal and path ahead feel foggy and unclear, and it takes effort for the developer to make progress. The developer is exploring, answering unknowns, clarifying their understanding, and figuring out a path to the goal. The early part of the task starts with more questions than answers, as P07 describes:

\begin{quote}
\textit{"So you definitely have to understand the requirements and have a sense of scope. Like how much work are we talking about here? What are the thorny problems that lay ahead? What am I going to have to figure out? How straightforward is this?"}
\end{quote}

As developers work to get momentum going, progress can be very non-linear, involving trial-and-error and frequent changes in direction.  P05 describes the challenges that can arise:

\begin{quote}
\textit{"When you're going up the hill, starting off your task, it's trying to figure out what is the best solution, right? [...] And it can be frustrating, things aren't working the way you want. You might have to step back and try another approach. You might have to do some research, hit some stack overflows, get a better idea of what you're doing.  Or try something, go down one road, take it as far as you can, and then throw it out the window cause it's absolutely no good. Not what you wanted. That's all very frustrating. But then you get to a point where everything starts to make sense."}
\end{quote}

P07 highlighted that when you are not as familiar with the language you are working on, everything can feel like an uphill struggle, but AI can make it much easier to get going.  P02 also highlighted the benefits of using AI for tasks where you were less familiar, and how you could use your experience in one language to help you work in another. P05 described a transition at the crest of the hill that corresponded to a shift in clarity and visibility, where the pathway to the end goal becomes clear.  They started with a fuzzy sense of the goal, then by the time they reached the top of the hill, everything became clear, even if the solution was not yet fully implemented.  P06 further described this moment as emotionally significant, noting a sense of elation when everything is working and the pathway to the goal is clear:

\begin{quote}
\textit{"It's initially like, how am I gonna make this work? And then I hit that light bulb moment at the top of the hill, right? It's working, it all makes sense. [...] As I'm going down, I'm feeling elated because it's working. It's working, it's working, everywhere it's working. Because I did all that initial work, right? I did that big push going up the hill."}
\end{quote}

During the "downhill'' portion of the task, there is still coding to do, but the path to the goal is clear. With the major uncertainties resolved, progress becomes more straightforward and easier to sustain, in contrast to the effortful and exploratory nature of the uphill phase. The developer feels confident they can reach the goal. In contrast, during the uphill phase, several developers (P01, P06, P07, P10, and P11) reported feelings of anxiety and self-doubt from not knowing whether they would be able to find a pathway through the terrain.  P03 and P06 also described a shift in the process toward the end of the task where the developer decides how much to polish the solution or cut corners to save time.  P03 described making decisions by imagining their future selves and teammates trying to extend the functionality, then optimizing to make those future experiences better:

\begin{quote}
\textit{"You've got a lot of other things to do and you're like, 'Do I have time to do this? Maybe it's best practice to do it this way, but maybe I'll cut a corner right now.' [...] Anyone who's worked with me knows that I have this concept of 'do the type of work that future you wants to give past you a high five for.' [...] There was a co-worker's pull request---they were building on something I had done and I had multiple bits of praise in there, 'this was so easy to add to, you wrote this knowing it needed to be extended and I came and extended it and it was super easy, thank you!' [...] A little extra effort in design and careful organization pays dividends later."}
\end{quote}

Even after the transition to the downhill phase of the task, not all uncertainty is eliminated.  Both P05 and P06 described encounters during the downhill part of the task where they would discover part of the solution was not going to work, and have to return to an uphill effort. P02 highlighted how AI does not always produce correct results, which means needing to double-check things and can be frustrating when code does not work as expected. P05 described the subtleties of this experience, noting that even with increased clarity and confidence, some uncertainty remains:
\begin{quote}
\textit{"You're at the crest of the hill, you can see where your goal is and realize, yes, I can get there, the end is in sight. But at the same time, there's still the unknowns of that terrain.  It's like, oh, I didn't think about this on the way up. [...] But you do recognize that you're on the downslope, the unknowns have been answered, essentially. [...] Once you are on that downslope, it's much easier to say, I think there's this much time left before I'm finished with this task. So you could almost think about it in terms of a completion duration confidence."}
\end{quote}
Taken together, these accounts describe momentum as a dynamic and evolving experience of progress, shaped by shifts in clarity, effort, and confidence over time. Developers move from an initial uphill phase of uncertainty and path-finding, through a transition where the pathway to the end goal becomes clear, into a downhill phase of implementation where progress is easier to sustain and estimation of the remaining work is more reliable.  However, this progression is not strictly linear. Even after reaching the crest of the hill, developers may discover that their plan does not fully work, requiring a return to more effortful, uphill-like work. 

In the feedback interviews, developers confirmed that this uphill-downhill experience was still present with agentic AI workflows, and much of the work still involved gaining clarity and coming up with a good plan even when generating code with AI.  P08 described the downhill portion as being more compressed---once clarity is reached, code can be generated rapidly. P08 wondered if this compression of the downhill phase might explain their experience of increased fatigue with agentic AI coding, since the compressed downhill phase was immediately followed by a new task and going uphill again.  These shifting experiential dimensions are a rich area for future research.

\subsection{Strategies to Get Momentum Going}

In addition to actively exploring and constructing a viable path forward, developers also described stepping away from the task as a way to make progress when the path was not yet clear. In these cases, developers would first build up an understanding of the problem through focused effort, then disengage from the task—taking a break or sleeping on it—allowing a potential solution or new perspective to emerge later. These accounts suggest that progress during the uphill phase is achieved through two complementary modes: \textit{active incremental construction}, where developers gradually build clarity through exploration and externalization, and \textit{non-linear breakthroughs}, where new insights emerge after periods of disengagement. In this section, we examine the strategies developers use to get momentum going during this early uphill phase of the task—when uncertainty is highest and the path forward is unclear.

\subsubsection{Active Incremental Construction of a Path}

At the beginning of the task, the developer starts with zero momentum, and is just beginning to understand the task's requirements.  Developers described a range of strategies for actively constructing a path forward through the uncertain terrain of the task. These strategies included research and context gathering, scaffolding ideas in the code with comments, sketching diagrams, writing out TODOs, and setting up feedback loops.  In a feedback interview, P03 highlighted that scaffolding ideas in the code with comments was no longer used as a strategy in agentic AI workflows, but the challenge of constructing a path forward remained. 

P01, P06, P07, P08, P11, and P12 all described a research and exploration phase at the beginning of the task to gather context and better understand the requirements.  P03 described how a substantial amount of the work, even with agentic AI coding, is still this "problem scoping" and figuring out a good plan.  P01 described what starting a new task is like:
\begin{quote}
\textit{"Most of the tasks I am assigned, my initial thought is like, I have no idea what's going on. Let me front load as much research and context-gathering as I can. That's usually the reading intensive part, reading through all the Slack history, messaging history, ticketing history, et cetera. [...] The slowest part is always the beginning, the data gathering and figuring out how things work... and filtering signal from noise. [...] I basically let myself explore for a bit, and go look at the code, look at logs, and make sure I understand what's going on."}
\end{quote}

Several developers (P03, P05, P06, P08, P11, and P12) described strategies of organizing and externalizing their thinking by writing notes or documents to clarify a more concrete plan of what to build.  P03 described their process of thinking through the breakdown of the task by externalizing their thoughts in a notepad, then once they had clarity, immediately jumping into coding:
\begin{quote}
\textit{"So I literally just write down the breakdown of the task. I have another section [of notes], which is notes for my handoff, because as I'm doing development work, stuff comes up where I need to let the next person know about X. So I have all this sketched out in my notepad.  And then I end up fiddling with it and rearranging what makes sense to do first, like what's the smallest unit. And then I usually pop in and start coding. [...] I try to do the simplest thing, whether it's just a spec'd out endpoint that returns nothing, just to prove that it's running. And I do this kind of iterative test-driven development thing."}
\end{quote}

Both P03 and P05 described setting up a feedback loop early in the process as a way to get momentum going. P03 highlighted how tight feedback loops—where you can run tests quickly or do quick interactive tests—help to speed up progress, and that enabling developers to quickly spin up a system and get feedback can be particularly beneficial.  P05 described their approach of getting momentum going, and also emphasized the importance of a feedback loop:
\begin{quote}
\textit{"If I have a mechanical task, where I just want to replace A with B, that's easier to do. I can just dive in and do it. If it's something where I know what the goals are but I'm not exactly sure what the implementation is going to look like, it's probably gonna be a little more writing, and starting to think about what I want to have the code do, and thinking about where the right places to interject that are. [...] Then I need to be able to start to prove what I'm doing, to get that feedback loop [to see] it is actually making the change, and I'm starting to see the behavior that I want, without needing to run all the tests."}
\end{quote}

During the uphill phase of the task, when the plan remains fuzzy and uncertain, it can be difficult to begin coding.  Setting up a feedback loop early helps developers gain clarity about what the system is doing and whether their changes are having the intended effect, making it easier to move forward. P07 offered advice for overcoming the initial inertia of having to write code before the full plan is clear:
\begin{quote}
\textit{"I've caught myself doing this mantra a few times. Just keep coding. Just keep doing it. Don't stop. Don't think about it forever. Find your mistakes or find the ways that the design doesn't work through coding and seeing where it doesn't."}
\end{quote}
Taken together, these strategies show how developers actively construct a path forward by externalizing their thinking, and iteratively refining their approach through feedback. Rather than waiting for a complete plan to emerge, developers begin with partial understanding and progressively build clarity through action, enabling momentum to take hold.

\subsubsection{Non-Linear Breakthroughs}

In contrast to the active incremental construction of a path, developers also described moments where progress emerged through non-linear breakthroughs.  Developers described \textit{active strategies} they used for making these breakthrough insights more likely.  By first building up an understanding of the problem through focused effort, then disengaging from the task by taking a break or sleeping on it, the developer could ensure their cognitive background processing was puzzling over it.  This strategy increased the likelihood of improved clarity or a new perspective arriving later as an insight.
\begin{quote}
\textit{"As a more senior developer, I've learned that sometimes if you're stuck and you're starting to panic, and it's becoming this kind of urgent thing, like I have to figure this out before I go home---there's a lot of utility in just being like, don't panic, just stuff everything you've learned in your head and go home and unwind, sleep on it. And when you come back the next day and you look over everything, and this is where the notes thing comes into play too, if you've made notes about what you've learned, and given yourself some time to really just sit with it, sometimes things become much more clear after a bit of rest."}
\end{quote}
P08 further highlighted how it can feel a bit weird and difficult to just sit there in frustration, trying things and not knowing what to do, but if the aim is to facilitate a non-linear breakthrough, taking a break prematurely can be counterproductive. P11 also shared how these moments of eureka and breakthrough insight can renew the developer's momentum with motivating energy:
\begin{quote}
\textit{"And there's those eureka moments where you're asleep and then you wake up and somehow the answer to the problem is in your head miraculously. And that's really nice because then you wake up and you're stoked and really want to get at it, try your new approach. Put everything else aside and just jump into it."}
\end{quote}
P08 also noted that sometimes going for a walk or taking a break at the office is not enough to have a breakthrough insight, and you really need to sleep on it, or try another strategy such as talking with colleagues which P09, P11, and P12 also found effective.  P04 described how it is much easier to get momentum going in the morning when you are fresher and have indirectly been processing the task from the day before. 

\subsection{Building Up Working Context}

In this section, we describe how developers build up working context in memory to enable momentum and flow during a task session, and how this working context must be rebuilt through a re-entrant looping process when completing a task requires multiple sessions. P07 describes the experience we refer to as \textit{building up working context}:

\begin{quote}
\textit{"You have to build up a context. I've got a model I'm working with in another service. But I want it to work a little bit differently, as always happens. So I'm building up what are the architectural layers, what are the models involved here? And I'm doing all this without any diagram in front of me. So I have to build up this context in my head, and then I map that into different tabs in my IntelliJ or Cursor or whatever I'm using at the time, and just kind of get it all into my head---then decide what my next steps are.  That, for me at least, is not preserved overnight. Because I've usually done other things. I've lost my context completely because I've gone out with friends, or I've watched a three-hour movie, or who knows what's happened. And you just have to---we have to rebuild it."}
\end{quote}

\begin{figure*}[h]
\centerline{\includegraphics[scale=0.41]{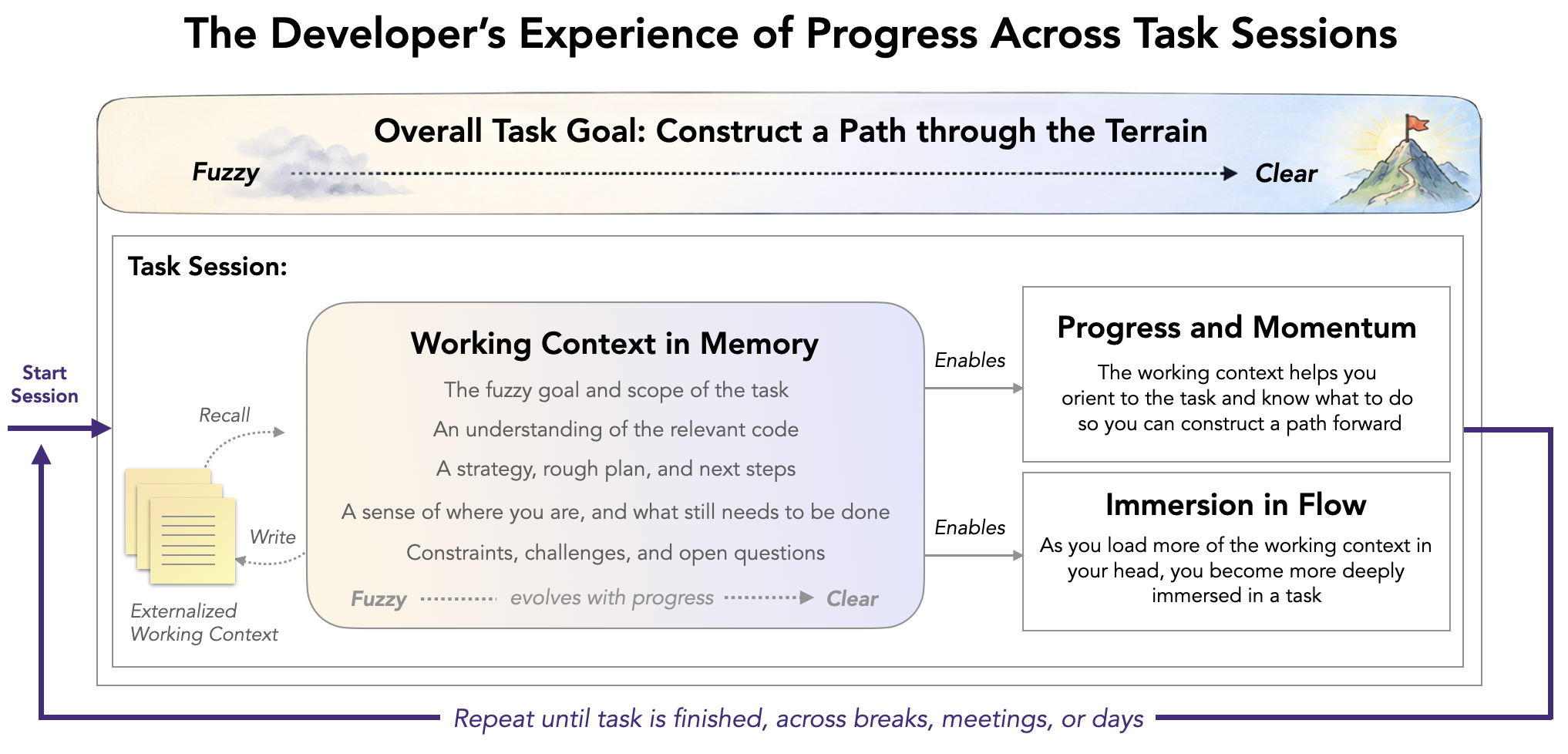}}
   \caption{Building up working context in memory enables both momentum and flow. The working context evolves from fuzzy to clear and must be rebuilt for each task session when there are breaks across meetings, interruptions, or days.  Externalization supports this process so that the required working context can be rebuilt in subsequent task sessions.}
  \label{fig:working_context}
\end{figure*}

This account highlights several key aspects of building up working context. P07 described constructing an internal mental model of the system that enables the developer to reason about the task and decide on next steps.  Without building up this working context in memory, the developer is unable to make progress. As P07 describes, the working context is also fragile and is not typically preserved across days.  Several developers (P01, P03, P04, P05, P07, P08, P09, P12) also described the need to build up working context at the beginning of a task session to get momentum going or to get in flow, and having to rebuild it in subsequent task sessions.  P06 used AI to help them rebuild the working context, asking it to explain the code they just wrote to help them remember.  In the feedback interviews, developers confirmed that building up working context was still a requirement in agentic AI workflows. P08 further described having to build up working context for the agents as well to get momentum going with the AI collaboration. Supporting the agents in this way required more of their own working context to be externalized.  Figure \ref{fig:working_context} illustrates how working context enables both momentum and flow within a task session, and must be rebuilt when a task spans multiple sessions.

P01, P04, P05, P06, P07, and P08 also highlighted the fragility of these mental models, describing how interruptions and distractions could "pull down the castle" (P07) that the developer had built up in their head.  The timing of these interruptions also affects their impact. P05 described how interruptions during the deeper work of the task, when the developer is still trying to understand what the nature of the change is (uphill phase), can be especially disruptive.  By contrast, P08 described how when the path to the end goal is clear (downhill phase), interruptions are not necessarily a problem.  P07 also highlighted an important nuance: once significant momentum is established, the context tends to stay in the developer’s head more easily.

\subsection{The Craft of Maintaining Motion}

Software tasks are commonly done across multiple task sessions, requiring developers to rebuild context multiple times. Several developers (P03, P06, P07, P09) described the experience of progress in terms of having a speed, reflecting a felt sense of motion.  We relate momentum to this sense of speed, of moving fast or moving slow.  In this section, we focus on strategies developers use to maintain forward motion across task sessions.

\subsubsection{The Craft of Note-Taking to Maintain Motion}

Nearly all the developers we talked to (P01, P03, P04, P05, P06, P07, P08, P10, P11, P12) described note-taking as a strategy to maintain motion across breaks, meetings, or days.  In a feedback interview, P08 confirmed they still relied on note-taking to externalize their thoughts, even with their agentic AI workflow.  P01 described their process of reserving time at the end of the day to prepare notes for tomorrow:
\begin{quote}
\textit{"If I am at the end of the day, and I'm not going to be able to finish what I'm currently working on, when I see it's close to five or six p.m., I will start wrapping up my day, writing down notes for tomorrow. [...] I find it almost like writing a prompt, not to an LLM, but to myself. It feels like prompting myself for tomorrow."}
\end{quote}
P03, P06, P07, P10, and P12 all described how the quality of notes determines how easy or difficult it is to resume the task. P10 describes what the experience is like the next morning, when good notes are left behind, and also the benefit of not switching to other tasks during the break:
\begin{quote}
\textit{"It was relatively straightforward [to resume the task] because I had left myself some good breadcrumbs, a failing test and some expanded notes and some diagrams, and left myself in a place where it was relatively easy to pick up.  I think it was easy because I was not working on any other tasks in between. It was still the main thing in my head.  And so it's pretty easy to recover and resume."}
\end{quote}
These accounts illustrate how developers use notes to externalize context and create a bridge between task sessions to maintain forward motion. While these examples focus on task sessions across days, developers P01 and P06 also described using similar note-taking strategies to support shorter breaks in work.  P01 described their practice of specifically writing down what they were currently working on and planning to do next before disengaging:
\begin{quote}
\textit{"If I'm at a point where I'm kind of brain dead, I'll write stuff down in a document related to that one task or thing I was doing. [...]  Or if it's something I have to get up for and focus on for a little bit, I try really hard to write down what I was currently working on, and what I was gonna do next. [...] So before I get up and go feed the dogs or take them for a walk or whatever, I will just brain dump everything I have in my head."}
\end{quote}
Note-taking emerged as a central strategy used by developers for maintaining motion on a task. The craft and care put into the quality of notes shape the experience of resuming work, determining whether progress can continue smoothly or whether it is difficult to rebuild the working context.

\subsubsection{Designing the Restart Across Task Sessions}

In addition to taking notes, developers considered how the state of the code and their stopping point in the task would affect their ability to restart the next day.  Developers described a range of strategies for supporting restart, including organizing their environment through dedicated desktop spaces, Jira ticket comments, diagrams, open browser tabs, and leaving code diffs and files open in the IDE. These were often used in combination with notes and arranged in ways that made it easier to resume forward motion. P03 described this as designing a "hook": leaving the work at a point where the next step is clear and it is easier to get going again:
\begin{quote}
\textit{"So it's always best if I've got a hook so I know what's coming next. I find it easiest if I had completed part A and B of a feature set, and I knew what part C was, and I just had to dig into that. Or sometimes I leave little notes for myself where I was writing integration tests or unit tests and I have some comments that you need to do these three test cases, and I'm like, "great I know where to start off!"  It's always best if past me has given me a hint of what the next step is."}
\end{quote}
Leaving the code in a stable working state across task sessions was also emphasized by P01, P06, and P07. P06 highlighted how picking up from a stable working place was much easier than resuming a feature that was left mid-thought. P07 emphasized stopping when you know what you need to do next, rather than when you are stuck. P03 similarly described the difficulty of restarting from an unsolved "mystery." P05, by contrast, practiced intentionally leaving the code in a broken state with a failing test and TODOs to make it easier to recall the context:
\begin{quote}
\textit{"If I've been working on something continually, I will leave things in a broken state with some to-do notes that I left myself that tells me where to pick up from, where I left off. So I kind of leave things in a purposefully jagged place. I find that easier to reestablish context and momentum rather than starting from scratch again."}
\end{quote}
Another strategy developers used to make restarting easier was being intentional about how they ended their day.  P04 shared their strategy for boosting initial momentum by ending the day with a new task, starting with investigation and reading up on it.  Because of the benefits of background processing when sleeping on it, this strategy made it easier to get momentum going on the task the next day.  P06, by contrast, suggested avoiding taking on a new task at the end of the day, and that it is better things are stable, finished, and working.  Both strategies reflect different ways of designing the next task session—either by giving yourself a head start on the next task, or ensuring a stable and reliable place to resume.  These contrasting approaches highlight that the goal in designing the restart is to leave the work in a state where the developer has enough context and clarity of next steps to re-orient themselves and regain traction. A variety of strategies—from how the code is left, to the stopping point in the task, to how the environment is organized—are designed to support this goal.

\subsubsection{Clarity as a Foundation for Progress}

One of the foundational themes that emerged across interviews was the central importance of clarity in enabling progress. Clarity  allows developers to convert understanding into directed action, rather than figuring out how to proceed. Developers identified a variety of important aspects of what needed to be clear: a clear understanding of the goal and scope of the task, a clear direction and plan for what to build, and clear next steps. P06 highlighted how, when you know where you are going, you can quickly gain momentum, and that clarity also helps sustain momentum across days. In contrast, P11 described the experience of an unclear task:
\begin{quote}
\textit{"Another thing that's hard is when tasks are unclear and I have to kind of—I feel uncertain. If I'm feeling uncertain about the direction I'm headed, but I'm headed in it anyways—like I don't have any way to make anything clearer... then I start to feel a bit tentative about what I'm doing."}
\end{quote}
This account highlights that in the absence of clarity, developers may still act, but their actions become more tentative, making it more difficult to build momentum. When the developer has clarity, it is much easier to resume the task and keep going. P03 highlighted what clarity feels like: 
\begin{quote}
\textit{"If I had good notes and I'm starting from a place of progression rather than starting from an unsolved mystery that you still have to figure out... With this task, I had a nice progression. So it felt good. I was clear on my task, like jump in, see where I left off, amazing, keep going. Like it was very smooth and that felt very good."}
\end{quote}
This account highlights how clarity enables smooth continuation of the task across sessions, allowing developers to re-engage quickly and maintain forward motion.  These dynamics highlight a fundamental characteristic of how progress is enabled by clarity: as the goal, plan, and next steps become clearer, developers move from tentative, uncertain exploration of the terrain to confident directed action.

\subsubsection{Staying in Sync with the Team} \label{sync}

Even though we did not ask directly about the team, several developers (P02, P03,P05, P06, P08, P09, P11, P12) highlighted their activities of synchronizing with the team as part of maintaining motion.  P08 described the importance of getting feedback as part of maintaining long-term momentum:
\begin{quote}
\textit{"When I was a junior developer, or this could still happen today--- [...] When I'm allowed to just quickly do a lot of work and there's not any oversight or somebody else having to give feedback into it, it often feels like there's this massive burst of progress early on, but then you start to get mired in all these little messes you've created because you didn't have to vet the work with anybody else. [...] That kind of feedback is really important to maintaining long-term momentum. I feel like a lot of teams get caught by that specific trap. They're like, we're just gonna let people run ahead and do it the way they want, and we'll make so much rapid progress. But then you hit this wall of, nobody knows how to change or fix anything anymore because there was never any cleanup involved."}
\end{quote}
P08 described two scenarios: one where the junior developer is running ahead without feedback, and another where the team members are running ahead without integration and cleanup. In both cases, lots of progress is happening with code being written and momentum appears high—but the work is not being shaped by others.  Without feedback into the work, integration of understanding, and cleanup of the code, problems accumulate invisibly until the system becomes too hard to understand, change, or fix. P08 emphasized the importance of teams including this type of feedback to ensure long-term momentum. As a platform engineer, P12 had several projects involving collaboration with multiple people often across teams.  P12 highlighted strategies to optimize team momentum and synchronize more effectively:
\begin{quote}
\textit{"It's good to make sure things are still on board and people are still speaking about the same thing when they talk about the problem. That also can drift---what people expect, and the outcome that people are hoping to see, can change day to day. Checking back in and making sure that everything is still headed in the right direction is sometimes useful especially if it's a project where a lot of people are trying to contribute to the same thing."}
\end{quote}
P12 highlighted that coordination is not a one-time agreement—it requires ongoing re-synchronization because goals, assumptions, and expectations drift over time.  Synchronization preserves the \textit{usefulness of motion} so that it is headed in a direction that is coherent at the team level. P12 further highlighted the tension between individual flow and team coordination, describing how the communication required to stay synchronized can feel like it slows down progress, while also being necessary for the work to succeed:
\begin{quote}
\textit{"If you're directly collaborating with somebody who is trying to either adopt or utilize what you're building, getting them on board and actually partnering with them can be a really useful way to accelerate things.  But it also requires a synchronized collaboration of going back and forth, and checking in, and making sure people are still on the same page and still working on the same thing. You could have a conversation with somebody and both agree on what you want to happen, but if you went away for a month, I don't think the two things that you built would ever work together, because that's just not how this works. And so delivering things in increments, and actually communicating with people simultaneously impedes my ability to go fast and do flow things, but also is a requirement of the work. Like it would not be successful without that."}
\end{quote}
P12 highlighted that maximum motion at the individual level is not necessarily the path that produces meaningful team-level progress.  Work that feels slower because of coordination and feedback may ultimately move faster at the team level by preventing divergence and downstream failure. Before starting the work, P12 described a process of aligning with stakeholders and establishing a shared understanding of what should be built as a key strategy to keep momentum going smoothly:
\begin{quote}
\textit{"It can be really helpful to get stakeholders that actually own the product informed and on board, before you just start making the change. Typically, they have additional context that it's good to get before you start doing the thing.  Because a lot of the times, if you start the project first, you'll find out later that there's a reason that it was doing the thing that it was doing. And then your thing won't work. So figuring out what those are, and getting everybody to agree on what the thing should be first, also helps keep things in a flow state, kind of moving forward."}
\end{quote}
P12’s strategies suggest a longer-term, team-level view of momentum, where synchronization activities are practiced to maintain motion across people and over time.  By creating shared clarity about what should be built up front, the developer can ensure progress is happening in the right direction, avoid a restart, and maintain long-term momentum. 

Across these strategies the differences between individual momentum and longer-term, team-level momentum become clear.  A developer can feel like they are moving fast by writing lots of code and staying in uninterrupted flow, but this does not necessarily equate to meaningful progress from a longer-term, team-level view of momentum.  Maintaining motion at the team level requires maintaining coherence of the overall team direction and ensuring that work remains understandable, aligned, and integrable across people and over time.  The craft of maintaining motion starts with recognizing the dynamics of momentum at both the individual and team levels, then supporting the conditions that make meaningful progress possible.

\subsection{What Developers Mean by "Flow"}

Rather than assuming developers were using Csikszentmihalyi’s definition of flow state when they talked about flow, we made space for developers to characterize their own experiences of flow, and we listened for what they meant when using this term.  This gave us an opportunity to illuminate a richer phenomenology of the developer’s experience of flow over time. Developers did not describe an instantaneous transition, but rather a process of \textit{getting in flow} that took time and effort to "ramp up" (P03) from partial to full immersion in the task.  The notion of being ``in a flow state'' corresponded to the state of full immersion, where attention is stabilized on the task and competing distractions fall away.  Developers also described an additional dimension of \textit{flow depth} (P03, P05, P08, P12), reflecting how deeply they were absorbed in the task.   In this section, we first examine the dynamics of immersion in the task, then return to the experience of flow depth.

\subsubsection{Getting In Flow: Establishing Immersion in the Task}

\begin{figure*}[h]
\centerline{\includegraphics[scale=0.40]{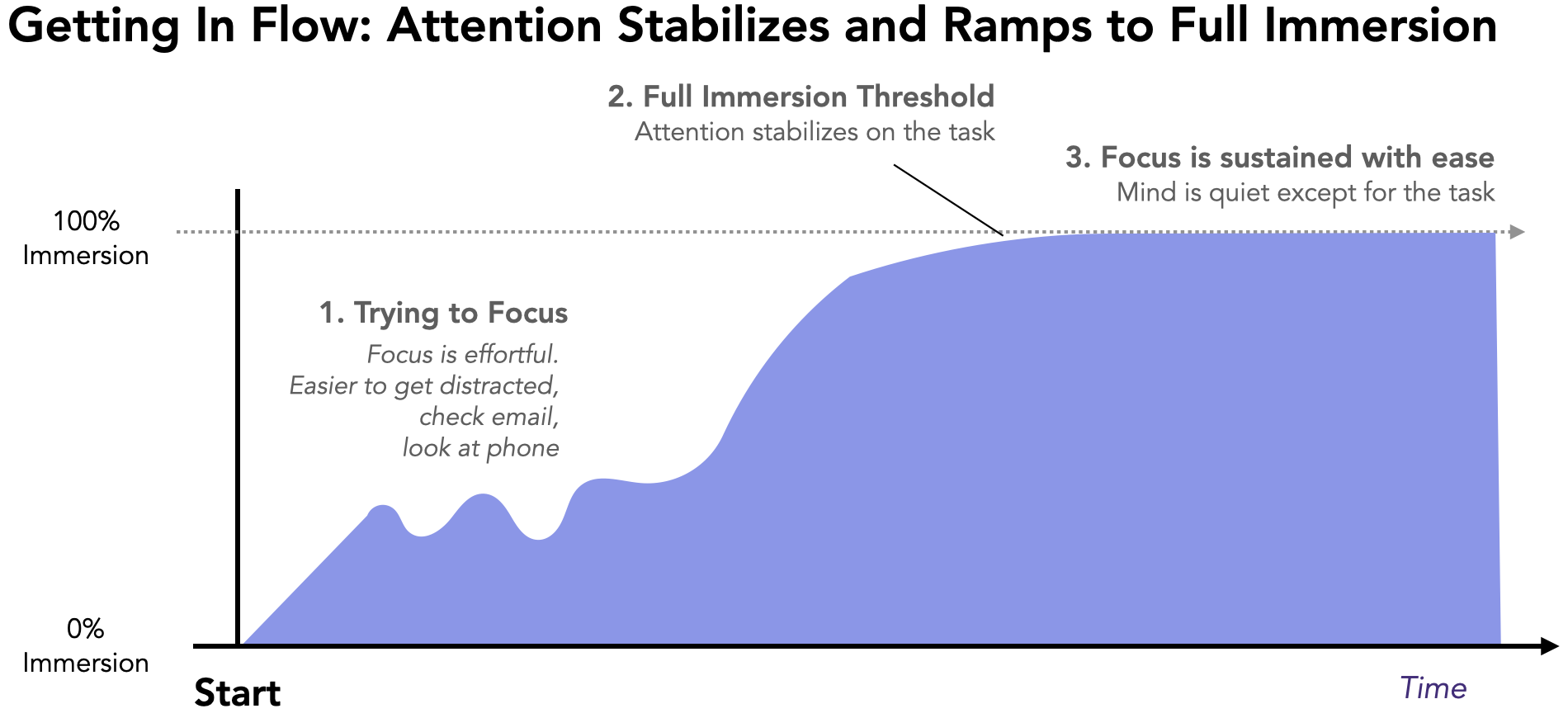}}
   \caption{“Getting in flow” is a process in which attention stabilizes over time. Prior to this stabilization, focus is effortful, immersion is partial, and it is easy to get distracted. Once attention stabilizes on the task, the developer can sustain focused attention with ease.}
  \label{fig:flow_immersion}
\end{figure*}

Developers described a transitional period before becoming fully immersed in the task, when it was easy to become distracted.  P08 described the experience of being "half into it":
\begin{quote}
\textit{"When I'm in a really proper flow state, it doesn't even occur to me, but when I'm in this kind of transitional half-into-it state, [...] there's this temptation to just pick up your phone and kind of flip through content."}
\end{quote}
In a similar account, P09 described a mental urge to check email or look at chat, which went away once they got going on the task. Several developers (P01, P02, P03, P08, P09, P10, P11, P12) described the initial phase of a task session as ``trying to focus'' which, depending on a variety of circumstances, could be difficult. In contrast, once full immersion was established, staying focused did not require so much effort.  P03 described the difference in the felt experience after this threshold was reached:
\begin{quote}
\textit{"When I'm in the middle of [flow], it's really nice because I'm focused and my brain is quiet except for the thing that I want to be thinking about. It feels nice and really great to be making progress. You lose track of time, 
[...] you're just in the zone, working on your focus task undisturbed---it feels nice once you're in it. Just sometimes ramping up into it is difficult, so I'm always trying to minimize the ramp up time if at all possible."}
\end{quote}
Together, these accounts suggest a cognitive mechanism underlying the experience of getting in flow in which attention stabilizes over time through sustained focused effort. Initially, focus is effortful, immersion is partial, and it is easy to become distracted. Once attention stabilizes on the task, the developer crosses a threshold into full immersion (flow state): urges toward distraction fall away, the mind becomes quiet except for thoughts related to the task, and focused attention can be sustained with ease (Figure~\ref{fig:flow_immersion}). However, this transition into full immersion does not always occur. In a feedback interview, P01 described rarely getting into a flow state, not reaching the immersion threshold, and focus remaining effortful for the duration of a task. They connected this sustained cognitive effort to their experience of fatigue.  Further research is needed to investigate the implications when developers are unable to reach full immersion.

\subsubsection{Immersion Across Enjoyment and Frustration}

While Csikszentmihalyi's concept of flow state is often associated with enjoyment and intrinsically rewarding experiences, developers highlighted that the state of full immersion did not always correlate with positive affective states.  Namely, during troubleshooting, it is common for the developer to feel frustrated and stuck, while simultaneously being fully immersed with attention stabilized on the troubleshooting task \cite{starr2026theory}.  P01 described this duality as two different kinds of absorbed engagement (full immersion) involving fundamentally different emotional experiences:
\begin{quote}
\textit{"I would say there's two kinds of engaged or absorbed. One is out of interest,  the subject matter is interesting. The other is out of anger or frustration. I don't know what the heck is going on. I'm really frustrated. I want to fix the thing. And there's a million things that could separate what I do out of frustration versus what I do out of pure interest. It's either, there's something really painful going on that I want to fix, or there's something like, ooh, I want that. [...] The more there is on either one of those spectrums, the more engaged I am."}
\end{quote}

This account suggests that full immersion is not inherently tied to positive affect, but can also be sustained during frustrating experiences.  This does not contradict Csikszentmihalyi's theory; rather, it suggests that the peak experience he described as the "flow state" includes full immersion, but also encompasses additional conditions, and represents a narrower set of experiences than how developers more commonly use the term. While these peak experience conditions are less common, full immersion was the most salient quality by which developers identified being in flow. Our theory describes this more common variety of flow experience over time. 

Developers also highlighted how the experience of progress differs across these two forms of immersion.  P01, P06, P08, and P09 all described a felt loss of momentum during troubleshooting.  P08 described feeling trapped in repetitive debugging with very little sense of progress:
\begin{quote}
\textit{"When you get stuck on something, having to try lots of different configuration combinations to find the one that works, you're just sort of like, will it be happy if I do this? No. Will it be happy if I do that? No. That's when the feeling of progress is really low. [...] It feels like a lot of wasted trial and error where you're trying a thing and it didn't work. And that feels like a waste of your time because you didn't necessarily have to try that if you just knew the right thing to do."}
\end{quote}

This account highlights a fundamental aspect about the experience of forward motion toward the goal of a task, that not all motion feels like progress. If the developer is not gaining traction with their troubleshooting efforts, it no longer feels like they are moving toward the goal in a meaningful way.  The developer is still fully immersed, but forward progress has stalled.  As the developer remains stuck and unable to regain traction, the previously built-up sense of momentum erodes.  Together, these accounts distinguish the immersion experience of being in flow from momentum. While immersion reflects stabilized attention on the task, momentum reflects the developer’s experience of meaningful progress toward the goal of the task. During troubleshooting, immersion can remain high even as the developer struggles to gain traction and momentum erodes.  This highlights how supporting developer effectiveness is not simply a matter of maximizing time in flow; it also requires supporting the developer in maintaining momentum.

\subsubsection{Continuity of Orientation and Sustaining Immersion}

Csikszentmihalyi’s theory highlights the importance of clear goals and immediate feedback in supporting flow experiences \cite{csikszentmihalyi1990flow}. Developers  similarly described clear goals (P04,P05,P06,P07,P10,P11) and feedback loops (P03,P05,P08,P09) as important to sustaining flow and momentum, but their accounts also revealed a more nuanced experiential process underlying why these aspects are important.  P06 described their emerging sense of orientation as their understanding became clearer:
\begin{quote}
\textit{"There's a fogginess and a doubt, because I knew what the scope of the changes were, but I didn't know every change that needed to be made. And as I worked with that initial change, the understanding of what I was doing became more clear. I started to be able to predict what the next changes were going to be or what would be required or what the potential changes could be."}
\end{quote}
As clarity emerged, the next step became visible, and the developer could orient and respond.  P08 further described how this seeing of the next step, then the next step, enabled them to stay in flow:
\begin{quote}
\textit{"If I know I need to make these changes, but don't necessarily understand why these are the changes, that's really hard to reach a state of flow.  When it feels like I understand why these changes work, and why we're doing it this way, then I get really absorbed because it feels like you're always chasing the next thing like, oh now I need to do this, now we need to do this. I can't get in that state if the next step doesn't occur to me, because I didn't really understand why we did the last step."}
\end{quote}
These accounts suggest a causal process chain: clarity enables the developer to see the next step, then the next step, and this continuity of orientation is what enables the developer to sustain immersion.  In contrast, when the developer is unclear on the changes, the next step does not naturally emerge, which disrupts the continuity of orientation and the developer's ability to stay in flow. 

The importance of feedback loops becomes clearer when viewed through the lens of continuity of orientation.  In Csikszentmihalyi's theory, "immediate feedback" is not framed in the narrow software engineering sense of having a fast build or fast tests, but rather is  about how the flow activity continuously reveals whether your actions are working and how to adjust.  A climber feels whether their movement works.  A musician hears whether their note fits the song.  The responsiveness of the activity to the person's actions enables them to continuously orient themselves as conditions unfold.  In a software context, developers are similarly orienting themselves continuously to the code and the task at hand, but software work introduces unique constraints because the environment is not directly perceptible.  A developer cannot tell whether their understanding of the system matches reality until they run the code and see the system respond.  P03 described how even relatively small delays in obtaining feedback disrupt this continuity:
\begin{quote}
\textit{"I feel like I haven't talked enough about iteration cycle time, and how smoothing that out can be really beneficial. [...] Even just making unit tests run five minutes faster would be huge, making it so you can quickly spin up or do something to get feedback. [...]  Having those [feedback loop] delays, it's the worst when it's like five minutes or ten minutes, because it's not really enough time to do anything else, but it's just enough time to be incredibly inconvenient and slow to get feedback on the test cycle running." }
\end{quote}
When feedback is delayed, especially with 5–10 minute delays, the developer is stuck in an awkward state of trying to maintain their orientation through the delay, while also being unable to fully disengage.  If orientation is lost during this period, not only does this break immersion, but the developer must re-orient themselves once the feedback arrives to continue making progress.  In a feedback interview, P08 described breaks in continuity of orientation related to their agentic AI workflow.  After prompting an agent, they could wait 15 minutes for the agent to return, at which point they needed to re-orient themselves and found this disruptive to their flow.  P07 described no longer achieving flow states in their agentic AI workflow and identified similar workflow patterns as possible causes. P05 described the loss of flow states as a widespread experience reported across their team that was now using AI. P02 highlighted that the degree of re-orientation differed depending on whether the agent was surfacing a decision that needed clarification along the intended path, or had gone down a wrong path and the developer needed to understand what happened to intervene. Further research is needed to better understand how these changing workflows and breaks in continuity of orientation affect the flow experience.

\subsubsection{Flow Depth and the Cognitive Puzzle Required to Make Progress} \label{flowdepth}

While continuity of orientation helps explain how immersion is sustained over time, developers also described important differences in the intensity and depth of immersion itself.  Several developers (P03,P04,P05,P06,P08,P11,P12) used phrases like “flow depth,” “depth of focus,” being “deep in it,” “deeply absorbed,” or “deeply engrossed” to describe the experience. “Deep” refers phenomenologically to the degree to which awareness becomes absorbed in the problem-space and the surrounding world recedes from attention.  P03 described how the complexity of a task affects the flow depth required:

\begin{quote}
\textit{"There's a lot of tasks that I'm doing that are very complex and have a lot of context that I have to keep in my head simultaneously in order to make progress. But once I'm [in flow] and it's all in my head and I'm focused, and I can stay focused, it feels actually quite good.  [...] On this particular task, I think [my flow depth] was at a pretty solid 3.5 out of 5. It wasn't crazy complex. So I didn't have to use 100\% of my processing power on it." }
\end{quote}

\begin{figure*}[h]
\centerline{\includegraphics[scale=0.40]{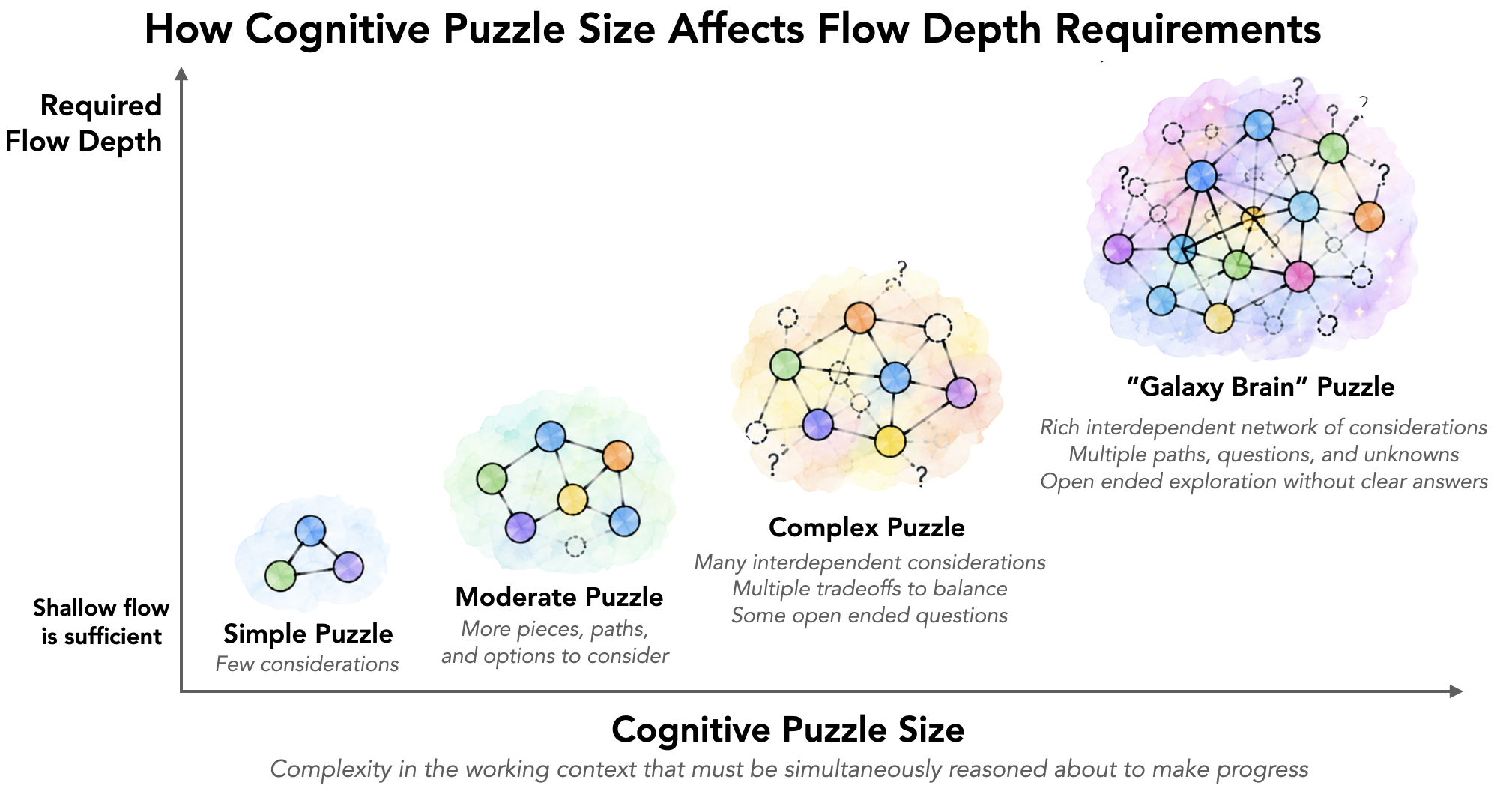}}
   \caption{Tasks with larger cognitive puzzle sizes require deeper flow states in order to make progress.}
  \label{fig:flow_depth}
\end{figure*}

P03's account suggests that the flow depth required for a task is related to the size and complexity of working context needed to make progress.  P05 expanded this notion and highlighted that it is the "puzzle-y" quality of the work that creates variation in flow depth:
\begin{quote}
\textit{"The more puzzle-y the change might seem, the more absorbed I would be, versus something that is a very rote change.  There was a change I did the other week, and it broke some unrelated tests. [...] It was all fine, just, it was grunt work, essentially. So it wasn't rewarding work to do. [...]  [In another task], I was working through this meatier portion of changing some of our internal APIs to be, rather than a single API that does one thing, having more specific APIs for each of the three use cases that we have.  I didn't know exactly how I was going to end up achieving that. This is a case where I had taken everything into pieces. [...] It ended up being a lot cleaner than I thought it would be. It's a crisper representation of the domain concepts that we have. I'm very pleased and happy with the end result. [...] The coding tasks I take on tend to be more of these more complex things that the others on the team may not necessarily have the full vision to do."}
\end{quote}
Taken together, P03 and P05's accounts suggest that larger, more complex or open-ended puzzles require a deeper flow state in order to make progress, whereas simpler and more routine tasks can be advanced with a shallower flow state.  P08 characterized these very large puzzles as "galaxy brain" puzzles that required the developer to load the "galaxy" in their mind in order to do the work. Figure \ref{fig:flow_depth} illustrates how cognitive puzzle size affects the flow depth required by the task.  P05's account also highlights how different types of contributions are made possible by engaging in these more complex puzzles—achieving a "crisper representation of domain concepts" has broad implications on the ease of future work.  Both P05 and P12 found larger, complex puzzles more satisfying and rewarding.  This relates to Csikszentmihalyi's concept of challenge-skill balance with developers enjoying the challenging work.  In a feedback interview, P07 described not doing many simple or moderate puzzles, and when they did encounter them, challenging themselves to automate them away. This further suggests a relationship between cognitive puzzle size and experienced challenge.

P03, P04, P05, P08, and P12 also described how deeper flow states more broadly shaped the flow experience, including the cost of entering and exiting this state. P05 described how they considered these tradeoffs:
\begin{quote}
\textit{"You could look at this through the lens of, "how deep of a flow state is required for the work at hand?"  If it's a shallower state, it's easier to come in out of, but then it's also simpler work, rote, grunt, whatever adjective somebody wants to put on it.  But if it is this deeper thinking that's required, then there's more that is needed to help that entry and exit.  The deeper entry and exits, if it's gonna take me more time to enter/exit, then I need a longer block of time to do something, so once I've entered, I can spend more time." }
\end{quote} 

Several participants (P03, P05, P07, and P12) also described exit costs associated with flow states. P03 described having to "go outside and stare at a tree" in order for work tasks to leave the mind, P07 described strategies for "clearing the brain" with music or a breathing exercise, and P12 described needing an explicit break before being able to context switch. Taken together, these accounts suggest that deeper flow states influence not only the effort required to build up the required context, but also the time required to mentally disengage and transition to another activity. P03 also considered these tradeoffs, but since their role emphasized being responsive to teammates, they preferred tasks that could be accomplished with a more moderate flow depth:
\begin{quote}
\textit{"In a lot of ways, that 3.5 [flow depth] is more desirable than the five, because the five is intense.  The five is exhausting when you're thinking of 12 things and aware of their subtle interplay together.  Having a bit of a gentler task is nicer, because I know it's not gonna drain me, and then I can make really good progress on it.  Then if someone pings me on Slack and is like, "Oh, hey, do you know the answer to da da da?" it's not as disruptive when my focus is more on that 3.5 rather than the intense 5.  It's still engaged, it's still focused, and I can move through it, but it was a gentler focus where I could pull out a little bit here and there and that was fine and it didn't ruin the path. [...] I've been at the five level of focus, and if I get interrupted, I almost never can get back in, usually because I'm too tired, like I'm just wiped."}
\end{quote}
P03's account highlights how a moderate flow depth allowed them to remain engaged with their task while still being able to temporarily shift attention when needed.  When deeper flow states were disrupted, P05, P06, P07, and P08 all described a more extreme impact from the interruptions where the "castle in their head" (P07) came "crashing down and was lost" (P08). P06 similarly described a difficulty of ramping back up after a deeper flow state was disrupted. While several participants discussed the phenomenon of flow depth, P03 was the only participant who explicitly described deeper flow states as tiring.  P03's account of "thinking of 12 things and aware of their subtle interplay together" suggests that maintaining deeper flow states may place greater demands on attentional capacity, contributing to an experience of cognitive fatigue \cite{starr2026theory}. Additional research is needed to better understand the factors that shape these experiences.

\subsection{The Motivating Energy That Supports Forward Motion}

While momentum describes the experience of progress and forward motion on a task, and flow describes the experience of immersion and attention stabilizing on the task, developers also described a feeling of motivation and motivating energy, and whether this energy was available to support putting effort into the work.  Unlike momentum or flow, motivation describes the developer's experience of motivating energy, willingness to engage with the task, and the available resource capacity to support forward motion.

\subsubsection{Caring as the Source of Motivating Energy}

When asked about momentum, several developers (P03, P04, P06, P07, P09, and P11) talked about the dynamics of feeling motivated and the importance of the task being connected to something they cared about. P04 highlighted how seeing the connection to "the why" affected motivation:
\begin{quote}
\textit{"If I'm working on something that doesn't feel important, the why, if it seems like it's a ridiculous task, [...] maybe it's not gonna have value for anybody, or is just keeping people busy? Then sometimes it'll be a little hard to get motivated on. [...] If it feels like a grind, then there'll be a lot less [motivation]. But I can find motivation intrinsically in other areas, like, I like the company I work with, I like the people I work with, I'll still get it done and it'll be fine, I'm just not as excited about it. [...] If there's a well-defined direction to go in, and being able to see the value of that at the end, like why am I doing this? What's the bigger picture? How does this little task fit into a bigger picture? That's important." }
\end{quote}

While not all participants cared about the same things, there were common themes in what developers cared about and a similar relationship between caring and how motivated they felt toward a task.  P03 found that high visibility high impact work was more enjoyable than work that felt invisible like "nobody cares", and how hard it could be to feel motivated in those circumstances.  P07 felt very motivated and enjoyed doing a task when it enabled functionality they were wanting.  P06 highlighted the benefits of giving the team "new options for doing cool things with ease" as their why and reason for caring.  When developers could see why the work mattered and how it contributed to something they valued, these carings (what is cared about) became the source of their motivation.  P04's account also highlights the way that carings can be substituted in circumstances where a specific task might not be very enjoyable---liking the company and the people provided another reason for the developer to care.  When asked about the "feeling" of momentum, P09 described feeling momentum at the beginning of the task, before any work had started, contrasting with other developers' accounts that momentum starts at zero:
\begin{quote}
\textit{"There was definitely kind of a feeling [of momentum] at the beginning, because of this particular task, the level of importance that I gave it. It was like, it's useful for our team and it's useful for this other team, so I really want to focus, I'm interested in this, let's figure out how to put it together."}
\end{quote}
One plausible explanation for why motivating energy might feel like momentum is that motivating energy is typically transformed into forward motion on a task. In this sense, the developer can predict and feel the future momentum before any actual progress has occurred.  P09 and P11 also discussed an aspect of motivation they described as \textit{emotional investment} that occurred over a longer-term engagement on a project and with a team. P11 described how this longer-term emotional investment affected the energy available to work from day to day:
\begin{quote}
\textit{"A lot of it has to do with how emotionally invested I am in getting the work done. [...] There's a place that I worked that I felt super enthusiastic about, like I'd come to work in the morning and say yeah let's go, and the momentum was there. Everybody was working really fast and really supportive of each other and supportive of their work, and clear about requirements as far as they could be.  And that emotional energy made it really easy to just dive into stuff. You didn't have questions about whether it was going to be worth your while to spend all these hours on it."}
\end{quote}
P11's account suggests that caring can deepen over time with emotional investment and longer periods of engagement.  Putting these accounts together, motivation, and the developer's cognitive experience of motivating energy at its center, arises from the developer's carings and deepens with long-term emotional investment.  Phenomenologically, these carings can be understood as the source of motivation and motivating energy, varying in strength and shaping the direction and intensity of the developer's efforts.

\subsubsection{Feeling Fresh and the Daily Capacity to Engage}

While the developer's carings provide a stable source of motivation and wanting to engage over a longer term, developers also described variability in their available capacity to engage that followed a daily cycle.  P04 and P07 both described a feeling of "freshness" in the morning where mental energy was more readily available and focusing was easier, then gradually becoming depleted throughout the day.  P04 described the experience of freshness, and how doing work with high cognitive load depleted their available capacity:
\begin{quote}
\textit{"I'm fresher in the morning, to be honest. In the morning, if I'm working on a particular task, I can just sit down and get focused. The first 20 minutes or so, would be getting into that context and mindset, and then after that, I'd be able to be in that flow [state]. [...] It also depends on how much work I'm having to do that takes a lot of cognitive load and thinking. If I'm doing a bunch of coding that day, it might be near impossible for me to do something on this in the evening, for example. So I might try to do something more mundane if I do anything at all, or I just set it aside.  What I choose to work on, would reflect my mental state. So if I'm feeling pretty fresh, I might take on a big chunk of something."}
\end{quote}
P04's account suggests that freshness corresponds to an available capacity to engage that is depleted by cognitively demanding work and is restored by sleeping. It also suggests that developers are aware of this changing capacity and can strategically adjust their choice of task to fit their available capacity. P01, P06, and P07 similarly found it was easier to focus at the start of the day, and P01 found mornings were a better time to do more cognitively demanding work, suggesting a similar experience of freshness. P07 further highlighted the value of this period of freshness and reflected on how it could be used more strategically:
\begin{quote}
\textit{"I'm thinking of trying to ignore [email] even more than I do, because I work for a big company, and I don't really need to know all initiatives that are going on and all that, but just reading all that stuff takes time. [...] There's this backlog of communication at the beginning of the day that [makes it hard] to get revved up first thing, when your mind is really fresh at the beginning of the day, right? It seems almost a shame to waste it on that. [...] I feel like I could benefit if I did block out the first couple of hours and just closed Slack, closed my email, and just did my task."}
\end{quote}
P07's account suggests that developers recognize periods of freshness as a valuable cognitive resource, and intentionally plan their work around its anticipated availability. Protecting periods of freshness for cognitively demanding work by deferring communication until later represents one way developers can operationalize this understanding.

\subsubsection{Becoming Energized By Meaningful Progress}

Several developers (P03, P04, P06, P08, P10, P11) talked about the positive feelings that accompanied making meaningful progress, describing the experience as "feeling good", "rewarding", "satisfying", or "joyful". Research on subjective vitality connects positive affective experiences with feelings of energy, aliveness, and vitality \cite{ryan1997energy, nix1999revitalization}, providing a theoretical bridge for understanding how the positive feelings associated with progress can be energizing. P03 described the experiences they associated with these positive feelings: 
\begin{quote}
\textit{"I was succeeding, so that felt good. [...] When I'm in the middle of it, it feels nice and it feels really great to be making progress.  And you sort of lose track of time, which is a neat effect, to not really be aware of time. [...] With this task, I had a nice progression, so it felt good. I was clear on my tasks, like jump in, see where I left off, amazing, keep going. It was very smooth and that felt very good."}
\end{quote}
As we discussed in section \ref{experience_progress}, during the uphill phase of the task, the developer can experience anxiety, uncertainty, and self-doubt as they struggle to find a pathway through the terrain of the task.  P03's account suggests that as developers begin making meaningful progress, the emotional quality of the experience can shift dramatically toward one that feels increasingly positive and energizing.  P08 further described how it is these successes and "wins" that help motivate the developer to keep going: 
\begin{quote}
\textit{"It feels good when you get over the hump, when you've actually accomplished something, I think about getting wins. [...] When you're seeing a lot of quick wins, there's a lot of feedback, this feeling that you're making progress, [then it] is really easy to maintain momentum, that feeling of momentum that you're making progress really rapidly. [...] I think for a lot of developers, what keeps them going is that feeling of, maybe you struggled all week to get anything done, but then there was this one day that things clicked and you actually did something and it felt like that piece of work made a difference." }
\end{quote}
P08's account suggests that meaningful progress does more than create good feelings. As developers receive ongoing evidence that they are succeeding and moving toward their goal, these experiences also reinforce motivation and become a source of motivating energy to keep going.  Taken together, these accounts suggest that motivating energy is not only invested at the beginning of a task, but can also be renewed through the experience of progress.  As illustrated in Figure~\ref{fig:theory_of_momentum}, momentum itself reinforces motivation.

\subsection{How the Habitat Affects Internal Motivation}

In the previous section, we discussed the developer's experience of motivating energy---where it comes from, the factors that affect available capacity, and how developers can become energized by meaningful progress.  In this section, we turn outward to consider how external environment factors of the habitat can affect the developer's internal motivation.

\subsubsection{Disheartening Forces that Injure or Rupture Caring} \label{dishearten}

\begin{figure*}[h]
\centerline{\includegraphics[scale=0.45]{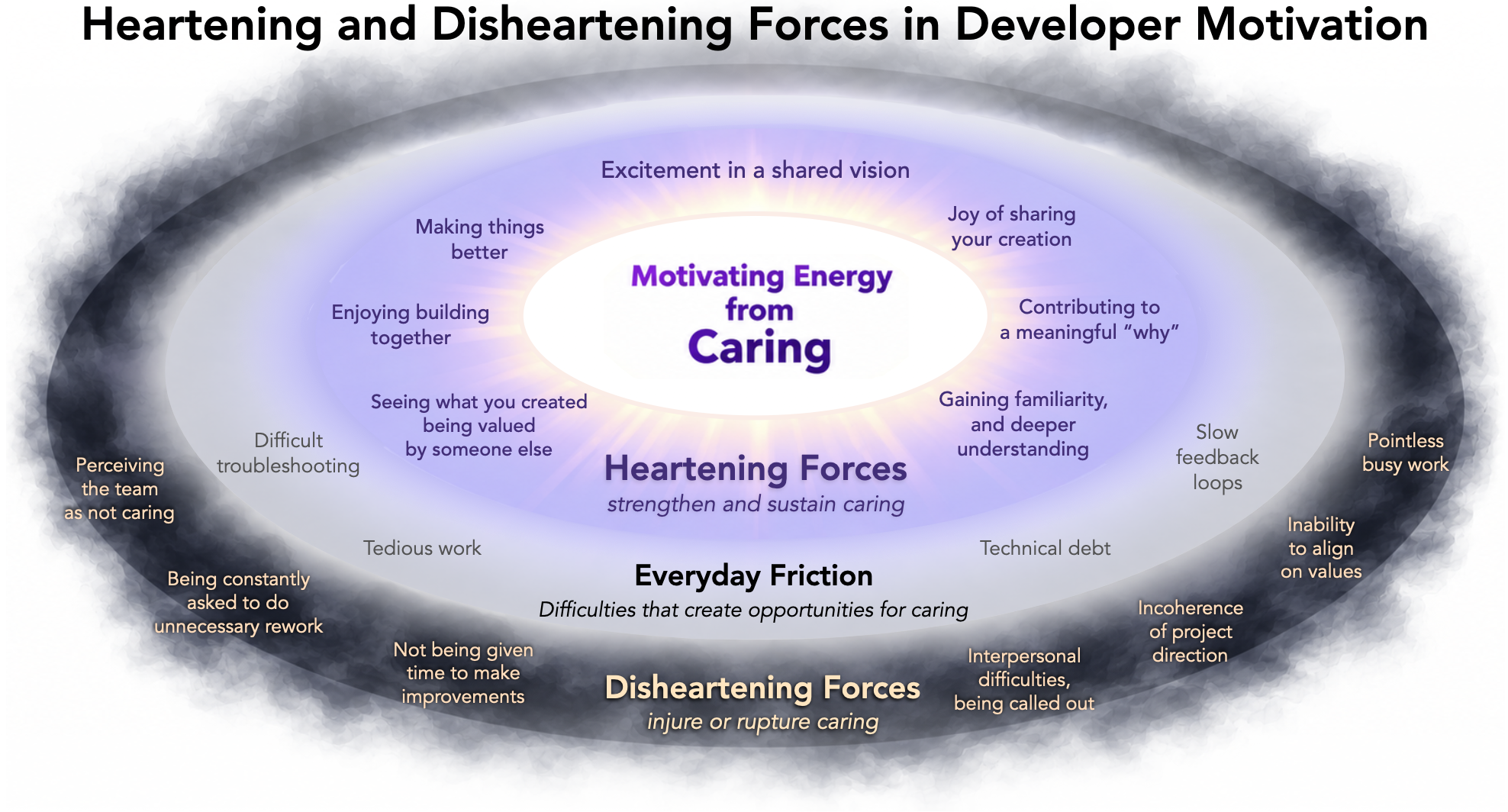}}
   \caption{With an understanding of caring as the source of motivating energy, our theory describes the external environment factors that can strengthen and sustain caring or injure and rupture caring.  These heartening and disheartening forces are in contrast with difficulties of everyday friction that create opportunities for improvement, but do not impact whether developers care. }
  \label{fig:motivation_forces}
\end{figure*}

With an understanding of caring as the source of motivating energy, our theory helps explain why factors that affect the developer's relationships with people, their sense of purpose, and connection to "the why" can have an outsized effect on motivation. We classify these factors into two groups: "disheartening forces" that injure or rupture caring, and "heartening forces" that strengthen or sustain caring (Figure~\ref{fig:motivation_forces}).

The language several developers (P01, P04, P06, P07, P08, P11) used to describe disheartening experiences such as "demoralizing", "discouraging", "emotionally draining", or "burning out", suggests that these experiences were qualitatively different from the everyday frictions of software development. The impact of disheartening experiences can be severe.  P11 and P08 described how being disheartened led to them eventually leaving their jobs. P01 described how a negative social experience and interpersonal issues destroyed their motivation:
\begin{quote}
\textit{"The worst experiences are pretty much always involving strict deadlines, unreasonable expectations, miscommunications, just interpersonal issues of people being mean or rude or whatever. One example, this was a year ago, but I still remember it almost as a traumatic experience, but it sticks in my head. [...] I made a mistake at one point in some code and the bug was originally my fault from months prior. Someone at work posted in a public Slack message, calling me out saying, "Hey, this is your fault. We shouldn't have ever merged this code." [...] I told my boss shortly after, that this completely [destroyed] my interest on this thing, and this project, I don't ever wanna touch it again. It was just a really negative experience."}
\end{quote}
The severity of the impact in P01's experience suggests that disheartening experiences can rupture caring, the source of motivating energy, causing motivation to suddenly drop to zero.  However, not all disheartening experiences result in immediate rupture. For example, P11 felt discouraged and demoralized by being repeatedly asked to rework code during code reviews that was not broken, and felt they were being treated unfairly when judgments were not applied consistently. P11 continued to care, but after several disheartening experiences found it increasingly difficult to feel motivated, and eventually burned out and quit their job.  P11's account suggests that as caring is repeatedly injured, continuing to care may eventually become too painful, leading developers to withdraw their caring as a protective response and coping strategy \cite{greiler2022actionable}.  Both of these examples highlight how caring is deeply relational, and a person's relationships with the people around them can impact their willingness to invest themselves in the work.

Another major environment factor that affected the developer's motivation related to their connection to the meaning and purpose of the work.  Developers became disheartened when it was difficult to see their effort as meaningful, valued, or contributing to something worthwhile.  P06 shared a story of how incoherence in project direction led to the developers becoming disheartened:
\begin{quote}
\textit{"On that project, it was not a clearly defined scope, so we spent a lot of time investigating one thing to be told it wasn't the right thing, and then pivot back to something else. There was a lot of circular direction from the project owners. That got very difficult and very frustrating very quickly. It seemed like every time we got a bit of momentum going, the carpet got pulled out from under us. [...] It was the lack of direction, we didn't know what we were building. [...] So that was demoralizing. It was just being stuck in never gaining momentum." }
\end{quote}
P06's account suggests that maintaining a coherent sense of direction is not simply important for project execution, but it also helps developers connect their day-to-day efforts to a purpose that they care about that helps them sustain motivation.  P08 described a remarkably similar experience attributing burnout to poorly defined projects, a lack of meaningful work, and month after month of struggle while being unable to gain momentum.  Similarly, P04 and P07 described becoming disheartened when they felt like they were wasting their time on busy work that did not have value, or efforts that would just be thrown away.

Developers also felt disheartened when they felt the shared space of the code and surrounding environment that they "lived in" day-to-day (the habitat)\cite{miles_sovereign_2026} was not being adequately cared for.  P07 described feeling demoralized because nobody would spend the time to pay down technical debt and it made it a lot harder to build up momentum.  This account suggests that the discouragement did not arise from the technical debt itself, but from what it communicated about the team's caring about the shared space.  The habitat had become a visible expression of how the work was collectively cared for.  When developers experienced the shared space as no longer being cared for, it became more difficult to continue investing their own caring into it.  P05 and P06 also described experiences with technical debt where they "enjoyed making things better", suggesting that even in a poor quality codebase if people are actively caring and making improvements, it does not feel demoralizing, instead it is actually energizing.

\subsubsection{Heartening Forces that Strengthen and Sustain Caring}

Once we understand caring as the source of motivating energy, the developers' descriptions of joyful, exciting, rewarding, satisfying, and "really good" feelings take on new significance. Rather than simply describing enjoyable moments, these experiences reveal the kinds of heartening forces that strengthen the "heart in it" that sustains caring itself.  One kind of heartening experience developers described involved seeing the effects of their caring having a meaningful impact.  P11 described such an experience:
\begin{quote}
\textit{"The most satisfying moment is when you've designed something, and built it, and then it's presented to people and they go, 'Oh, look at that, that's cool! It's so awesome! I've been wanting to be able to do that forever.' That moment when you have a sense of pride because you did something that improved somebody else's day—that's very satisfying." } 
\end{quote}
P11's account suggests that the heartening experience comes through having invested themselves in caring, then seeing their creation being valued by someone else. This example highlights how caring is deeply relational---it is heartening to see our efforts in creating something becoming valuable in another person's life.  P06 described the rewarding feelings after having done all the hard work on a task that brought so much value to the product and to the customers, feeling it was beautiful and being able to share it with others. P03 described the really good feelings after they put extra effort into an extensible design and it was noticed and appreciated by a coworker.  It is through these human relationships, mediated by the creation itself, that caring can be experienced, strengthened, and renewed.

In Section~\ref{dishearten}, we described how the software habitat becomes a visible expression of how the work and shared space that developers live within is collectively cared for.  P03, P04, P05, and P06 all described substantial efforts in caring for the habitat itself. P03 described testing tools that made it fast and easy to spin up APIs and write tests as you go, and code generation scripts that simplified repetitive workflows. P05 considered the ergonomics of how the APIs would feel to work with, and improved the semantics and understandability for future developers. P04 and P06 described refactoring efforts that would clean up the code and leave it in a better state. When the developers could see their caring reflected in a habitat that was clearer, more understandable, and more supportive of future work, they described feeling proud, happy, pleased, satisfied, and even joyful that they had made things better.  P11 described the satisfying feelings that could also come from gaining a deeper understanding of the software:
\begin{quote}
\textit{"[I like] getting an in-depth perspective on why things work the way they do.  As you work deeper into [a problem], and spend more time on a body of code, then you understand more about the little areas where there might be something kind of quirky and unexpected.  I find it very satisfying when you feel like you can be certain about why things are working the way they are, and how it's working." }
\end{quote}
P11's account suggests that coming to know the habitat deeply can also be heartening.  As developers spend time inhabiting a body of code, understanding its quirks, they gain a confidence in navigating the code that feels satisfying. 

Developers also described heartening experiences that emerged through building together with others, and invested significant efforts into supporting the people on their team. P03 enjoyed collaborating with their colleagues, and the experience of building together. P11 described how a positive supportive environment made a big difference in feeling enthusiastic about the work. P08 found it easy to enjoy tasks when their teammates had a high sensitivity to each other and could collaborate easily. P02, P05, and P09 all highlighted their primary responsibility toward the team, prioritizing unblocking others and helping their fellow engineers over their own independent work. These accounts suggest that developers come to see the team's long-term success as something they collectively care for. Through everyday acts of helping, coordinating, and supporting one another, the habitat becomes a place where people care. 

Further, when caring becomes aligned around a shared vision for the product, excitement can grow and creativity can become expansive as people begin imagining possibilities together.  P06 described how their work moved beyond just coding into a shared creative experience:
\begin{quote}
\textit{"When we got the MVP working, people saw the potential, and that's when everyone got excited. 'Can we do this? Can we do that?' And I was like, 'yes, we can do this! Yes, we do that!' And it just became this.. really, it moved beyond just coding. It moved into this team experience—where people were contributing, who wouldn't normally contribute to the code.  They weren't coding, but they were having a direct influence on what we were building. And as the people building it, we could rely on them to be part of that process—to be a positive contributor to making it a better product."}
\end{quote}
P06's account suggests that once the team collectively saw the potential in what they were building, the work itself changed character. As everyone got excited about bringing the vision to life, they began imagining new possibilities that no individual person could envision alone. Through shared caring by the team and the alignment of shared vision, the habitat became a place that invited creativity, imagination, and wholehearted participation in what they were creating together. Through these accounts, we begin to recognize what the shape of a thriving habitat might look like—where developers not only have the motivating energy to maintain momentum, but are energized by the experience.  A thriving habitat is a place where people's contributions are seen and valued, where people care for one another and for the habitat they share, where they come to know that habitat deeply, and where excitement in a shared vision invites creativity, imagination, and wholehearted participation in creating together.

\section{Discussion} \label{discussion}

Our theory reveals a rich craft that developers already use to get momentum going and sustain forward motion in their work. Developers experience progress as moving through the terrain of a task toward a fuzzy goal, and gaining clarity through that motion.  They have to build up working context in order to get immersed and get momentum going, and rebuild this working context sufficiently with every task session.  The motivating energy that drives the process forward comes from developers caring---and is shaped by daily freshness, strengthened by heartening forces, and energized with momentum.  By disentangling momentum, flow, and motivation, we can see the relationships between them and how they interweave to support progress.  By seeing these dynamics with greater resolution, our theory expands the operational surface area for developers to exercise skilled judgment and craft in maintaining motion.

In this section, we consider the implications and contributions of our theory in light of the present challenges in the AI era. We begin by examining how the theory helps illuminate both continuity and change in the developer's cognitive experience as the craft of developing software is being reshaped. We then turn to the predominant way of seeing through the manufacturing lens, the blind spots created by this lens, and the need for another way of seeing. Next, we demonstrate what the lens of developer momentum affords in seeing the motion of developing software. We then consider how the habitat's living capacity supports this motion. Finally, we return to new opportunities for craft and our agency in creating a habitat where teams can thrive.

\subsection{An Experiential Lens for Changing Cognitive Workflows}

Even before AI, developers struggled to articulate the challenges they were experiencing. Dockter described a communication breakdown where developers had a strong intuitive sense of when their systems were creating drag, but struggled to explain this ``gut feel'' to others \cite{gradle_dpe_2022}.  He argued that the software industry urgently needed better conceptual foundations that could explain the developer's experience in a way that did not require living it firsthand. Research and industry have made substantial progress in developing frameworks to understand and improve Developer Experience with DevEx, DPE and SPACE \cite{noda2023devex, peerless_making_2023, forsgren2021space}.  Yet despite these advances, understanding what is changing within the developer's experience and making developers' experiences and intuitions more legible to others remains a fundamental challenge in industry---a gap that becomes increasingly consequential as the activity transforms.

While the activity of developing software has shifted dramatically in the AI era, foundational concepts in our theory remain relevant. Whether the developer is prompting and reviewing AI-generated code, orchestrating agents, or writing code by hand, there is a human trying to make progress toward a goal.  We can examine how clarity is gained, how the developer stays oriented, and what working context the developer needs to make progress. The activity still unfolds across time and multiple task sessions where developers have to put the task down, return the next day, then get momentum going once again.  These experiential dimensions give us a way to examine how the developer's experience is changing without assuming the cognitive workflow is the same. 

After reading the paper, P02 sent us detailed reflections on what the theory revealed about their experiences, and they were struck by how well the theory captured the essence of their experiences across most of their career.  P02 also demonstrated its usefulness as a lens to examine how their cognitive experience was changing with agentic AI:
\begin{quote}
\textit{"First, I want to say that I think this is fantastic and remarkably thorough work. [...] One of the strongest validations of the theory for me is the distinction between flow and momentum. [...] Previously, implementation could put me into a deep, almost meditative state. The outside world disappeared, and I could spend a long time completely immersed in bringing something across the finish line.  That happens much less now. I may have several agents working simultaneously, move between tasks, review results, make decisions, and redirect agents when necessary.  And yet my sense of forward motion is stronger.  [...]  The theory holds up surprisingly well even when the shape of the development activity changes dramatically.  What changes is where its different components appear.  Flow can decrease while momentum increases.  Cognitive load can move from complexity within a task to context switching between tasks.  Caring can move from implementation toward outcome. Agency can move to a higher level of abstraction.  Competence can move from detailed implementation knowledge toward problem understanding, judgment, evaluation, and direction.  And knowledge that was previously a passive part of the habitat can become operational input that directly shapes how autonomous an AI agent can be.  To me, that makes the theory more interesting, not less. It suggests that Developer Momentum may provide a useful way of reasoning about the transition to agentic software development precisely because it lets us describe these changes without reducing the development experience to output or productivity alone."}
\end{quote}

Our findings revealed several consequential aspects of the developer's cognitive experience that may help explain the loss of flow reported in agentic AI workflows.  Sustaining immersion depends on continuity of orientation and the continual unfolding of a problem, where the developer sees the next step, then the next.  Yet when an AI agent works asynchronously and the developer must re-orient when it returns, this can break the continuity in which flow normally arises.  If these breaks in continuity prevent the developer from reaching the immersion threshold where attention stabilizes, maintaining focus may remain effortful, potentially contributing to greater cognitive fatigue.  The compressed downhill portion of the task, and the consequent experience of "always going uphill," may further compound these fatigue effects.  While future studies are needed to investigate these possibilities, our theory provides a rich experiential lens for recognizing what is changing in the developer's experience and opens new possibilities for inquiry.

\subsection{Unflattening Productivity: From Productivity to Momentum}

There is an intended aim of our software projects--a fuzzy goal that is often beyond words---and there are many lenses that can help us see and navigate more effectively. Yet the way we see software development shapes what we recognize as valuable, and what we risk overlooking. In this section, we examine the blind spots of seeing software development primarily through the lens of production, and why we need another way of seeing that brings the motion of the activity into view.  In particular, we consider the problem of missing understanding in the AI era, and precisely what type of understanding is needed to support developers' capacity for skillful action as the activity unfolds.

\subsubsection{The Production of Software}

Manufacturing has given us many successful ways of reasoning about production systems: we see flow of work, bottlenecks, waste, queues, throughput, cycle time, quality, feedback, constraints, and continuous improvement.  Every lens also highlights some details while hiding others.  This lens gives us our notion of "productivity" that foregrounds the production of work artifacts while backgrounding the design process.  Important contributions that may not “look like” productivity become harder to justify \cite{storey2019towards}.  SPACE, DX Framework, and others have already been trying to unflatten our aim. At Bridges Summit, we explored reframing developer productivity as a research and industry community and asked: \textit{"What are we aiming for?"} \cite{starr_reframing_2025}

If we are trying to get somewhere, producing more is not the same thing as motion toward where we are trying to go.  In this AI era, code can be produced really fast---but producing more does not mean having more impact, enabling more value, or bringing in more customers.  Producing software also creates responsibilities that do not disappear when AI takes on more of production.  Once people depend on the software, the team becomes accountable for its reliable operation, security, safety, maintenance, and commitments made to customers.  The team needs sufficient understanding to exercise good judgment, yet producing code with AI does not automatically bring a corresponding understanding of it.  Developing software requires more than production.  Teams need to understand the software, be able to exercise skilled judgment, and navigate toward the project's intended aim.

\subsubsection{The Problem of Missing Understanding}

Cognitive debt gives a name to the situation where developers are missing understanding and trying to make decisions. In order to operate and support systems in production, to diagnose and troubleshoot failures, to exercise judgment and know what to do, developers need to understand things. The need for understanding has not gone away just because the process that used to generate that understanding has changed.

"Understanding" is a broad term.  Everyone can agree it is important while imagining quite different things by it.  Our Theory of Troubleshooting gives a higher resolution lens to what, precisely, is missing that supports developers in knowing what to do \cite{starr2026theory}.  When the developer gets confused by a surprising system behavior, they rely on \textit{experiential intuition} to give them a hunch about where to look and what to try.  These hunches help the developer orient to the problem and narrow the space of possibilities. In our theory, these hunches draw on both general knowledge accumulated across a career and specific knowledge of a particular code base.  With cognitive debt, these hunches can be missing.  When developers can no longer resolve their confusion and make sense of what the system is doing, this loss of understanding can lead to a project-level loss of control and an inability to maintain the software.

The difficulties developers faced in understanding, editing, and debugging AI-generated code in Vaithilingam et al.'s study \cite{vaithilingam_expectation_2022} suggest that reading the code may be insufficient to restore the missing hunches. Sweller's formulation of schema acquisition from Cognitive Load Theory helps explain why \cite{sweller1988cognitive}.  Problem-solving expertise depends on \textit{schemas} that organize knowledge into long-term memory structures that allow a person to recognize similar problem states and know what to do---a concept similar to experiential intuition from our Theory of Troubleshooting.  In Sweller's maze studies, participants who were shown the goal could solve the problem, but failed to acquire the schemas learned by participants who had to discover the goal themselves.  They had succeeded at the task without acquiring the expertise for future problem-solving.  When AI generates the code, developers may achieve a solution without acquiring the schemas they would have in writing the code themselves, and reading the code afterward may not quite put them back.  If paying back cognitive debt means restoring the schemas and hunches, then reviewing the code being produced is not necessarily solving the problem.

\subsubsection{Understanding Needs to Be in the Foreground}

The long-term capacities we need for maintaining the software need to become a first-class consideration, rather than an overhead of production.  For example, Habitat Thinking values sustained human understanding over throughput, asking how we build places where “the understanding that keeps a system alive does not quietly leak away” \cite{miles_habitat_manifesto}. But what exactly is the understanding that keeps a system alive?

Ryle highlights the difference between \textit{knowing-that} a particular fact is the case, and \textit{knowing-how} to do an activity and having the skills to perform it well \cite{ryle1949concept}.  Ryle uses the example of a mountaineer walking through high winds over ice-covered rocks that must think what he is doing, test his steps, interpret new evidence, and innovate his strategy as the activity unfolds.  Knowing-how is expressed as a capacity to act intelligently---judging, adapting, innovating---in response to what is encountered. Ryle describes this capacity of knowing-how as possessing a theory that allows a person to perform and also explain, justify, and answer questions about it.

Naur builds on Ryle's ideas, describing how software developers build and possess a \textit{theory of the program} that supports their knowing-how \cite{naur1985programming}. In Naur's examples, he describes developers knowing how to diagnose faults and clear them up easily, and knowing how to recognize when design modifications would be problematic and suggest better alternatives.  When developers had access to all the code and documentation, but did not possess the theory of the program, they were unable to recognize these problematic designs and struggled to troubleshoot faults.  The understanding that keeps the system alive, then, that we need to bring to the foreground, is the \textit{living theory} that supports developers' capacity for skillful action.

\subsubsection{Supporting the Motion Within a Larger Activity}

Similar to the mountaineer finding a path through the terrain, the developer must navigate through the terrain of a task---assessing, judging, and making decisions as the activity unfolds.  The challenges that arise happen within these moments.  Does the developer have a hunch about what to do?  Fagerholm's conceptualization of Developer Experience centers the activity of developing software and the developer's experiencing of that activity \cite{fagerholm2015software}.  We operationalized this lens in our study as \textit{cognitive experience}---the developer's internal experience of thinking, feeling, and striving over time as an integrated whole---enabling us to see the richness of this activity in motion.  When we aggregate the experiences, the understandings, or the team as a whole, it is easy to lose sight of what is needed in these moments.

The developer's activity is also situated within the context of a team and a project with longer-term goals.  Fagerholm highlighted that this activity should be considered a \textit{design process} rather than a manufacturing process, describing it as a collaboration involving decision-making among many stakeholders \cite{fagerholm2015software}.  Even though we did not ask questions about the team, several developers in our study described synchronizing with their team as an inseparable part of what momentum means at the team level and over a longer term.  The social and technical environment is an integrated part of the experience \cite{fagerholm2015software}.  Naur also highlighted how once software is in operation, its very use inspires ideas for new features it ought to provide, and the need to continue modifying the software is inevitable \cite{naur1985programming}. As Storey described, no single individual needs to understand all aspects of the system, what matters is that sufficient understanding exists across the team to support making changes \cite{storey_technical_2026}.  The understanding that keeps a system alive at a team level, then, is the \textit{shared living theory} that supports the team's capacity for skillful action.

To support the activity of developing software, we need to consider multiple scales and resolutions without collapsing one into the other---the moments of the activity experienced by the individual, the coordination of the team, and the team's longer-term capacity to maintain the software.  The shift from productivity to momentum, then, is a shift from producing software to seeing and supporting a living activity in motion, finding its way toward its intended aim.

\subsection{Seeing the Motion of Developing Software}

In this section, we demonstrate how our theory helps us see the activity of developing software in motion at multiple scales and resolutions.  To make the lived experience more legible, we start with a first-person vantage point using the metaphor of a traveler.

\subsubsection{The Traveler: The Lived Scale of Motion}

At the closest experiential scale, I am a developer, a traveler on the journey of developing software, trying to accomplish a task.  At first: \textit{? in the fog.}  I have some orientation---a reason why I am here, an impact I am trying to achieve---but I have not constructed the route clearly enough yet. I do not enter empty-handed. I carry a \textit{repertoire}: technical knowledge, experiential intuition, problem-solving strategies, ways of poking and seeing, externalizing, researching, asking others, or breaking things apart. I know some of this terrain---I have wandered this area before. I have a living theory of this place and come equipped with all my knowing-how---so even though where I am going is fuzzy, I still feel confident that I can get there.

Some progress consists of making the path knowable.  I have to move in order to gain some clarity---encounter something that helps me better see the terrain.  Once I clear a bit of the fog, I will have a better idea of the path.  As I journey, I build up a working context of the relevant places and construct the path forward.  Eventually, I have to rest for the night, so I can come back fresh---which means rebuilding the working context so I can keep on going.  Sometimes, I encounter something unexpected: \textit{Wait... what?}  All my attention becomes focused on the confusing thing, and I poke and see until I figure it out: \textit{Ohhhh... that's why.}  My maps were wrong, but it is okay, I fixed them. \textit{What was I doing again?} I need to refresh my memory of the working context. I am finding my way. \textit{Oh! I see how to get there now.} I can see how far it is.  I can see the steps.  I feel excited and at ease---as it is all downhill from here.

\subsubsection{Gaining Clarity Through Motion and Finding Our Way}

There are a few things to notice about the traveler's experience. First, the journey involves a traversal through fog, and gaining clarity is fundamental to progress.  Motion enables clarity, and clarity enables motion. Before the developer has words, they have a fuzzy sense of direction---a \textit{compass} that helps them recognize a situation and know what to do---hunches inspired by their knowing-hows. Through motion, they gain enough clarity to see the next intelligent move. Eventually, as the developer reaches the crest of the hill, they can see all the way to the end---the fog has cleared.  This is when estimation becomes more reliable.

The moment of something unexpected---\textit{Wait... what?}---is the encounter with a \textit{puzzling event} from our Theory of Troubleshooting \cite{starr2026theory}.  The puzzling event triggers an orientation response that shifts the developer's attention to resolving the confusion.  This creates a cognitive context switch to a temporary goal of making sense of what is happening to repair the mental model.  Once the developer figures it out (\textit{Ohhhh... that's why.}), they have to recover their orientation toward the main task goal but some amount of working context is lost: \textit{What was I doing again?}. The traveler example illustrates how these encounters with puzzling events weave into the overall developer's journey.  These goal switches happen while the developer remains immersed.  Immersion can span progress or stuckness, joy or frustration, as the developer finds their way.

Seeing the activity in motion also repositions flow as a temporal process in which attention stabilizes on a task, rather than a binary state.  Getting in flow takes time and effort to ramp up from partial to full immersion, at which point distractions fall away and focus is sustained with ease.  The depth of flow corresponds to the cognitive puzzle size of a task, and deeper flow states are more heavily impacted by interruption.  Our study revealed an existing craft of managing flow depth where developers considered the puzzle size of the task, how responsive they needed to be, and whether sufficient time was available to get to an adequate flow depth for progress. 

Our theory opens new opportunities for researchers to investigate how the developer's cognitive experience is being reshaped in the AI era.  New agentic AI paradigms are rapidly evolving with spec-driven \cite{github_spec_2026}, loop engineering \cite{osmani_loop_2026}, graph engineering \cite{feng_graph_2026}, and AI factories \cite{wang_factory_2026} among the most recent.  To understand how the theory maps to these changing workflows, we can begin by recognizing the phenomena that the theory describes.  Rather than relying on workflow labels that are rapidly changing, we can recognize tasks by their experiential qualities.  From the perspective of the developer's cognitive experience, our theoretical notion of \textit{task} is the experiential process of building up working context to enable flow and momentum toward a fuzzy goal.  Researchers can investigate how developers build up working context, orient to a task, gain sufficient clarity to move, re-orient when agents return, become immersed, and get momentum going.  The uphill portion of a task can be distinguished from the downhill portion, and the fog and path visibility can be assessed along the way. Restarts are still designed across days to maintain motion. Because our theory centers the motion of problem-solving toward a fuzzy goal, these concepts can translate across changing workflows.

\subsubsection{The Team: The Project Scale of Motion}

Momentum is not simply an individual experience---it is interwoven with team momentum and connected to the overall aim of the project.  What impact is the project trying to achieve?  What value is it trying to enable?  There is a fuzzy vision of the project's aim and a fuzzy sense that it is even the right thing. P12 discussed how people's expectations and goals could change day to day, and highlighted the importance of checking in.  Stakeholders contributed context.  Customers participated through feedback.  Team members synchronized changes to maintain coherence in the evolving design.  P02 described how in agentic AI workflows very high individual momentum could outpace the team's capacity to verify and integrate the resulting changes.  Across many tasks, over the lifetime of a project, as developers come and go, the team develops a shared living theory.  All the distributed knowing-how gives the team hunches and helps them know what to do. There is knowledge about customers, and knowledge about the software, and knowledge about who knows what.  In this AI era, we have to externalize our intentions to support AI agents in making intelligent decisions too \cite{storey_technical_2026}.  At the project scale, the team is moving and gaining clarity, and through greater clarity moving more intelligently as they find their way toward the intended aim. Through this motion---through this continuous learning---the team "stumbles toward the right thing" \cite{miles_path_2026}.

\subsection{The Habitat's Living Capacity for Motion}

What kind of place sustains the team's living capacities such that motion toward the right thing becomes possible, enabled, and supported? Our multi-resolution concept of momentum makes room for the meaning to remain complex so we can recognize what the habitat needs to sustain living activities.  Habitat Thinking expands the design surface to the whole place \cite{miles_sovereign_2026}.  As inhabitants, we are designing the place that we live and create within.  We ask three questions about the habitat's living capacity for motion:

\begin{enumerate}
    \item \textbf{Possible} — Is there a path—a strategy that could move us in the direction we are aiming toward?
    \item \textbf{Enabled} — Do we have the tools, agency, time, and know-how to move intelligently toward our goals?
    \item \textbf{Supported} — What happens as we try? Do we express and experience caring, and do we sustain our shared living theory so we can continue moving intelligently?
\end{enumerate}

\subsubsection{From Fixed Capacity to Living Capacity}

In a software production frame, human capacity is represented as a quantity of time available for work.  A team of eight developers working eight hours has 64 developer-hours of capacity that can be allocated.  The underlying metaphor sees capacity as fixed: time is available, work consumes it, and the management problem is how to allocate this efficiently. Our theory reveals caring as the source of motivating energy and demonstrates how 8 hours of energized and excited time---when a developer feels heartened and starts with "yeah let's go"---is fundamentally different than 8 hours when a developer feels disheartened. These are living capacities.  Experiences that hearten or dishearten can change the motivating energy available to get stuff done.  Energies can also spread as humans get one another excited.  Unlike a fixed resource that is consumed through use, our living capacity can be strengthened, energized, or ruptured through the activity itself.  This is where the ecological nature of the habitat becomes clear.  It is ecological in that the living relationships between people, software, and the habitat continuously shape one another and affect the capabilities of the whole.

By recognizing our own living nature, we can design opportunities for heartening---make them possible, enabled, and supported by the design of the habitat.  Demos create opportunities for developers to encounter people affected by their work and see their contributions being valued.  We can learn about customers, connect with the why, make efforts to notice and help people feel appreciated.  We can make room in our process for caring to be expressed.  Developers refactoring, improving API ergonomics, and smoothing feedback loops all showed caring for the place.  The heartening was not about the state of the code, but whether people were able to express, experience, and see traces of care.

\subsubsection{Maintaining the Capacity for Skillful Action}

The habitat needs to support both the opportunity to care and the living theory needed to exercise that care skillfully. Developers remain accountable for the consequences of software they create with AI and need sufficient understanding to make responsible decisions. The big challenge in the AI era is the missing hunches.  There are many possible interventions that could support human understanding. One strategy is for AI agents to help maintain human hunches as a design goal. The \textit{diagnostic legibility} agent based on our Theory of Troubleshooting constructs diagnostic affordances as part of the deliverable that enables the developer to poke and see as a way to learn and gain hunches \cite{starr2026theory, starr_miles_diagnostic_legibility}.  As the developer learns, they share their hunches about how the system works, and the agent revises the design for conceptual clarity--an approach we call \textit{hunch-oriented telemetry}.  Since hunches are easier to observe than internal schemas \cite{sweller1988cognitive}, they give an empirical basis to evaluate whether interventions are having the intended effect.

\subsubsection{Designing Affordances for Stumbling}

One of the essential qualities of the experience of progress is the fuzzy goal and not knowing the path.  If we take this into account as an essential aspect of the experience, our habitat could be designed to support us in stumbling skillfully. "Vibe coding" is a way of coding with LLMs that Karpathy described as "fully giving in to the vibes" and "forgetting the code even exists" \cite{pimenova2025good}.  This keeps the activity in motion with imagining, encountering the idea, then iteratively refining---a powerful affordance for stumbling.  Pimenova et al. found joy and flow to be salient aspects of the vibing experience \cite{pimenova2025good}.  Yet the very thing that preserves this motion---not stopping to make sense of the code and letting AI make decisions---leaves developers with a missing living theory.  

One strategy to address this gap is to deliberately stumble twice. In the first pass, the developer clarifies the vision of what they want to create.  Once clarity is gained, the code changes are thrown away.  Next, the developer recreates the code changes, this time co-designing the conceptual internals of the software, stumbling toward a clear domain model while gaining hunches about what the system is doing.  The diagnostic legibility agent could support this process by exposing the domain model and higher-level conceptual abstractions, giving the developer a new surface to apply craft.

\subsection{New Opportunities for Craft}

One of our interview questions was about joy. We asked developers about a time they remembered experiencing genuine joy while programming.  This led to several beautiful stories on software projects---when everything felt aligned, alive, and full of energy. The stories helped us see the shape of a thriving habitat.  Once we can see the motion, and see the place, we have new opportunities.  With a fuzzy shape to guide our compasses, equipped with our knowing-hows to make new things possible, enabled, and supported---what if we were to stumble skillfully toward this aim?

Throughout the history of Developer Experience, people have aspired toward a better human experience and believed doing so would support productivity and better project outcomes.  Ruby was designed for programmer happiness, Gradle for build happiness, and DPE aspired to an alignment where "joy == productivity." Csikszentmihalyi's study of optimal experience also shows that the most enjoyable experiences do not arise from a restful state, but when the body or mind is stretched to its limit as the person tries to accomplish something both difficult and worthwhile \cite{csikszentmihalyi1990flow}.  We can create environments in which people do extraordinary work and have deeply satisfying experiences while doing it.

As AI accelerates the production of software, however, something significant in the developer's experience is changing. Vella et al. found that developers generally perceived productivity benefits from using AI, but also a drift toward a more negative developer experience, with worsening flow state as the predominant issue \cite{vella2026impact}. Participants described a shift from directly producing code to \textit{supervisory engineering work}---directing the AI and evaluating its output.  Industry thought leaders Kent Beck and Keith Adams described how programming did not have the flow state that it used to \cite{beck_air_2026}. Adams described grieving an ideal state of deep immersion and "communing with the problem" and how working with AI agents felt more like being an air-traffic controller---very interrupt driven.  Beck highlighted that what he loved about the flow state before AI was feeling so oriented that when he made a change in one place, he knew what would happen.  Beck acknowledged that more gets done with AI, but something real also went away.

With a production lens we are well equipped to see when more gets done, but less equipped to see what went away. Our theory brings resolution to Beck's account of "feeling so oriented" as deep familiarity, immersion, and continuity of orientation in the moment.  He knows this place. He has a living theory of it and can anticipate what will happen when he makes a change.  The experience of “communing with the problem” also conveys a deep relationship with what is being created. These experiences resonate with Csikszentmihalyi's concept of flow state and the sense of mastery that brings rewarding feelings \cite{csikszentmihalyi1990flow}. They also illustrate how experiences that feel deeply satisfying can overlap with doing extraordinary work.

The craft has not gone away; it has changed shape.  Recognizing what mattered in these experiences frees us to ask how we might make those same qualities possible, enabled, and supported within the emerging ways of developing software.  What does a place look like from within when its inhabitants are thriving?  The grounded accounts from developers help us see a shared vision worth aiming toward.  A thriving habitat is a place where people's contributions are seen and valued, where people care for one another and for the habitat they share, where they come to know that habitat deeply, and where excitement in a shared vision invites creativity, imagination, and wholehearted participation in creating together.  The particular acts of caring may change as the activity changes, but our caring still leaves traces that others encounter.  Caring still circulates through relationships, being expressed and experienced as heartening.  By recognizing the shape of a thriving habitat, we can attune our compasses toward the types of places we want to create. We need enough clarity to move, and through motion we can gain clarity, and find our way to a great place to live. 

\subsection{Evaluating the Theoretical Contribution} \label{evaluate}

In closing, we reflect on the contribution of this study using evaluation criteria from CGT \cite{charmaz_constructing_2014}. These criteria—credibility, originality, resonance, and usefulness—are particularly well-suited for evaluating theories grounded in lived experience and aimed at meaning-making in complex practice.

\vspace{0.3em}
\textbf{Credibility:} The study builds on prior grounded theory work with in-depth interviews from a theoretically sampled subset of 12 professional developers, representing 268 years of experience. The theory was developed using a rigorous process of initial grounded coding with high-resolution descriptive codes, constant comparison, and memoing to construct the emerging theory. 

\vspace{0.3em}
\textbf{Originality:} The Theory of Developer Momentum offers a novel paradigm that helps us see the activity of developing software in motion at multiple scales and resolutions, and recognize the living capacities that support it.  By disentangling momentum, flow, and motivation, our theory reveals the relationships between them and expands what becomes available for skilled judgment and craft.

\vspace{0.3em}
\textbf{Resonance:} Eight feedback interviews were conducted to evaluate resonance, gather feedback, and refine the models.  Participants indicated that the theory resonated with them and meaningfully captured their experience of progress, flow, motivation, and getting momentum going, even across changing AI-assisted workflows.  Feedback indicated that the terminology was relatable, practically useful, and supported reflection and decision-making. 

\vspace{0.3em}
\textbf{Usefulness:} Our theory illuminates a skilled craft of maintaining motion that developers already use and strategies that can be readily adopted. It helps teams and organizations to consider dynamics of momentum across multiple scales, recognize opportunities for improvement, and shape conditions for thriving.  Our theory also provides researchers with a rich experiential lens to investigate how developers' experiences are shifting in emerging AI workflows.

\vspace{0.3em}
Taken together, these criteria affirm the strength and grounding of the theory. At the same time, we recognize that every theory is partial, situated, and evolving. In the following section, we reflect on opportunities for future work.

\subsection{Future Work}

The rapid shift toward agentic AI paradigms creates an immediate opportunity to investigate how the developer's cognitive experience is changing.  Our research program has involved interviewing and building relationships with many of the same developers from 2023 to 2026 during this period of profound industry change.  Future work could build on this emerging longitudinal dataset to investigate how developers' experiences have changed across these shifts. Together, our Theory of Troubleshooting and Theory of Developer Momentum illuminate the activity of developing software in motion, and give us a way to investigate and map how the cognitive experience of problem-solving is shifting across these changing workflows.

Our theory also inspires new opportunities for designing tools to support developers.  A central challenge in the AI era is finding new ways to maintain human understanding and the living capacities to support skillful action.  We proposed the diagnostic legibility agent as a strategy to maintain human hunches as an explicit design goal, using hunches as an empirical basis to evaluate whether interventions have the intended effect. A follow-on design science study could evaluate whether this intervention works in practice.  New dashboards could also be designed to orient teams toward what needs tending to support forward motion, acknowledging that some progress means gaining clarity and making the path knowable. Future work could explore dashboard designs that center a motion-oriented paradigm. 

\section{Conclusion}

Our hope is that this theory helps readers to see the world differently. It brings into view the motion that a production lens flattens: the software developer traveling through fog, building and rebuilding working context, gaining clarity through motion, becoming immersed, stumbling, caring, becoming heartened or disheartened, and finding their way. At project scale, we see the team coordinating decisions and synchronizing changes to maintain coherence in the evolving design.  We see team momentum over a longer-term, the fuzzy aim of the project, and how the team stumbles toward the right thing.  When we look around us at the place---the habitat---we see the living capacities and shared living theory that support this motion.  We see people coming to know the habitat deeply, developing hunches and knowing-hows, caring for what they create and for one another, and leaving traces of that caring in the place.  Once we can see the motion, and see the place, greater resolution expands what becomes available for craft.  As inhabitants, we are shaping the place in which we exercise our own cognition and craft, and have agency in crafting the type of place we want to live.  We can shape what becomes possible, enabled, and supported and find our way to a place where we can thrive.

\section{Acknowledgments}

We would like to thank Russ Miles for the rich conversations at the intersection of momentum and habitats that inspired many ideas, the participants in the study for their valuable contributions, and Amber Horvath, Nathan Cassee, and the reviewers for their time, engagement, and thoughtful feedback that made this manuscript better.

\bibliographystyle{ACM-Reference-Format}
\bibliography{references}

\pagebreak

\section{Appendix} \label{appendix}

\raggedbottom
Included in the Appendix are: 

\begin{enumerate}
    \item the demographics of the participants in the study;
    \item a figure example of our initial grounded coding process;
    \item the 12 emerging questions we explored before narrowing our theoretical focus, corresponding to the 12 reports in the data archive with 2059 initial grounded codes sorted by emerging question;
    \item links to the Miro boards with the 98 theoretical categories prior to narrowing our focus and the 1469 unique initial grounded codes related to notions of momentum, flow, motivation, and strategies for maintaining motion;
    \item a diagram of data analysis terminology used in Constructivist Grounded Theory;
    \item the interview protocol and revised interview protocol from the iterative rounds of developer interviews, with the revised interview protocol appearing in full with changes highlighted in blue; and
    \item the feedback interview protocol, which shows an early revision of the model diagrams before any feedback was received to demonstrate how the models evolved during the study and show our work.
\end{enumerate}

We also provide a data archive on Zenodo (\underline{\url{https://doi.org/10.5281/zenodo.23022001}}) which includes the 12 emerging question reports with 2059 codes.  If a researcher wants to understand the variety of ways a developer experiences the progress and momentum, they can consult a dedicated report that includes 123 initial grounded codes, each labeled with corresponding participant numbers.  Similar reports exist for the flow state, motivation, interruptions, strategies used to focus, and other related topics.  These materials can help researchers deepen their understanding of the theoretical constructs when using the theory for research coding in their own studies.

\pagebreak

\subsection{Demographic Information By Participant} 

\begin{center} \begin{tabular}{l l r l l r l} \hline Participant & Organization Size & Team Size & Work Env. & Education & Years Exp. & Location \\ \hline P01 & 201-1000 people & 6 & Remote & Bachelor's degree & 11 & US \\ P02 & 51-200 people & 5 & Remote & Bachelor's degree & 30 & Europe \\ P03 & 1-50 people & 7 & Remote & Bachelor's degree & 15 & Canada \\ P04 & 51-200 people & 6 & Remote & Bachelor's degree & 32 & US \\ P05 & \textgreater{} 1000 people & 7 & Colocated & Some college, no degree & 27 & US \\ P06 & \textgreater{} 1000 people & 20 & Hybrid work & Bachelor's degree & 20 & Canada \\ P07 & \textgreater{} 1000 people & 4 & Remote & Post-graduate degree & 30 & US \\ P08 & \textgreater{} 1000 people & 8 & Remote & Bachelor's degree & 23 & Canada \\ P09 & \textgreater{} 1000 people & 7 & Remote & Post-graduate degree & 17 & US \\ P10 & 1-50 people & 1 & Remote & Bachelor's degree & 35 & US \\ P11 & 1-50 people & 1 & Remote & Bachelor's degree & 18 & Canada \\ P12 & \textgreater{} 1000 people & 5 & Remote & Bachelor's degree & 10 & US \\ \hline \end{tabular} \vspace{1ex} \parbox{1.0\linewidth}{ \footnotesize \textit{Gender information is summarized to preserve participant anonymity. Four participants were identified as women and eight as men, based on publicly available information on LinkedIn (unconfirmed).} } \end{center}

\subsection{Initial Grounded Coding Process Detail}

\begin{figure}[H]
\centering
\includegraphics[scale=0.44]{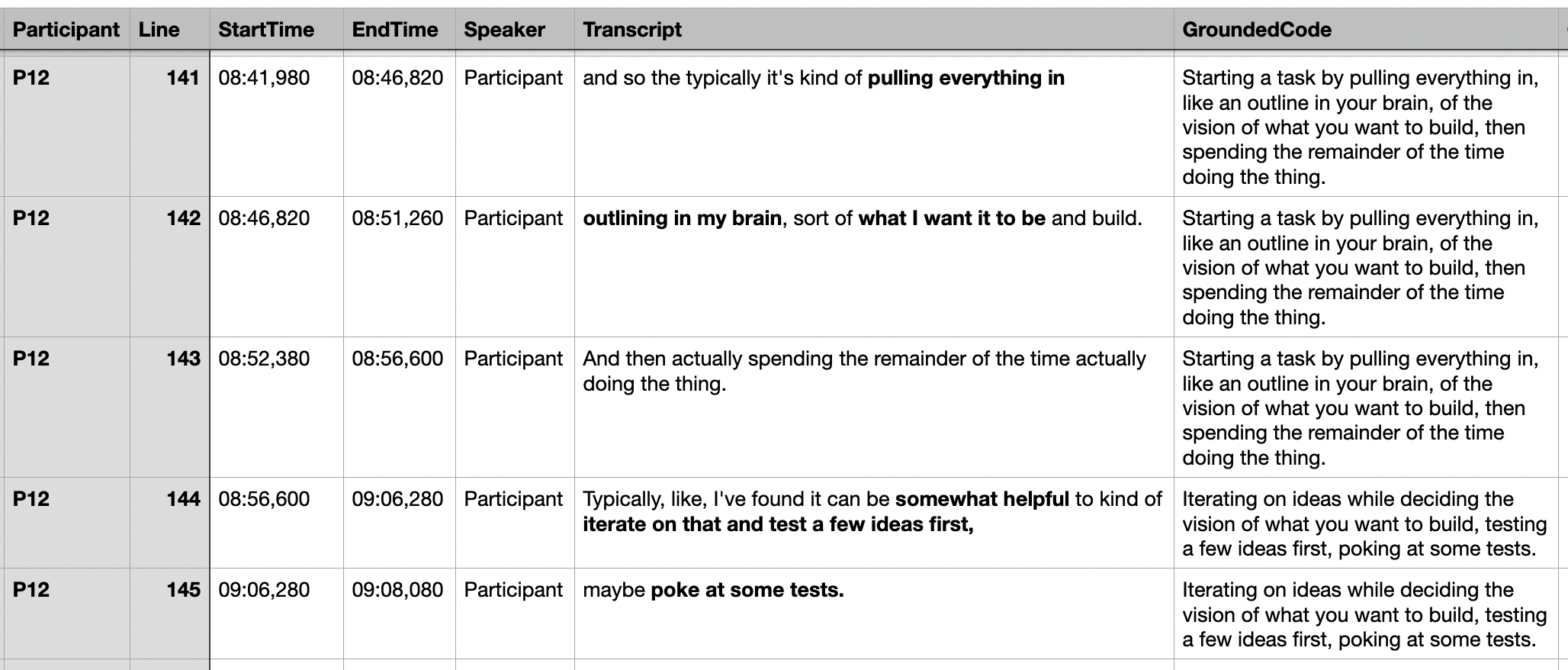}
\caption{Example of the initial grounded coding process from P12. We first bolded key ideas while listening to the interview to make the transcripts easier to scan during subsequent reviews of the transcript.  Then we went line-by-line creating high-resolution initial grounded codes that started with gerunds to interpret what was happening in the scene.  These codes were intended to be atomic so that we could sort them on Miro boards without having to revisit the transcript to interpret them. }
\label{fig:grounded_coding}
\end{figure}

\pagebreak

\subsection{Emerging Questions Related to Momentum, Flow and Motivation} \label{rq}
\vspace{1em}

\begin{itemize}
\item {How do developers start their day?  What difficulties arise with getting started?  How does the experience in the morning contrast with other parts of the day? 
}
\item {What strategies do developers use to help them focus and get started? What challenges do developers experience with getting focused?
}
\item {What is the experience like when first starting a task, and then eventually moving into coding?  How does the task progress over time?
}
\item {What strategies do developers use to maintain momentum when taking breaks or across days?  What strategies do developers use to boost momentum?
}
\item {How do developers collaborate with their peers to maintain momentum?
}
\item {When do developers experience context-switching, and how do they think about this notion?
}
\item {What interruptions or distractions do developers experience?  How do developers describe the impact of an interruption?  How are these effects modulated by different contexts, such as where you are within the progress of the overall task?
}
\item {How do developers describe and experience the flow-state?  What factors seem to affect how absorbed they feel, or enjoy what they are doing?
}
\item {What factors influence the developer's feeling of motivation toward working on a task?  (social, technical, other)
}
\item {How are the developer’s meetings organized and how does that impact momentum?
}
\item {How do developers define or describe the notion of momentum?
}
\item {How do developers use AI in their workflow to support momentum, and what challenges and benefits do they describe?
}
\end{itemize}

\pagebreak

\subsection{Miro Boards to Support Constant Comparison} 

\vspace{1em}
\textbf{Theoretical Categories Miro Board}

\vspace{1em}
\noindent This board was used to sort our 98 initial theoretical categories into 13 themes, and a temporal perspective of progress through a task from start to finish began to emerge.  This helped us narrow our theoretical focus and identify gaps that still needed more data, leading to a revised interview protocol to answer our emerging questions and fill in these gaps.

\noindent  \href{https://miro.com/app/board/uXjVJAugMh8=/?share_link_id=776072223414}{\underline{Miro Board}}

\vspace{1em}
\noindent \textbf{Developer Notions of Momentum}
\vspace{1em}

\noindent This board was used to sort 123 initial grounded codes into 58 focus codes related to our emerging question: "How do developers define or describe the notion of momentum?"

\noindent  \href{https://miro.com/app/board/uXjVG6UEpEs=/?share_link_id=101392945584}{\underline{Miro Board}}

\vspace{1em}
\noindent \textbf{Developer Notions of Flow State}
\vspace{1em}

\noindent This board was used to sort 207 initial grounded codes into 76 focus codes related to our emerging question: "How do developers describe and experience the flow-state?  What factors seem to affect how absorbed they feel, or enjoy what they are doing?"

\noindent \href{https://miro.com/app/board/uXjVG6VMvuQ=/?share_link_id=434801008833}{\underline{Miro Board}}

\vspace{1em}
\noindent \textbf{Developer Notions of Motivation}
\vspace{1em}

\noindent This board was used to sort 156 initial grounded codes into 86 focus codes related to our emerging question: "What factors influence the developer's feeling of motivation toward working on a task?  (social, technical, other)"

\noindent \href{https://miro.com/app/board/uXjVG6Vb5rU=/?share_link_id=8998568219}{\underline{Miro Board}}

\vspace{1em}
\noindent \textbf{Strategies for Maintaining Motion}
\vspace{1em}

\noindent This board was used to sort 341 initial grounded codes into 167 focus codes related to our emerging question: "What strategies do developers use to maintain momentum when taking breaks or across days?  What strategies do developers use to boost momentum?"

\noindent \href{https://miro.com/app/board/uXjVG6Vclt4=/?share_link_id=20891643076}{\underline{Miro Board}}

\vspace{1em}
\noindent \textbf{How Developers use AI to Support Momentum}
\vspace{1em}

\noindent This board was used to sort 37 initial grounded codes into 21 focus codes related to our emerging question: "How do developers use AI in their workflow to support momentum, and what challenges and benefits do they describe?"

\noindent \href{https://miro.com/app/board/uXjVG6Ux9UE=/?share_link_id=380255128668}{\underline{Miro Board}}



\section*{Supplementary material}

# Data Analysis Terminology in Constructivist Grounded Theory

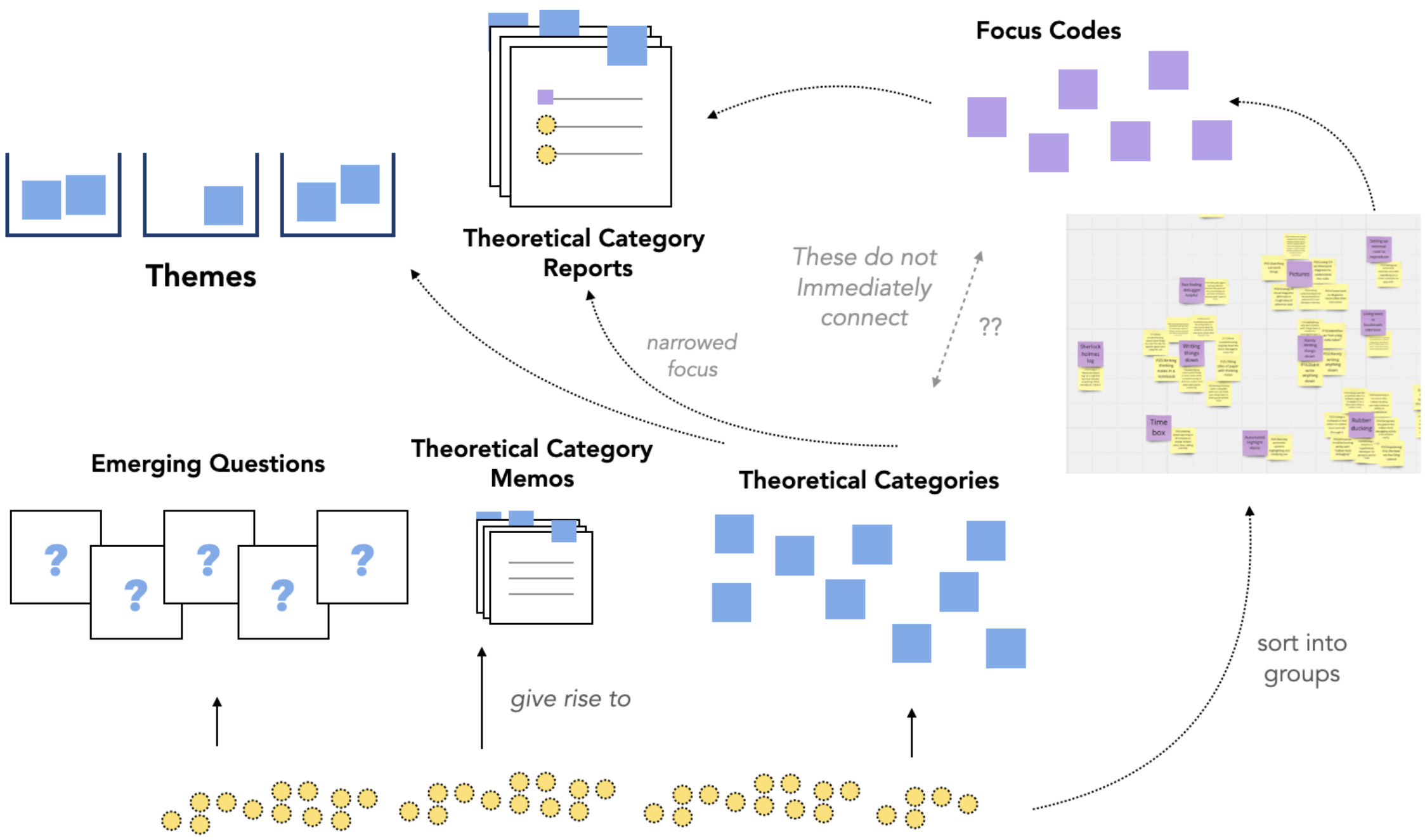


**Initial Grounded Codes** = high-resolution descriptive codes that stay as grounded as possible in interpretation, that start with gerunds and capture atomic units of what is happening in the scene

# Developer Momentum Study Interview Protocol

**Interview goals:** Collect qualitative data *to understand the developer's experience of getting momentum going in software development, including what the flow experience feels like at different stages of the task, what's required to make progress, barriers to getting in flow, impact of meetings or interruptions, and strategies to get back in flow and maintain momentum more easily. We also explore situational factors that might affect the experience such as different types of tasks, personal motivations, and amount of feedback. Momentum is a relatively new paradigm and unexplored area of largely tacit knowledge. This study illuminates what's happening in Developer Experience as the developer gets going with a task, and reveals developer wisdom related to momentum.*

**Introduction:** Interviewer describes the context, the motivation and the goal of the study. The participant is also informed that all information collected will be anonymized and it won't be shared outside the team of researchers directly involved in the project. The interviewer asks the participant for his/her consent to record the interview as well.

*Hi [NAME], and welcome. I am Arty Starr, a researcher from the computer-human interaction and software engineering lab, CHISEL, at University of Victoria, working in collaboration with my supervisor, Dr. Margaret-Anne Storey.*

*In this study, we would like to learn about your experiences of getting in flow and getting momentum going with software development. This includes challenges to getting in flow or staying in flow, how you get momentum going at the beginning of a task, what the flow experience feels like for you at different stages of the task, how meetings and interruptions affect your ability to get momentum going, and any strategies you might use to get back in flow more easily or that help you maintain momentum. My PhD thesis is focused on constructing a Theory of Developer Flow as two cognitive processes of Momentum and Troubleshooting, "Momentum" being this nuanced process of "getting momentum going".*

*I also want to inform (or remind) you that I've personally been a software engineer in industry for about 20 years, so it's okay if you speak in familiar developer terms in answering the questions, I'll be able to understand what you mean.*

*Only the researchers involved in this study will see your responses, and we will summarize the responses so they are fully anonymized. Your participation in this study is voluntary, and you do not have to answer any question that makes you uncomfortable, and you can stop the interview whenever you want. We'd like to encourage you to refrain from sharing identifying information as much as possible, such as names of 3rd parties, and to refrain from sharing any confidential information during the interview.*

*The interview should take no longer than 60 min. During the interview we will ask several questions, and we would appreciate it if you respond only based on your personal experiences. If needed, feel free to take your time in answering the questions.*

*So, before we get started, do you have any questions?*

*If this is clear, and you have no more questions, do I have your permission to record the interview?*

**Context: Getting started first thing in the morning**

We're going to be talking a lot about this process of **getting in flow** and **getting momentum going**. The first context where I'd like to talk about this with you, is what happens first thing, when you're starting your day, and wanting to get momentum going.

1. When do you normally start your day? What does a typical morning (or other time) look like for you when you're trying to get started?
2. What are some of the challenges you experience with getting momentum going first thing in the morning?
3. Have you found any strategies that seem to help or make it easier to get started in the morning?

**Context: Getting momentum going with coding**

So I want you to think of a software development task you did recently, that was a fairly typical task for you. For these next questions, I want to talk to you about **what happens in between** starting the task and the point where you're deep into coding and going at a pretty good pace. So first, just think of a recent task you worked on, and then I'll ask you questions about it.

1. Can you walk me through what steps you went through to get going on the task? Where did you start? How did you get momentum going with coding?
2. What's required in order for you to be able to get going with coding on a task? What factors make getting momentum on a task easier, or more difficult?
3. How did you feel as you were working on the task? Did your feelings change as you got going with the task?
4. How absorbed did you feel in what you're doing on a scale from 1-5? Does that experience vary for you? What factors seem to affect how absorbed you feel?
5. How much would you say you were enjoying what you were doing, on a scale from 1-5? Does that experience vary for you?
6. If you think of a time, if any, you've experienced joy while coding, what memory comes up for you? How does that experience compare to what is typical for you?

**Context: Impact of Distractions, Interruptions, and Meetings**

So next, I want to talk about the impact on your ability to get momentum going when you also have **distractions, interruptions, or meetings** to attend.

1. If you reflect for a moment on what's happening in your internal world, what are some of other things you find contending for your attention as you try to get momentum going on a task? How do those things impact your ability to get momentum going?
2. How is your day typically broken up by meetings, and how does that impact your ability to get momentum going?
3. What other types of interruptions, if any, do you experience during the day, and how do you find those things affecting your momentum? What happens when you are resuming the task after a meeting or interruption?
4. Are there any strategies you've found that seem to help you rebuild momentum?

**Context: Sustaining Momentum and Bridging Flow to Tomorrow**

So often a software task will take multiple days to complete, and we have to **sustain momentum** over a period of multiple days, so this is the context I'd like to talk to you about next. So first, just think of a recent task you worked on that took multiple days to finish, and then I'll ask you questions about it.

1. So can you give me a little context for what the task was that took multiple days, and how what you were doing changed from day to day?
2. What was the experience like of trying to resume the task from where you left off the day before? Did the experience vary on different days? How did your feeling of momentum change?
3. What factors make it easier to sustain momentum over a period of time? Can you think of an experience where it felt easy for you?
4. Are there any strategies you've found that seem to help you sustain momentum from day to day?

**Closing**: Interviewer thanks the participant for joining the study and briefly describes the next steps in the study and the possibility to contact him/her again for a focus group session.

*That was my last interview question. Is there anything else you want to share about your experiences, or things that you think should be included in the study?*

*Thank you for participating in the study, this concludes the interview. If we need to do a short follow up interview to ask a few more questions, would it be alright if we contacted you about a follow up interview?*

*We're also considering adding an empirical data component to this study, where we have tools that generate metrics around momentum and flow, and you can learn about your patterns of momentum, and the impact of things like meetings, with empirical data. It is a more involved process though. There's an opportunity to learn a lot, and the data is anonymized, but I realize there's also challenges with trying to gather data in a real context. Would this be something you would be potentially interested in, that we could email you about?*

*Our next step is to process and transcribe this interview. We will securely process the recording and the transcript. If at any point in time, you decide you want to withdraw from the study, you can email us and we will remove your data. Also feel free to email us if you have any questions.*

*Do you have any final questions or any further comments?*

*Alright, I'm going to stop recording.*

*Thank you very much for your time.*

# Developer Momentum Study Interview Protocol (Revised)

**Interview goals:** Collect qualitative data *to understand the developer's experience of getting momentum going in software development, including what the flow experience feels like at different stages of the task, what's required to make progress, barriers to getting in flow, impact of meetings or interruptions, and strategies to get back in flow and maintain momentum more easily. We also explore situational factors that might affect the experience such as different types of tasks, personal motivations, and amount of feedback. Momentum is a relatively new paradigm and unexplored area of largely tacit knowledge. This study illuminates what's happening in Developer Experience as the developer gets going with a task, and reveals developer wisdom related to momentum.*

**Introduction:** Interviewer describes the context, the motivation and the goal of the study. The participant is also informed that all information collected will be anonymized and it won't be shared outside the team of researchers directly involved in the project. The interviewer asks the participant for his/her consent to record the interview as well.

*Hi [NAME], and welcome. I am Arty Starr, a researcher from the computer-human interaction and software engineering lab, CHISEL, at University of Victoria, working in collaboration with my supervisor, Dr. Margaret-Anne Storey.*

*In this study, we would like to learn about your experiences of getting in flow and getting momentum going with software development. This includes challenges to getting in flow or staying in flow, how you get momentum going at the beginning of a task, what the flow experience feels like for you at different stages of the task, how meetings and interruptions affect your ability to get momentum going, and any strategies you might use to get back in flow more easily or that help you maintain momentum. My PhD thesis is focused on constructing a Theory of Developer Flow as two cognitive processes of Momentum and Troubleshooting, "Momentum" being this nuanced process of "getting momentum going".*

*I also want to inform (or remind) you that I've personally been a software engineer in industry for about 20 years, so it's okay if you speak in familiar developer terms in answering the questions, I'll be able to understand what you mean.*

*Only the researchers involved in this study will see your responses, and we will summarize the responses so they are fully anonymized. Your participation in this study is voluntary, and you do not have to answer any question that makes you uncomfortable, and you can stop the interview whenever you want. We'd like to encourage you to refrain from sharing identifying information as much as possible, such as names of 3rd parties, and to refrain from sharing any confidential information during the interview.*

*The interview should take no longer than 60 min. During the interview we will ask several questions, and we would appreciate it if you respond only based on your personal experiences. If needed, feel free to take your time in answering the questions.*

*So, before we get started, do you have any questions?*

*If this is clear, and you have no more questions, do I have your permission to record the interview?*

**Context: Getting started first thing in the morning**

We're going to be talking a lot about this process of **getting in flow** and **getting momentum going**. The first context where I'd like to talk about this with you, is what happens first thing, when you're starting your day, and wanting to get momentum going.

1. When do you normally start your day? What does a typical morning (or other time) look like for you when you're trying to get started?
2. What are some of the challenges you experience with getting momentum going first thing in the morning?
3. Have you found any strategies that seem to help or make it easier to get started in the morning?

**Context: Getting momentum going with coding**

So I want you to think of a software development task you did recently, that was a fairly typical task for you. For these next questions, I want to talk to you about **what happens in between** starting the task and the point where you're deep into coding and going at a pretty good pace. So first, just think of a recent task you worked on, and then I'll ask you questions about it.

1. Can you walk me through what steps you went through to get going on the task? Where did you start? How did you get momentum going with coding?
2. What's required in order for you to be able to get going with coding on a task? What factors make getting momentum on a task easier, or more difficult?
3. How did you feel as you were working on the task? Did your feelings change as you got going with the task?
4. How absorbed did you feel in what you're doing on a scale from 1-5? Does that experience vary for you? What factors seem to affect how absorbed you feel?
5. How much would you say you were enjoying what you were doing, on a scale from 1-5? Does that experience vary for you?
6. Was there any part of the task where you felt you had to germinate on the ideas?
7. What types of activities did you do to help you germinate?
8. Is that typical for you, or are there any other types of activities you find yourself doing to help you germinate?

**Context: Impact of Distractions, Interruptions, and Meetings**

So next, I want to talk about the impact on your ability to get momentum going when you also have **distractions, interruptions, or meetings** to attend.

1. If you reflect for a moment on what's happening in your internal world, what are some of other things you find contending for your attention as you try to get momentum going on a task?
2. How is your day typically broken up by meetings, and how does that impact your ability to get momentum going?
3. What other types of interruptions, if any, do you experience during the day, and how do you find those things affecting your momentum? What happens when you are resuming the task after a meeting (or interruption)? (Interruption) Is that any different than coming back from a meeting?
4. Are there any strategies you've found that seem to help you rebuild momentum?
5. Are there any parts of the task that are more sensitive to interruption than other parts?

**Context: Sustaining Momentum and Bridging Flow to Tomorrow**

So often a software task will take multiple days to complete, and we have to **sustain momentum** over a period of multiple days, so this is the context I'd like to talk to you about next. So first, just think of a recent task you worked on that took multiple days to finish, and then I'll ask you questions about it.

1. So can you give me a little context for what the task was that took multiple days, and how what you were doing changed from day to day?
2. What was the experience like of trying to resume the task from where you left off the day before? Did the experience vary on different days? If momentum is a sense or a feeling, how did your feeling of momentum change across the different parts of the task?
3. What factors make it easier to sustain momentum over a period of time? Can you think of an experience where it felt easy for you?

**Closing**: Interviewer thanks the participant for joining the study and briefly describes the next steps in the study and the possibility to contact him/her again for a focus group session.

*That was my last interview question. Is there anything else you want to share about your experiences, or things that you think should be included in the study?*

*Thank you for participating in the study, this concludes the interview. If we need to do a short follow up interview to ask a few more questions, would it be alright if we contacted you about a follow up interview?*

*We're also considering adding an empirical data component to this study, where we have tools that generate metrics around momentum and flow, and you can learn about your patterns of momentum, and the impact of things like meetings, with empirical data. It is a more involved process though. There's an opportunity to learn a lot, and the data is anonymized, but I realize there's also challenges with trying to gather data in a real context. Would this be something you would be potentially interested in, that we could email you about?*

*Our next step is to process and transcribe this interview. We will securely process the recording and the transcript. If at any point in time, you decide you want to withdraw from the study, you can email us and we will remove your data. Also feel free to email us if you have any questions.*

*Do you have any final questions or any further comments?*

*Alright, I'm going to stop recording.*

*Thank you very much for your time.*

## Momentum Feedback Interview Protocol (Validate Resonance)

**Interview goals:** Collect qualitative data *to validate whether the theoretical models constructed during the first set of interviews resonate with developers, and discover whether there are gaps or misunderstandings and how the developers interpret the models.*

**Introduction:** Interviewer describes the context and purpose of the second interview, and recalls the motivation and goal of the study. The participant is also informed that all information collected will be anonymized and it won't be shared outside the team of researchers directly involved in the project. The interviewer asks the participant for his/her consent to record the interview as well.

*Hello [NAME], and welcome. We are Arty Starr, and Margaret-Anne Storey, researchers from the computer-human interaction and software engineering lab, CHISEL, at University of Victoria.*

*This is a follow up interview for the Developer Momentum Study that you were interviewed for last year as part of my PhD research, where we were focused on understanding your experiences of getting in flow or staying in flow, and how you get momentum going on a task. Since then, we have completed 12 developer interviews and constructed theoretical models of the cognitive experience of flow and momentum. Our goal today is to see if those models resonate with you, and to discover whether there's any gaps, misunderstandings, or clarifications needed in the models, and to get your feedback and thoughts on how you interpret the models.*

*Only the researchers involved in this study will see your responses, and we will summarize the responses so they are fully anonymized. Your participation in this study is voluntary, and you do not have to answer any question that makes you uncomfortable, and you can stop the interview whenever you want. We'd like to encourage you to refrain from sharing identifying information as much as possible, such as names of 3rd parties, and to refrain from sharing any confidential information during the interview, and we would appreciate it if you respond only based on your personal experiences.*

*This follow up interview should take no longer than 30 min. During the interview we will show you several diagrams that show aspects of the theoretical models, and ask you questions about what you see. I also want to make it clear that it's okay if you disagree or have different ideas. Our goal is to improve the clarity of our understanding, so all of this feedback is helpful. If needed, feel free to take your time in answering the questions.*

*So, before we get started, do you have any questions?*

*If this is clear, and you have no more questions, do I have your permission to record the interview?*

**Resonance Testing for Developer Momentum Models**

**1.1. Momentum/Flow/Motivation High-Level Model** - So I'm going to be showing you a series of diagrams that together help paint a picture of the cognitive experience related to Developer Momentum. So first, if you want to just take a look at the diagram, and tell me what you see represented.

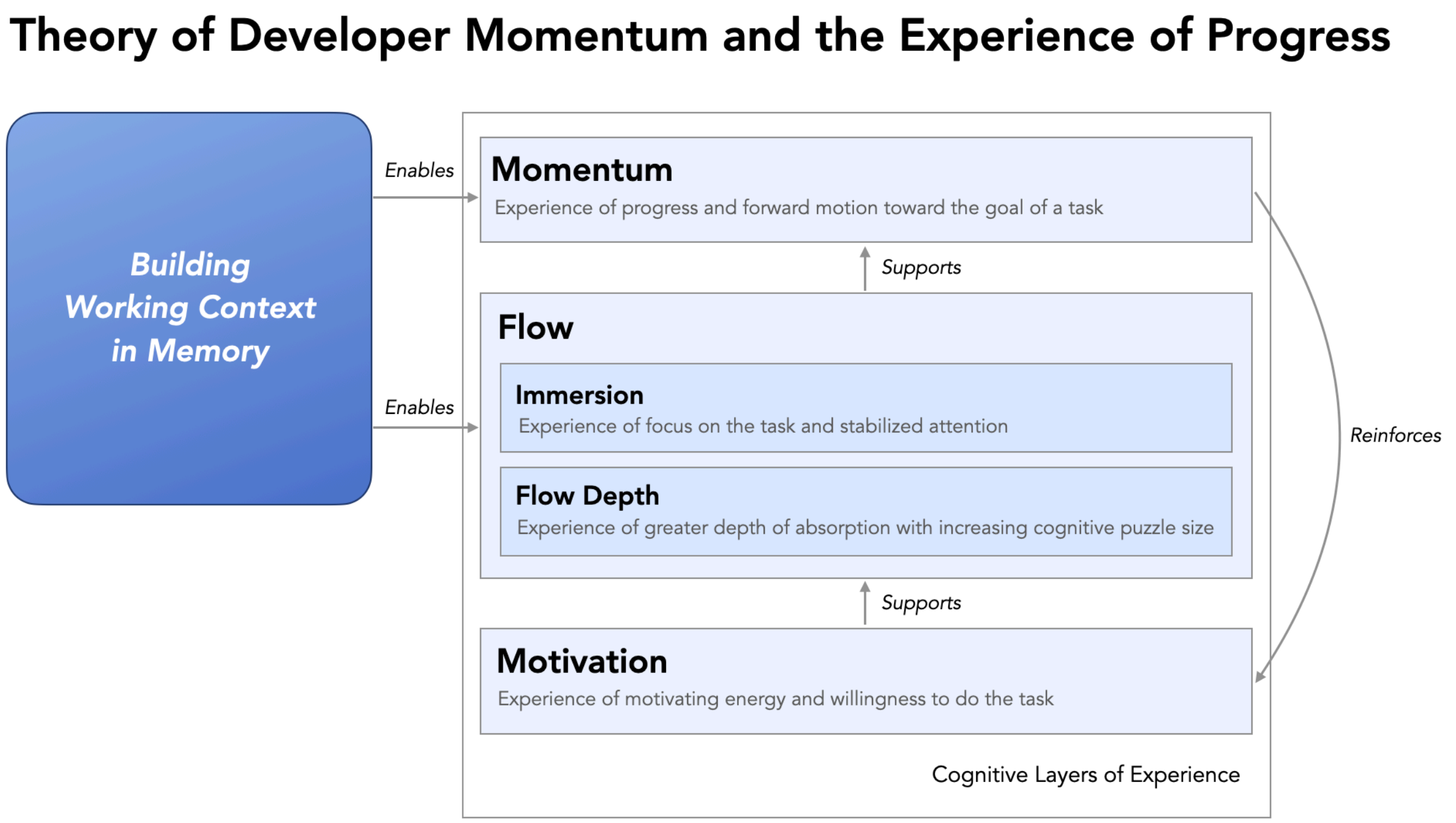


1.2. Do you feel like this resonates with you? Is there anything here that stands out to you that feels inaccurate or that you would change? (If needed, clarify the meaning of resonance, by rephrasing: "Does your experience match what I'm showing you? If it doesn't, I'd like to know...")

1.3. On a scale from 1 to 5, one being it doesn't resonate and feels very different than how you would think about it, and 5 being it resonates and fits with your own mental model, how would you rate this on a scale from 1 to 5?

**2.1. Experience of Momentum Hill** - So next diagram, if you want to just take a look and tell me what you see represented...

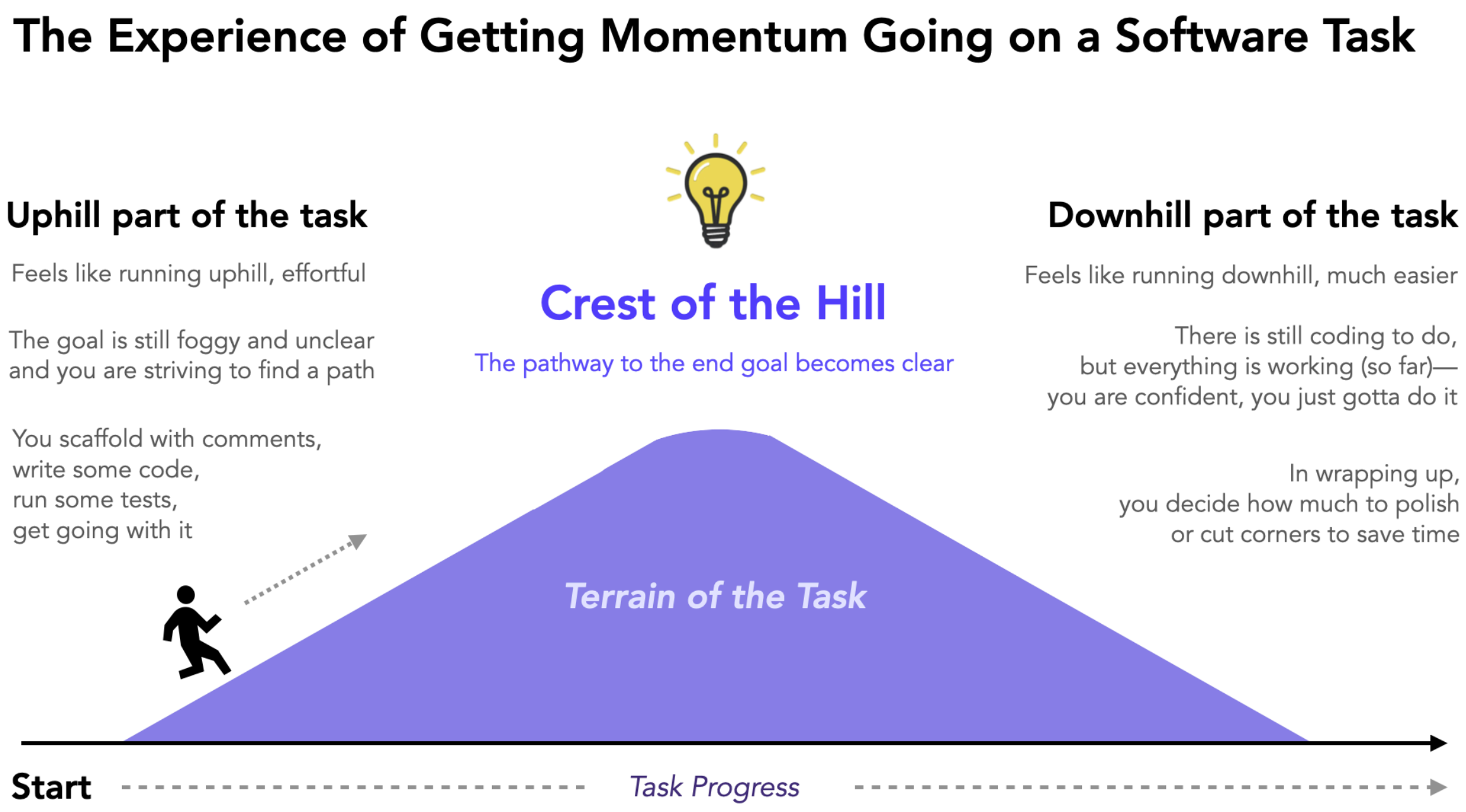


2.2. Do you feel like this resonates with you? Is there anything here that stands out to you that feels inaccurate or that you would change?

2.3. On a scale from 1 to 5, one being it doesn't resonate and feels very different than how you would think about it, and 5 being it resonates and fits with your own mental model, how would you rate this on a scale from 1 to 5?

**3.1. Fuzzy to Clear Experience** - So next diagram, if you want to just take a look and tell me what you see represented...

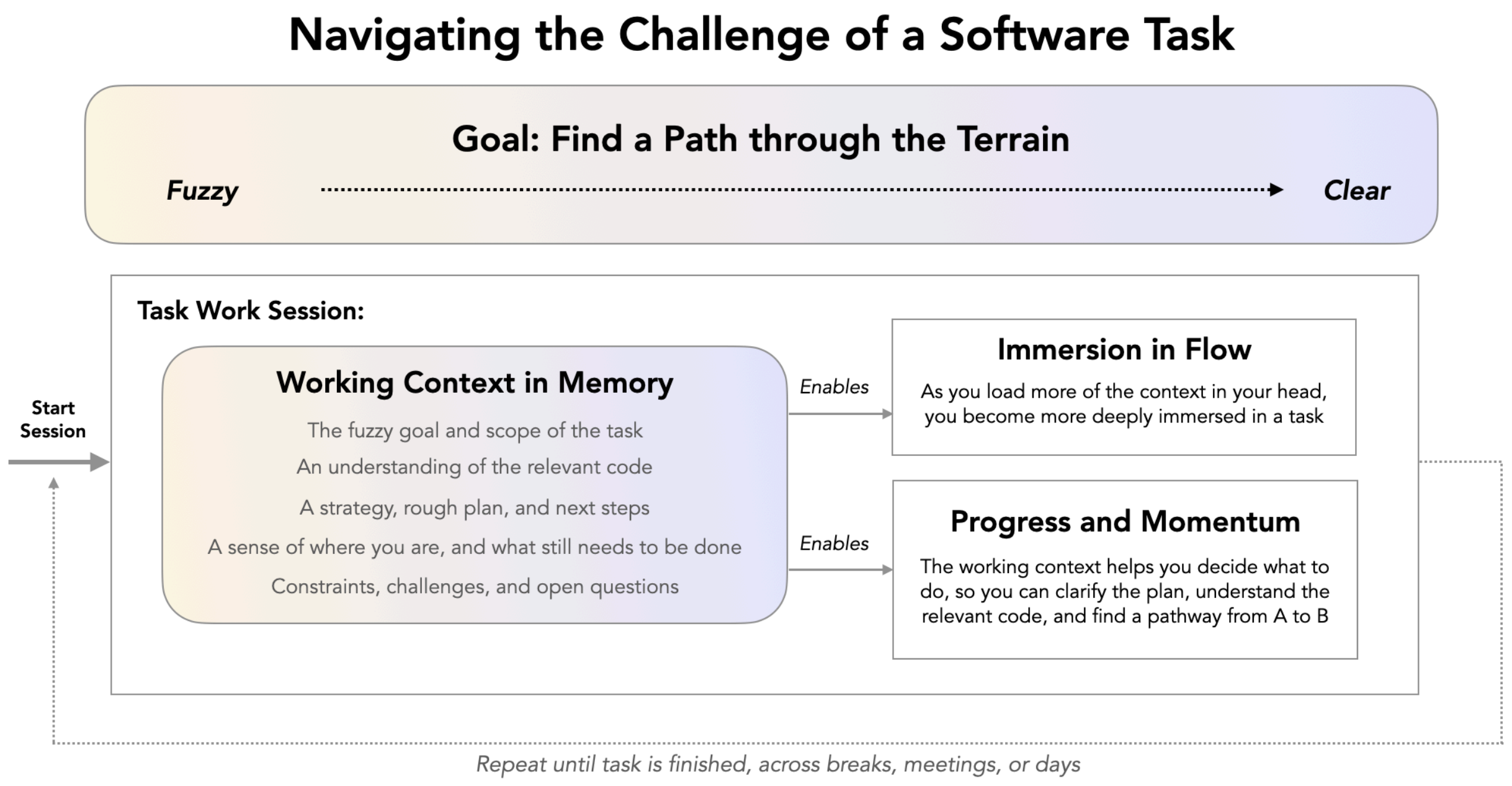


3.2. Do you feel like this resonates with you? Is there anything here that stands out to you that feels inaccurate or that you would change?

3.3. On a scale from 1 to 5, one being it doesn't resonate and feels very different than how you would think about it, and 5 being it resonates and fits with your own mental model, how would you rate this on a scale from 1 to 5?

**4.1. Immersion in Flow Experience -** So next diagram, if you want to just take a look and tell me what you see represented...

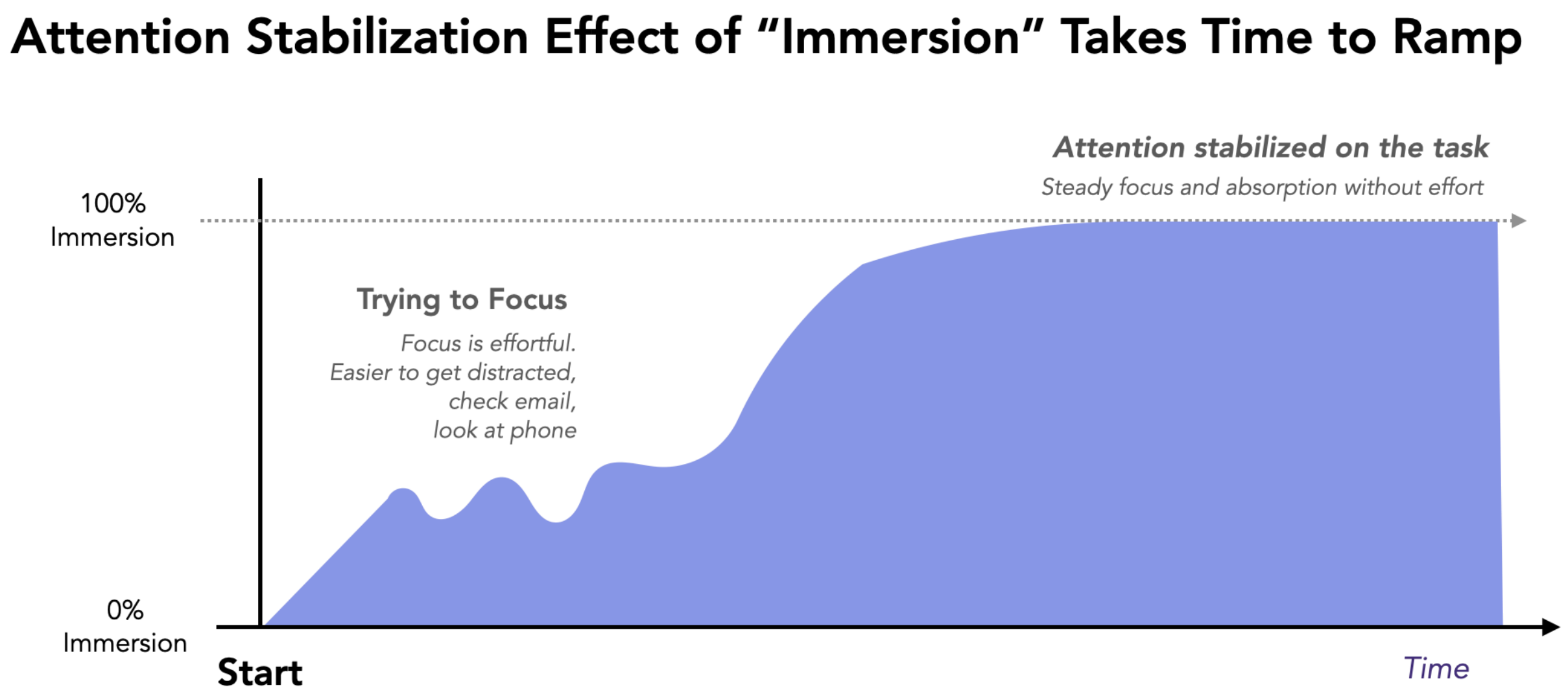

4.2. Do you feel like this resonates with you?  Is there anything here that stands out to you that feels inaccurate or that you would change?

4.3. On a scale from 1 to 5, one being it doesn't resonate and feels very different than how you would think about it, and 5 being it resonates and fits with your own mental model, how would you rate this on a scale from 1 to 5?

**5.1. Flow Depth Experience** - So next diagram, if you want to just take a look and tell me what you see represented…

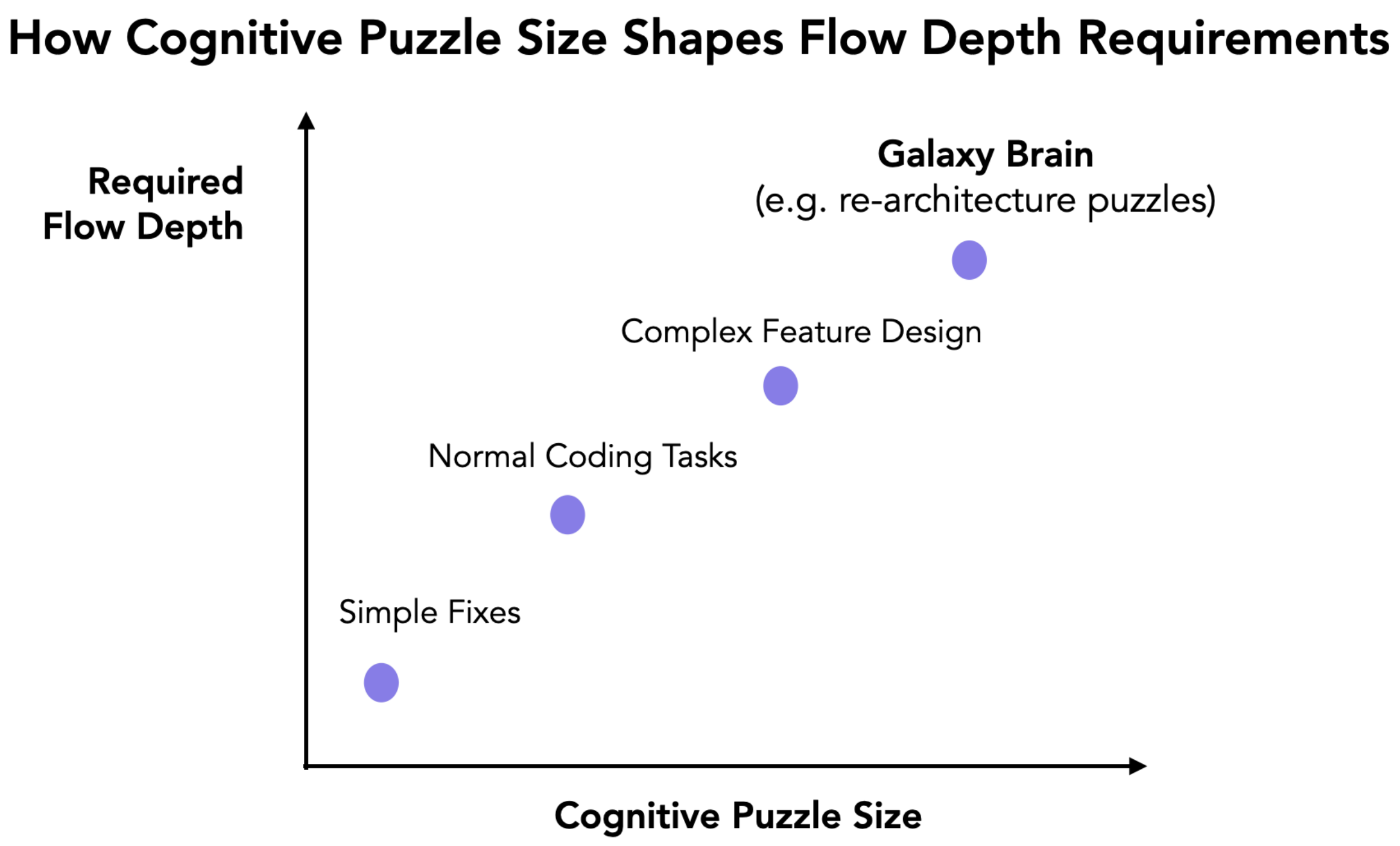


5.2. Do you feel like this resonates with you?  Is there anything here that stands out to you that feels inaccurate or that you would change?

5.3. On a scale from 1 to 5, one being it doesn't resonate and feels very different than how you would think about it, and 5 being it resonates and fits with your own mental model, how would you rate this on a scale from 1 to 5?

**6.1. Motivation Experience of Care-** So next diagram, if you want to just take a look and tell me what you see represented...

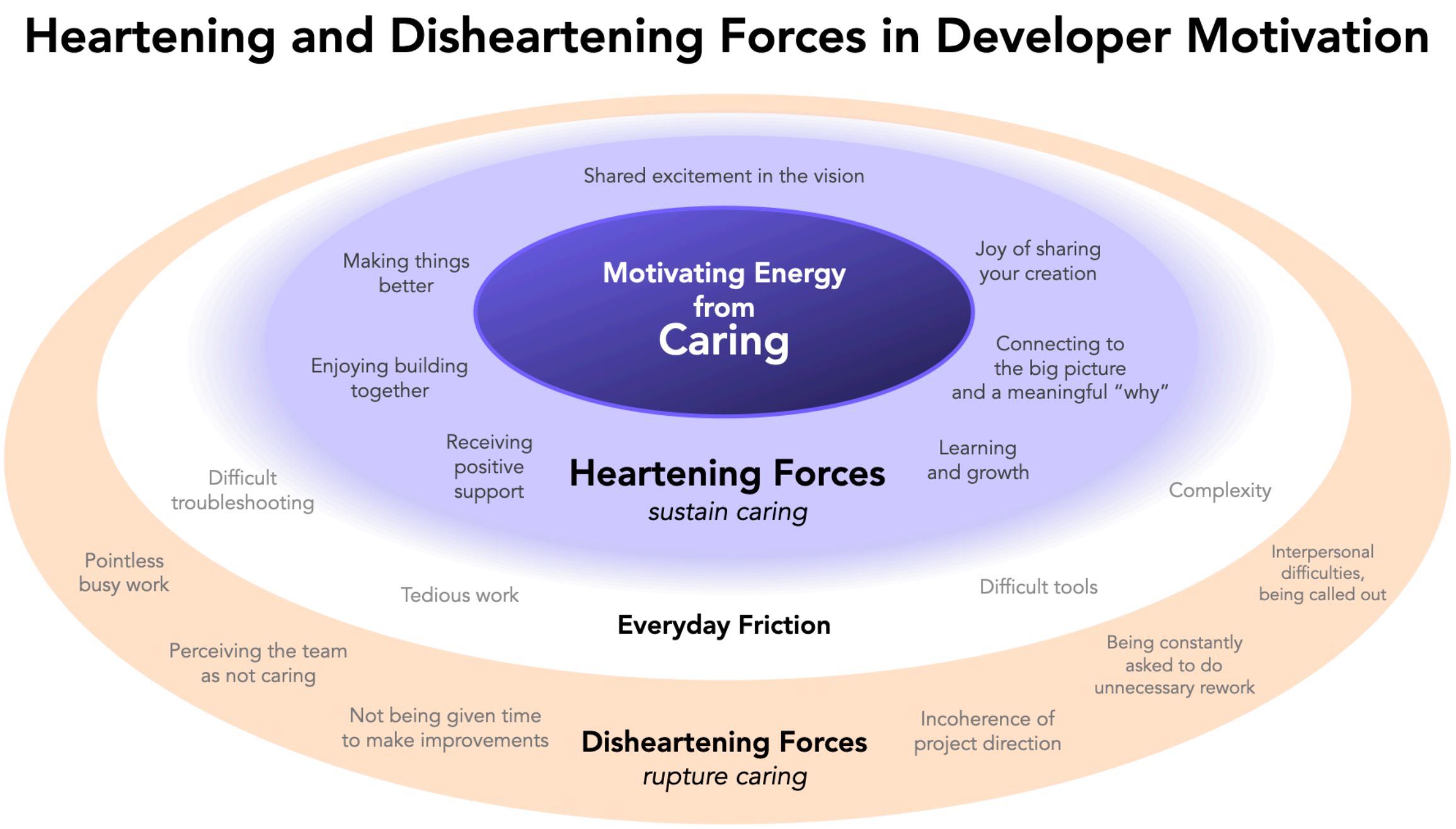


6.2. Do you feel like this resonates with you? Is there anything here that stands out to you that feels inaccurate or that you would change?

6.3. On a scale from 1 to 5, one being it doesn't resonate and feels very different than how you would think about it, and 5 being it resonates and fits with your own mental model, how would you rate this on a scale from 1 to 5?

**Final Reflections**

7. Do you have any final thoughts after seeing all the diagrams, or any other reflections?

**Final Wrap Up**

*So that concludes our follow up interview, thank you for participating in the study. We will securely process the recording and the transcript. If at any point in time, you decide you do not want to participate in the study, or if you have any questions, you can email us.*

*Our next step is to refine the models as needed until they resonate with the developers, and to finish the academic paper writeup of the findings. Would you be interested in reading the academic paper for the study when it's available? Would you be open to me emailing you about participating in another short interview if it's necessary to complete the study? For example, if we do another iteration on the diagrams you saw today, and we needed to get more feedback.*

*Do you have any final questions for us or any further comments?*

*Alright, thank you very much for your time, and have a great day.*



\end{document}